\documentclass[12pt]{article}

\usepackage{tablefootnote}
\usepackage{hhline}  
\usepackage{jheppub}
\usepackage{pstricks}
\usepackage{color}
\usepackage{amssymb,amsmath,bm,bbm}
\usepackage{epsf}
\usepackage{epsfig}
\usepackage{afterpage}
\usepackage{longtable}
\usepackage{latexsym,mathrsfs,dsfont}
\usepackage{graphics}
\usepackage{url}
\usepackage{paralist}
\usepackage{bbold}
\usepackage{hyperref}
\usepackage{diagbox}
\usepackage{multirow}
\usepackage{placeins}

\newcommand{\vcb}{|V_{cb}|}

\def\epe{\varepsilon'/\varepsilon}
\newcommand{\beq}{\begin{equation}}
\newcommand{\eeq}{\end{equation}}
\newcommand{\be}{\begin{equation}}
\newcommand{\ee}{\end{equation}}
\newcommand{\bi}{\begin{itemize}}
\newcommand{\ei}{\end{itemize}}
\newcommand{\ba}{\begin{array}}
\newcommand{\ea}{\end{array}}
\newcommand{\beqa}{\begin{eqnarray}}
\newcommand{\eeqa}{\end{eqnarray}}
\newcommand{\bea}{\begin{eqnarray}}
\newcommand{\eea}{\end{eqnarray}}
\newcommand{\beqn}{\begin{eqnarray}}
\newcommand{\eeqn}{\end{eqnarray}}

\newcommand{\re}{{\rm Re}}
\newcommand{\im}{{\rm Im}}

\definecolor{red}{cmyk}{0,1,1,0.4}

\usepackage{fancyhdr}
\advance \headheight by 3.0truept       

\title{\large\bf\boldmath{On the Interplay of Constraints from $B_{s}$, $D$, and $K$ Meson Mixing in $Z^\prime$ Models with Implications for $b\!\to\! s \nu\bar\nu$ Transitions}}

\author[a,b]{Andrzej~J.~Buras}
\author[c]{and Peter Stangl}

\affiliation[a]{TUM Institute for Advanced Study, Lichtenbergstr. 2a, D-85747 Garching, Germany}

\affiliation[b]{Physik Department, TUM School of Natural Sciences, TU M\"unchen,\\ James-Franck-Stra{\ss}e, D-85748 Garching, Germany}

\affiliation[c]{CERN, Theoretical Physics Department, CH-1211 Geneva 23, Switzerland}

\preprint{AJB-24-3}
\preprintadd{CERN-TH-2024-218}

\abstract{
  Within $Z^\prime$ models, neutral meson mixing severely constrains beyond the Standard Model (SM) effects in flavour changing neutral current (FCNC) processes.
  However, in certain regions of the $Z^\prime$ parameter space, the contributions to meson mixing observables become negligibly small even for large $Z^\prime$ couplings.
  While this a priori allows for significant new physics (NP) effects in FCNC decays, we discuss how large $Z^\prime$ couplings in one neutral meson sector can generate effects in meson mixing observables of other neutral mesons, through correlations stemming from $\text{SU(2)}_L$ gauge invariance and through Renormalization Group (RG) effects in the SM Effective Field Theory~(SMEFT).
  This is illustrated with the example of $B_s^0-\bar B_s^0$ mixing, which in the presence of both left- and right-handed $Z^\prime bs$ couplings $\Delta_L^{bs}$ and $\Delta_R^{bs}$ remains SM-like for $\Delta_R^{bs}\approx 0.1\,\Delta_L^{bs}$.
  We show that in this case, large $Z^\prime bs$ couplings generate effects in $D$ and $K$ meson mixing observables, but that the $D$ and $K$ mixing constraints and the relation between $\Delta_R^{bs}$ and $\Delta_L^{bs}$ are fully compatible with a lepton flavour universality~(LFU) conserving explanation of the most recent $b\to s\ell^+\ell^-$ experimental data without violating other constraints like $e^+ e^-\to\ell^+\ell^-$ scattering.
    Assuming LFU, invariance under the $\text{SU(2)}_L$ gauge symmetry leads then to correlated effects in $b\to s\nu\bar\nu$ observables presently studied intensively by the Belle~II experiment, which allow to probe the $Z^\prime$ parameter space that is opened up by the vanishing NP contributions to $B_s^0-\bar B_s^0$ mixing. In this scenario the  suppression of $B\to K(K^*)\mu^+\mu^-$
      branching ratios implies {\em uniquely}  enhancements of
$B\to K(K^*)\nu\bar\nu$ branching ratios up to $20\%$.
}
    
\begin{document}

\maketitle
\newpage

\section{Introduction}

In the search for flavourful new physics (NP), processes involving flavour-changing neutral currents (FCNCs) are of particular interest, as they are loop and CKM suppressed in the Standard Model (SM) and therefore highly sensitive to NP effects.
An important class of models that can generate FCNCs at tree-level are $Z^\prime$ models, in which the SM is accompanied by a new neutral vector boson.
Within $Z^\prime$ models, the same flavour-changing quark couplings that enter $\Delta F=1$ FCNCs also give rise to tree-level contributions to $\Delta F=2$ meson mixing observables.
As these can be measured with high precision, they impose stringent constraints on the parameter space of $Z^\prime$ models and limit their potential effects in FCNCs.
However, as emphasized in \cite{Buras:2014sba,Buras:2014zga,Crivellin:2015era}, the NP contributions to
$\Delta F=2$ processes are suppressed for a particular pattern of left-handed~$\Delta_L^{q_1 q_2}$ and right-handed~$\Delta_R^{q_1 q_2}$ flavour-violating $Z^\prime$ couplings to quarks $q_1$ and $q_2$.
The basic quantity for studying the region of $Z^\prime$ parameter space that features this suppression is the ratio $r_{q_1 q_2}$ of left-handed and right-handed couplings defined by
\begin{equation}\label{eq:r_q1q2}
 r_{q_1 q_2} = \frac{\Delta_R^{q_1 q_2}}{\Delta_L^{q_1 q_2}}
\end{equation}
For example, in the case of $B_s^0-\bar B_s^0$ mixing, a suppression of NP contributions is effective for $r_{bs}\approx 0.1$, while NP in $K^0-\bar K^0$ mixing is suppressed for $r_{sd}\approx 0.004$ and the one in $D^0-\bar D^0$ mixing
for $r_{uc}\approx 0.05$.
  In such cases, large flavour-changing $Z^\prime$ couplings to the quarks $q_1$ and $q_2$ are possible, allowing for potentially sizeable effects in FCNCs.
However, as we emphasize in this paper, the NP contributions to the $\Delta F=2$ observables in different meson sectors are not independent of each other, but are correlated.
In particular, $\text{SU(2)}_L$ gauge invariance and CKM mixing link $D^0-\bar D^0$ mixing with $B_s$, $B_d$, and kaon mixings, and Renormalization Group~(RG) effects in the SM effective field theory (SMEFT) link all meson mixing sectors to each other.
While these effects are usually negligible, since they depend on small CKM elements and loop factors, they become relevant in the case of large $Z^\prime$-$q_1$-$q_2$ couplings, which are possible due to the above described suppression of NP contributions in a given meson mixing sector.

To illustrate these dynamics, we consider a $Z^\prime$ scenario with negligible contributions to $B_s^0-\bar B_s^0$ mixing due to $r_{bs}\approx 0.1$.
This example is interesting as it allows for potentially large contributions to $b\to s\ell^+\ell^-$ and $b\to s \nu\bar\nu$ FCNC processes, which are otherwise strongly constrained in $Z^\prime$ models.
The former processes are at the centre of the so-called $B$-physics anomalies, experimental data of $b\to s\mu^+\mu^-$ transitions in disagreement with SM predictions.
While $Z^\prime$ models have been among the prime candidates to explain these anomalies, the most recent measurements of $b\to s\ell^+\ell^-$ lepton flavour universality violation~(LFUV) by the LHCb collaboration \cite{LHCb:2022qnv,LHCb:2022zom} pose serious challenges for these models.
Indeed, the presence of lepton flavour universality~(LFU) in $b\to s\ell^+\ell^-$ transitions requires $Z^\prime$ couplings to electrons that are constrained by LEP-2 measurements of $e^+e^-\to\ell^+\ell^-$ scattering~\cite{Greljo:2022jac}, so that sizeable $Z^\prime bs$ couplings are required to explain the $b\to s\mu^+\mu^-$ data.
The $b\to s \nu\bar\nu$ processes, on the other hand, are presently studied intensively through the $B\to K^{(*)}\nu\bar\nu$ decays by the Belle~II experiment~\cite{Belle-II:2023esi}, providing valuable complementary information giving some
hints for NP contributions. For selected recent analyses of these data see
\cite{Bause:2021cna,He:2021yoz,Bause:2022rrs,Becirevic:2023aov,Bause:2023mfe,Allwicher:2023xba,Dreiner:2023cms, Altmannshofer:2023hkn, Gabrielli:2024wys, Hou:2024vyw, He:2024iju, Bolton:2024egx, Marzocca:2024hua,Allwicher:2024ncl,McKeen:2023uzo,Hati:2024ppg,Ho:2024cwk,Altmannshofer:2024kxb}.

The large $Z^\prime bs$ couplings allowed by $r_{bs}\approx 0.1$ and necessary for sizeable effects in $b\to s\ell^+\ell^-$ and $b\to s \nu\bar\nu$ transitions have profound effects on other meson mixing sectors.
In particular, they lead to NP contributions to $D$ and $K$ meson mixing observables.
We study these correlated effects in detail and investigate their implications for a simultaneous  explanation of $B\to K(K^*)\mu^+\mu^-$, $B_{s}\to\mu^+\mu^-$ and the $B\to K(K^*)\nu\bar\nu$ decays experimental data
  with the latter investigated by the Belle~II.

  Our paper is organized as follows. In Section~\ref{sec:meson_mixing}
    we present in detail the steps from the NP scale $\Lambda_\text{NP}$ at which
    the $Z^\prime$ is integrated out down to hadronic scales at which the
    relevant decay amplitudes are evaluated. These steps include the RG evolutions in the SMEFT and the WET, the matching between the $Z^\prime$ model and the SMEFT and the matching between the SMEFT and the WET. We present the structure of the
    meson mixing amplitudes for $B_{s,d}$, $K$ and  $D$  mesons which allows us
    to determine for each of these systems the relations between left-handed
    and right-handed $Z^\prime$ couplings to quarks that allow to suppress $Z^\prime$
    contributions to mixing amplitudes. Subsequently we discuss the
    correlations between different mixing amplitudes implied by the
    $\text{SU(2)}_L$ gauge invariance and the mixing between various operators in the process of the SMEFT RG evolution. We also perform a numerical
      analysis of the suppression of $Z^\prime$ contributions to $B_s-\bar B_s$
      mixing and discuss their implications for other meson systems.

    In Section~\ref{sec:DeltaB=DeltaS} we present analogous steps for semi-leptonic transitions, in particular for $B\to K(K^*)\nu\bar\nu$, $B\to K(K^*)\mu^+\mu^-$ and $B_{s}\to\mu^+\mu^-$. We summarize their present experimental and theoretical status and discuss the correlations between them that follow from the $\text{SU(2)}_L$ gauge symmetry and the mixing between involved operators implied by the     SMEFT RG evolution from $\Lambda_\text{NP}$ down to the electroweak scale.     We perform a detailed numerical analysis. The outcome of this analysis, summarized at the end of this section constitutes one of the most important results of our paper.

    A brief summary of our paper is given in Section~\ref{sec:summary}.
    Several technical details are presented in appendices. In Appendix~\ref{app:epsK} we update {the} SM prediction for $\varepsilon_K$, analyzing  its dependence on $\hat B_K$, $\vcb$ and $\gamma$. In Appendix~\ref{app:matrixelements}
  we    summarize the $\Delta F=2$ matrix elements. In Appendix~\ref{app:rediag} re-diagonalization of the running quark Yukawa matrices is performed. In Appendix~\ref{app:SMEFT_matching} the matching of the simplified $Z^\prime$ model to the SMEFT is presented in detail, and in Appendix~\ref{AppC} SMEFT RGE in $\Delta F=2$ coefficients for all meson mixings in down- and up-bases are listed.

\boldmath
\section{Interplay of Meson Mixing Constraints}\label{sec:meson_mixing}
\unboldmath
This section can be considered as the anatomy of the top-down approach
  illustrated on the example of $Z^\prime$ models discussed by us. Usually these
  steps are hidden in computer codes but it is useful to exhibit them in explicit terms.
  \begin{itemize}
  \item
    We begin in Section~\ref{step1} at the high scale $\Lambda_\text{NP}$ of the order of $M_{Z^\prime}$ and define $Z^\prime$ couplings to quarks in various flavour bases.
  \item
    Next in Section~\ref{sec:ZpSMEFTmatching} we perform the matching of the $Z^\prime$ to the
    SMEFT by integrating out $Z^\prime$. This results in the WCs of the relevant
    operators at the NP scale~$\Lambda_\text{NP}$. Subsequently RG group     evolution from  $\Lambda_\text{NP}$ down to the electroweak scale~$M_Z$ is performed within the SMEFT.
  \item
    Having the WCs of the  SMEFT operators at the $M_Z$ scale we perform in Section~\ref{sec:SMEFTWETmatching} the maching     of the SMEFT to the WET so that the WCs of the
    WET are known at the electroweak scale. Subsequently RG group     evolution from  $M_Z$ down to hadronic scales within the WET is performed.
  \item
    Having the  WET WCs at the hadrionc scales we are in the position to
    calculate meson mixing amplitudes in Section~\ref{step4}.
  \item
    Subsequently in Section~\ref{sec:meson_mixing_suppression} we discuss the suppression of NP
    to the mixing amplitudes.
  \item
    In Section~\ref{sec:SU2_corr} we elaborate on the correlations between different
    meson systems due to the $\text{SU(2)}_L$ invariance.
  \item
    In Section~\ref{sec:SMEFTRGeffects} we discuss in some details the correlations
    between different meson systems resulting from the SMEFT RG evolution.
  \item
    Finally, in Section~\ref{sec:fit_df2} we perform a numerical
      analysis of the suppression of $Z^\prime$ contributions to $B_s-\bar B_s$
      mixing  and discuss their implications for other meson systems.
  \end{itemize}

\subsection{The $Z^\prime$ Model}\label{step1}

The couplings of the $Z^\prime$ to the SM fermions that are relevant for FCNC decays and meson mixing are given by
\begin{equation}\label{eq:Zp_couplings}
\begin{aligned}
 \mathcal{L}
 &\supset
 \sum_{i, j}\left(
 \tilde\Delta^{q_i q_j}_L\,Z^\mu\,\bar{q}_i\,\gamma_\mu\,q_j
 +
 \Delta^{d_i d_j}_R\,Z^\mu\,\bar{d}_i\,\gamma_\mu\,d_j
 +
 \Delta^{u_i u_j}_R\,Z^\mu\,\bar{u}_i\,\gamma_\mu\,u_j
 \right)\\
 &+
 \sum_k
 \left(
 \Delta^{l_k}_L\,Z^\mu\,\bar{l}_k\,\gamma_\mu\,l_k
 +
 \Delta^{e_k}_R\,Z^\mu\,\bar{e}_k\,\gamma_\mu\,e_k
 \right)\,,
\end{aligned}
\end{equation}
where $q$, $d$, $u$, $l$, and $e$ denote the left-handed quark doublets, right-handed down-type singlets, right-handed up-type singlets, left-handed lepton doublets, and right-handed lepton singlets, respectively, and $i,j,k\in 1,2,3$ are flavour indices corresponding to the three generations of SM fermions.
We are particularly interested in quark flavour changing couplings (i.e.\ $i\neq j$) and we only consider lepton flavour conserving interactions.

While we define all right-handed fields in the mass basis, and the left-handed lepton doublet in the mass basis of the charged leptons, the left-handed quark doublet and the couplings $\tilde\Delta^{q_i q_j}_L$ are defined in an arbitrary basis in flavour space, in which in general neither of the two components of the left-handed quark doublet is in the mass basis.
In this basis, the quark Yukawa matrices can be expressed as
\begin{equation}\label{eq:Yukawas_left-agnostic}
 Y_u = U_{u_L} Y_u^{\rm diag}\,,
 \qquad
 Y_d = U_{d_L} Y_d^{\rm diag}\,,
\end{equation}
where $Y_u^{\rm diag}$, $Y_d^{\rm diag}$ are diagonal matrices and $U_{u_L}$, $U_{d_L}$ are unitary matrices.\footnote{%
Note that the two additional unitary matrices $U_{u_R}$ and $U_{d_R}$ that would multiply the diagonal matrices from the right in a completely arbitrary basis are absent since we define all right-handed fields in the mass basis.}
Furthermore, the quark doublet expressed in terms of mass-eigenstates $u_L$ and $d_L$ takes the form
\begin{equation}\label{eq:qdoublet_left-agnostic}
 q = \begin{pmatrix}
      U_{u_L}^\dagger\,u_L\\
      U_{d_L}^\dagger\,d_L
     \end{pmatrix}\,.
\end{equation}
We can perform a unitary transformation in the flavour space of the quark doublets that turns one doublet component into a mass eigenstate, but since $U_{u_L}\neq U_{d_L}$ this cannot be done for both components simultaneously.

Starting from Eqs.~\eqref{eq:Yukawas_left-agnostic} and \eqref{eq:qdoublet_left-agnostic} and performing the transformation
\begin{equation}\label{eq:down_aligned_trafo}
 q \to U_{d_L} q\,,
\end{equation}
we end up in the so-called \emph{down-aligned} basis, in which we have
\begin{equation}
 Y_u = V_{\rm CKM}^\dagger\, Y_u^{\rm diag}\,,
 \qquad
 Y_d = Y_d^{\rm diag}\,,
 \quad
 \text{and}
 \quad
 q = \begin{pmatrix}
      V_{\rm CKM}^\dagger\,u_L\\
      d_L
     \end{pmatrix}\,,
\end{equation}
where $V_{\rm CKM} = U_{u_L}^\dagger U_{d_L}$ is the Cabibbo–Kobayashi–Maskawa (CKM) matrix.
In the down-aligned basis, the down-type Yukawa matrix is diagonal and the down-type component of the quark doublet is in the mass basis.
Applying the transformation in Eq.~\eqref{eq:down_aligned_trafo} to the $Z^\prime$ interactions in Eq.~\eqref{eq:Zp_couplings}, we find that the couplings $\tilde\Delta^{q_i q_j}_L$ transform as
\begin{equation}\label{eq:Deltaq_downaligned}
 \tilde\Delta^{q_i q_j}_L \to \Delta^{q_i q_j}_L =  \big(U_{d_L}^\dagger\big)_{ik}\, \tilde\Delta^{q_k q_l}_L\, \big(U_{d_L}\big)_{lj}\,,
\end{equation}
where we denote the couplings of $Z^\prime$ and left-handed quark doublets in the down-aligned basis by $\Delta^{q_i q_j}_L$.
From this equation, it is obvious that $\tilde\Delta^{q_i q_j}_L$ is invariant under the basis change only if it commutes with $U_{d_L}$, which is in particular the case when $\tilde\Delta^{q_i q_j}_L$ is flavour-conserving \emph{and} universal, i.e.\ if it is proportional to the unit matrix. Otherwise, e.g.\ if $\tilde\Delta^{q_i q_j}_L$ is a diagonal, apparently flavour-conserving matrix, but its diagonal entries are non-universal, this basis change reveals off-diagonal flavour changing couplings of $Z^\prime$ and the left-handed (mass eigenstate) down-type quarks.
In the down-aligned basis, flavour-changing interactions of left-handed down-type quarks are in one-to-one correspondence with the off-diagonal terms in their coupling matrices. E.g. flavour-changing $Z^\prime$ interactions between $b$ and $s$ quarks are always present for $\Delta^{q_2 q_3}_L\neq 0$ and always absent for $\Delta^{q_2 q_3}_L=0$, and similarly for the other down-type quarks.
This makes the down-aligned basis particularly convenient for studying flavour-changing processes of down-type quarks.

Alternatively, starting from Eqs.~\eqref{eq:Yukawas_left-agnostic} and \eqref{eq:qdoublet_left-agnostic}, we can performing the transformation
\begin{equation}\label{eq:up_aligned_trafo}
 q \to U_{u_L} q\,,
\end{equation}
and we end up in the \emph{up-aligned} basis, in which we have
\begin{equation}
 Y_u = Y_u^{\rm diag}\,,
 \qquad
 Y_d = V_{\rm CKM}\, Y_d^{\rm diag}\,,
 \quad
 \text{and}
 \quad
 q = \begin{pmatrix}
      u_L\\
      V_{\rm CKM}\,d_L
     \end{pmatrix}\,.
\end{equation}
For the $Z^\prime$ interactions, we find
\begin{equation}\label{eq:Deltaq_upaligned}
 \tilde\Delta^{q_i q_j}_L \to \hat\Delta^{q_i q_j}_L =  \big(U_{u_L}^\dagger\big)_{ik}\, \tilde\Delta^{q_k q_l}_L\, \big(U_{u_L}\big)_{lj}\,,
\end{equation}
where we denote the couplings of $Z^\prime$ and left-handed quark doublets in the up-aligned basis by $\hat\Delta^{q_i q_j}_L$.
This basis is convenient for studying flavour-changing processes of up-type quarks, whose flavour-changing interactions are in a one-to-one correspondence to the off-diagonal components of $\hat\Delta^{q_i q_j}_L$.

We will mostly work in the down-aligned basis, but change to the up-aligned basis when studying $D$ meson mixing. To this end, it is convenient to obtain relations between objects in the up-aligned basis, which we decorate with a hat, and the corresponding objects in the down-aligned basis (without hat). In particular, from Eqs.~\eqref{eq:Deltaq_downaligned} and \eqref{eq:Deltaq_upaligned}, we find
\begin{equation}\label{eq:LH_Zp_relation}
 \hat\Delta^{q_i q_j}_L = \big(V_{\rm CKM}\big)_{ik}\,\Delta^{q_k q_l}_L\,\big(V_{\rm CKM}^\dagger\big)_{lj}\,.
\end{equation}
This shows that unless $\Delta^{q_k q_l}_L$ is proportional to the unit matrix, the CKM matrix induces flavour changing effects for up-type quarks if flavour changing couplings are absent for down-type quarks, and vice versa.

{
In the following, we will often use a convenient notation for the left-handed couplings in the up- and down-aligned bases using the mass-eigenstate doublet components as indices,
\begin{equation}\label{eq:Zp_couplings_WET_notation}
\hat\Delta_L^{uc} = \hat\Delta_L^{q_1 q_2},\qquad
\Delta_L^{ds} = \Delta_L^{q_1 q_2},\qquad
\Delta_L^{db} = \Delta_L^{q_1 q_3},\qquad
\Delta_L^{sb} = \Delta_L^{q_2 q_3}.
\end{equation}
This notation will be in particular convenient in the Weak Effective Theory (WET) below the electroweak scale, in which $q_1$ and $q_2$ do not denote quark doublets, but are used as placeholders for any quark mass eigenstate (cf.\ footnote~\ref{fnt:q1q2}).
}

\subsection{Matching to the SMEFT and RG Evolution in the SMEFT}\label{sec:ZpSMEFTmatching}

The tree-level matching of the model defined by \eqref{eq:Zp_couplings} onto the SMEFT results in contributions to Wilson coefficients (WCs) of three classes of four-fermion dimension-six operators~\cite{deBlas:2017xtg}, which we express in the non-redundant Warsaw basis~\cite{Grzadkowski:2010es} defined by the WC exchange format (WCxf)~\cite{Aebischer:2017ugx},

\begin{itemize}
 \item flavour-violating four-quark operators coupling $i$th and $j$th generation quarks,
\item flavour-conserving four-lepton operators,
 \item quark flavour violating semi-leptonic operators.
\end{itemize}
The full tree-level matching expressions relating the couplings in Eq.~\eqref{eq:Zp_couplings} to the WCs of these operator classes are given in Appendix~\ref{app:SMEFT_matching}.
Here, we only list the matching relations of the flavour-violating four-quark operators, which are relevant for meson mixing observables,

\begin{equation}\label{eq:SMEFT_matching_4q}
\begin{aligned}{}
 [C_{qq}^{(1)}]_{ijij} &= -\frac{\big(\Delta^{q_i q_j}_L\big)^2}{2\,M_{Z'}^2}\,,
 \qquad
 [C_{qq}^{(1)}]_{ijji}  = -\frac{\big|\Delta^{q_i q_j}_L\big|^2}{M_{Z'}^2}\,,
 \\
 [C_{dd}]_{ijij} &= -\frac{\big(\Delta^{d_i d_j}_R\big)^2}{2\,M_{Z'}^2}\,,
 \qquad
 [C_{dd}]_{ijji}  = -\frac{\big|\Delta^{d_i d_j}_R\big|^2}{M_{Z'}^2}\,,
 \\
 [C_{uu}]_{ijij} &= -\frac{\big(\Delta^{u_i u_j}_R\big)^2}{2\,M_{Z'}^2}\,,
 \qquad
 [C_{uu}]_{ijji}  = -\frac{\big|\Delta^{u_i u_j}_R\big|^2}{M_{Z'}^2}\,,
 \\
 [C_{qd}^{(1)}]_{ijij} &= -\frac{\Delta^{q_i q_j}_L\,\Delta^{d_i d_j}_R}{M_{Z'}^2}\,,
 \qquad
 [C_{qd}^{(1)}]_{ijji}  = -\frac{\Delta^{q_i q_j}_L\,\big(\Delta^{d_i d_j}_R\big)^*}{M_{Z'}^2}\,,
 \\
 [C_{qu}^{(1)}]_{ijij} &= -\frac{\Delta^{q_i q_j}_L\,\Delta^{u_i u_j}_R}{M_{Z'}^2}\,,
 \qquad
 [C_{qu}^{(1)}]_{ijji}  = -\frac{\Delta^{q_i q_j}_L\,\big(\Delta^{u_i u_j}_R\big)^*}{M_{Z'}^2}\,,
 \\
 [C_{ud}^{(1)}]_{ijij} &= -\frac{\Delta^{u_i u_j}_R\,\Delta^{d_i d_j}_R}{M_{Z'}^2}\,,
 \qquad
 [C_{ud}^{(1)}]_{ijji}  = -\frac{\Delta^{u_i u_j}_R\,\big(\Delta^{d_i d_j}_R\big)^*}{M_{Z'}^2}\,,
\end{aligned}
\end{equation}
where $i < j$.
In the case of real couplings $\Delta_{L,R}$, we thus find the relations
\begin{equation}
\begin{aligned}{}
 [C_{qq}^{(1)}]_{ijij} &= \frac{1}{2}\,[C_{qq}^{(1)}]_{ijji}\,,
 \qquad&
 [C_{dd}]_{ijij} &= \frac{1}{2}\,[C_{dd}]_{ijji}\,,
 \qquad&
 [C_{uu}]_{ijij} &= \frac{1}{2}\,[C_{uu}]_{ijji}\,,
 \\
 [C_{qd}^{(1)}]_{ijij} &= [C_{qd}^{(1)}]_{ijji}\,,
 \qquad&
 [C_{qu}^{(1)}]_{ijij} &= [C_{qu}^{(1)}]_{ijji}\,,
 \qquad&
 [C_{ud}^{(1)}]_{ijij} &= [C_{ud}^{(1)}]_{ijji}\,.
\end{aligned}
\end{equation}
The relations in Eq.~\eqref{eq:SMEFT_matching_4q} hold at the UV matching scale $\Lambda_{NP}\approx M_{Z^\prime}$. In order to obtain the predictions for meson mixing observables, we use the SMEFT RG Equations (RGEs) to evolve the WCs down to the electroweak scale, where we match the SMEFT to the WET, which we then evolve further down using the WET RGEs to the scale at which the meson mixing matrix elements are evaluated.
The  dominant contributions to the SMEFT RG running and mixing of the Wilson coefficients in Eq.~\eqref{eq:SMEFT_matching_4q} in the first leading-log approximation {are} given by~\cite{Jenkins:2013zja,Jenkins:2013wua,Alonso:2013hga}
\begin{equation}\label{eq:SMEFT_RG_dominant}
 \begin{aligned}
      {[}{C}_{qq}^{(1)}]_{ijij}(M_Z)
    \approx&\
    [C_{qq}^{(1)}]_{ijij}(\Lambda_\text{NP})
    \,
    \times
    \left[1+
     [\beta_{qq^{(1)}}^{qq^{(1)}}]_{ij}
    \,\frac{\log\left(M_Z/\Lambda_\text{NP}\right)}{16 \pi^2}
    \right]\,,
    \\
    {[}{C}_{qq}^{(3)}]_{ijij}(M_Z)
    \approx&\
    [C_{qq}^{(1)}]_{ijij}(\Lambda_\text{NP})
    \,
    \times
    \left[
     [\beta_{qq^{(3)}}^{qq^{(1)}}]
    \,
    \frac{\log\left(M_Z/\Lambda_\text{NP}\right)}{16 \pi^2}
    \right]\,,
    \\
        {[}{C}_{dd}]_{ijij}(M_Z)
    \approx&\
    [C_{dd}]_{ijij}(\Lambda_\text{NP})
    \,
    \times
    \left[1+
     [\beta_{dd}^{dd}]_{ij}
    \,\frac{\log\left(M_Z/\Lambda_\text{NP}\right)}{16 \pi^2}
    \right]\,,
    \\
       {[}{C}_{qd}^{(1)}]_{ijij}(M_Z)
    \approx&\
    [C_{qd}^{(1)}]_{ijij}(\Lambda_\text{NP})
    \,
    \times
    \bigg[1+
    [\beta_{qd^{(1)}}^{qd^{(1)}}]_{ij}
    \,\frac{\log\left(M_Z/\Lambda_\text{NP}\right)}{16 \pi^2}
    \bigg]\,,
    \\
    {[}{C}_{qd}^{(8)}]_{ijij}(M_Z)
    \approx&\
    [C_{qd}^{(1)}]_{ijij}(\Lambda_\text{NP})
    \,
    \times
    \bigg[
    [\beta_{qd^{(8)}}^{qd^{(1)}}]_{ij}
    \,\frac{\log\left(M_Z/\Lambda_\text{NP}\right)}{16 \pi^2}
    \bigg]\,,
    \\
    {[}{C}_{uu}]_{ijij}(M_Z)
    \approx&\
    [C_{uu}]_{ijij}(\Lambda_\text{NP})
    \,
    \times
    \left[1+
     [\beta_{uu}^{uu}]_{ij}
    \,\frac{\log\left(M_Z/\Lambda_\text{NP}\right)}{16 \pi^2}
    \right]\,,
    \\
    {[}{C}_{qu}^{(1)}]_{ijij}(M_Z)
    \approx&\
    [C_{qu}^{(1)}]_{ijij}(\Lambda_\text{NP})
    \,
    \times
    \bigg[1+
    [\beta_{qu^{(1)}}^{qu^{(1)}}]_{ij}
    \,\frac{\log\left(M_Z/\Lambda_\text{NP}\right)}{16 \pi^2}
    \bigg]\,,
    \\
    {[}{C}_{qu}^{(8)}]_{ijij}(M_Z)
    \approx&\
    [C_{qu}^{(1)}]_{ijij}(\Lambda_\text{NP})
    \,
    \times
    \bigg[
    [\beta_{qu^{(8)}}^{qu^{(1)}}]_{ij}
    \,\frac{\log\left(M_Z/\Lambda_\text{NP}\right)}{16 \pi^2}
    \bigg]\,,
 \end{aligned}
\end{equation}
where we have defined
\begin{equation}
 \begin{aligned}
{[}\beta_{qq^{(1)}}^{qq^{(1)}}]_{ij} &=
    \Big(
      g_s^2 +\tfrac{1}{3}\,g^{\prime 2} + 2\,[\gamma^{(Y)}_{q}]_{ii}+2\,[\gamma^{(Y)}_{q}]_{jj}
    \Big)\,,
    \\
    [\beta_{qq^{(3)}}^{qq^{(1)}}] &=
    3\left( g_s^2+g^2\right)\,,
    \\
     [\beta_{dd}^{dd}]_{ij} &=
    \Big(
      4\,g_s^2 +\tfrac{4}{3}\,g^{\prime 2} + 2\,[\gamma^{(Y)}_{d}]_{ii}+2\,[\gamma^{(Y)}_{d}]_{jj}
    \Big)\,,
    \\
    [\beta_{qd^{(1)}}^{qd^{(1)}}]_{ij} &=
    \Big(
      \tfrac{2}{3}\,g^{\prime 2}
      + [\gamma^{(Y)}_{q}]_{ii}
      +[\gamma^{(Y)}_{q}]_{jj}
      +[\gamma^{(Y)}_{d}]_{ii}
      +[\gamma^{(Y)}_{d}]_{jj}
      \\
      &\qquad\qquad
      + \tfrac{2}{3} \left|[Y_d]_{ij}\right|^2
      + \tfrac{1}{3} \left|[Y_d]_{ii}\right|^2
      + \tfrac{1}{3} \left|[Y_d]_{jj}\right|^2
    \Big)\,,
    \\
    [\beta_{qd^{(8)}}^{qd^{(1)}}]_{ij} &=
    \Big(
      -12\,g_s^2
      + 4 \left|[Y_d]_{ij}\right|^2
      + 2 \left|[Y_d]_{ii}\right|^2
      + 2 \left|[Y_d]_{jj}\right|^2
    \Big)\,,
    \\
     [\beta_{uu}^{uu}]_{ij} &=
    \Big(
      4\,g_s^2 +\tfrac{16}{3}\,g^{\prime 2} + 2\,[\gamma^{(Y)}_{u}]_{ii}+2\,[\gamma^{(Y)}_{u}]_{jj}
    \Big)\,,
    \\
    [\beta_{qu^{(1)}}^{qu^{(1)}}]_{ij} &=
    \Big(
      -\tfrac{4}{3}\,g^{\prime 2}
      + [\gamma^{(Y)}_{q}]_{ii}
      +[\gamma^{(Y)}_{q}]_{jj}
      +[\gamma^{(Y)}_{u}]_{ii}
      +[\gamma^{(Y)}_{u}]_{jj}
      \\
      &\qquad\qquad
      + \tfrac{2}{3} \left|[Y_u]_{ij}\right|^2
      + \tfrac{1}{3} \left|[Y_u]_{ii}\right|^2
      + \tfrac{1}{3} \left|[Y_u]_{jj}\right|^2
    \Big)\,,
    \\
    [\beta_{qu^{(8)}}^{qu^{(1)}}]_{ij} &=
    \Big(
      -12\,g_s^2
      + 4 \left|[Y_u]_{ij}\right|^2
      + 2 \left|[Y_u]_{ii}\right|^2
      + 2 \left|[Y_u]_{jj}\right|^2
    \Big)\,.
 \end{aligned}
\end{equation}
Here
\begin{equation}\label{eq:gammaY}
 \gamma^{(Y)}_{q}=\frac{1}{2}[Y_u Y_u^\dagger + Y_d Y_d^\dagger],\quad
 \gamma^{(Y)}_{u}=[Y_u^\dagger Y_u],\quad
 \gamma^{(Y)}_{d}=[Y_d^\dagger Y_d]
\end{equation}
with the Yukawa matrices $Y_d$ and $Y_u$. Subleading contributions are collected in Appendix~\ref{AppC}.

We mostly work in the flavour basis in which the down-type Yukawa matrix is diagonal, $Y_d=\text{diag}(y_d,y_s,y_b)$ and $Y_u=V^\dagger_\text{CKM}\,\text{diag}(y_u,y_c,y_t)$.
In some cases, in particular when considering $D^0$ mixing, we will also work in the basis in which the up-type Yukawa matrix is diagonal. In this case, all objects carrying flavour indices will carry a hat, e.g.\ $\hat Y_d=V_\text{CKM}\,\text{diag}(y_d,y_s,y_b)$ and $\hat Y_u=\text{diag}(y_u,y_c,y_t)$.

\subsection{Matching to the WET and RG Evolution in the WET}\label{sec:SMEFTWETmatching}

We define the $\Delta F=2$ meson mixing observables in terms of the effective Hamiltonian of the WET,
\begin{equation}\label{eq:Heff_DF2_SM_NP}
 {\cal H}_{\rm eff}^{\Delta F=2} = {\cal H}_{\rm eff, SM}^{\Delta F=2} + {\cal H}_{\rm eff, NP}^{\Delta F=2}\,,
\end{equation}
where the first and second term contains the SM and NP contributions, respectively.
For the NP part, we consider\footnote{\label{fnt:q1q2}%
{
Note that in the WET, we use $q_1$ and $q_2$ as placeholders for any quark mass eigenstate, which should not be confused with the quark doublets used in the SMEFT.  To avoid confusion, we will use the notation for the $Z'$ couplings introduced in Eq.~\eqref{eq:Zp_couplings_WET_notation}.
}
}
\begin{equation}
 {\cal H}_{\rm eff, NP}^{\Delta F=2} =  \sum_{q_1 q_2\in\{cu,ds,db,sb\}} {\cal H}_{\rm eff, NP}^{q_1 q_2} \,,
\end{equation}
where the terms relevant for $Z^\prime$ models are
\begin{equation}\label{eq:Heff_q1q2}
  {\cal H}_{\rm eff, NP}^{q_1 q_2}
=
-  C_{VLL}^{q_1 q_2}\, O_{VLL}^{q_1 q_2}
-  C_{VRR}^{q_1 q_2}\, O_{VRR}^{q_1 q_2}
-  C_{VLR}^{q_1 q_2}\, O_{VLR}^{q_1 q_2}
-  C_{SLR}^{q_1 q_2}\, O_{SLR}^{q_1 q_2}
~+~ {\rm h.c.} ~,
\end{equation}
which contribute to meson mixing in the up-type sector for $q_1 q_2 = c u$ and in the down-type sector for $q_1 q_2 \in \{ds,db,sb\}$.
The operators are defined as
\begin{equation}\label{eq:O_DF2}
 \begin{aligned}
  O_{VLL}^{q_1 q_2} &= (\bar q_1 \gamma_\mu P_L q_2)(\bar q_1 \gamma^\mu P_L q_2)\,,
  &\qquad
  O_{VRR}^{q_1 q_2} &= (\bar q_1 \gamma_\mu P_R q_2)(\bar q_1 \gamma^\mu P_R q_2)\,,
  \\
  O_{VLR}^{q_1 q_2} &= (\bar q_1 \gamma_\mu P_L q_2)(\bar q_1 \gamma^\mu P_R q_2)\,,
  &\qquad
  O_{SLR}^{q_1 q_2} &= (\bar q_1 P_L q_2)(\bar q_1 P_R q_2)\,.
 \end{aligned}
\end{equation}
In order to connect the SMEFT Wilson coefficient in Eq.~\eqref{eq:SMEFT_RG_dominant} to the WET Wilson coefficients in Eq.~\eqref{eq:Heff_q1q2}, we match the SMEFT to the WET at the electroweak scale and find the relations
\begin{equation}\label{eq:SMEFT_to_WET_matching}
 \begin{aligned}
C_{VLL}^{d_i d_j} &= [C_{qq}^{(1)}]_{ijij}+[C_{qq}^{(3)}]_{ijij}\,,
&\qquad
C_{VLL}^{cu} &= [\hat C_{qq}^{(1)}]_{1212}^* + [\hat C_{qq}^{(3)}]_{1212}^*\,,
\\
C_{VRR}^{d_i d_j} &= [C_{dd}]_{ijij}\,,
&\qquad
C_{VRR}^{cu} &= [C_{uu}]_{1212}^*\,,
\\
C_{VLR}^{d_i d_j} &= [C_{qd}^{(1)}]_{ijij} -\frac{1}{6} [C_{qd}^{(8)}]_{ijij}\,,
&\qquad
C_{VLR}^{cu} &= [\hat C_{qu}^{(1)}]_{1212}^* -\frac{1}{6} [\hat C_{qu}^{(8)}]_{1212}^*\,,
\\
C_{SLR}^{d_i d_j} &= -[C_{qd}^{(8)}]_{ijij}\,,
&\qquad
C_{SLR}^{cu} &= -[\hat C_{qu}^{(8)}]_{1212}^*\,,
 \end{aligned}
\end{equation}
where we denote SMEFT Wilson coefficients in the flavour basis in which the up-type Yukawa matrix is diagonal with a hat.

As the hadronic matrix elements entering meson mixing observables are evaluated at a scale $\mu=\mathcal{O}(\text{GeV})$, the coefficients at the scale $M_Z$ in Eq.~\eqref{eq:SMEFT_to_WET_matching} have to be evolved down to $\mu$ using the WET RGEs.
We express the WCs at the scale $\mu$ in terms of those at $M_Z$ and the RG evolution matrix $U^{q_1 q_2}(\mu,M_Z)$,
\begin{equation}\label{eq:WET_RG_evolution}
 \begin{pmatrix}
  C_{VLL}^{q_1 q_2}(\mu)\\
  C_{VRR}^{q_1 q_2}(\mu)\\
  C_{VLR}^{q_1 q_2}(\mu)\\
  C_{SLR}^{q_1 q_2}(\mu)
 \end{pmatrix}
 =
 U^{q_1 q_2}(\mu,M_Z)
 \begin{pmatrix}
  C_{VLL}^{q_1 q_2}(M_Z)\\
  C_{VRR}^{q_1 q_2}(M_Z)\\
  C_{VLR}^{q_1 q_2}(M_Z)\\
  C_{SLR}^{q_1 q_2}(M_Z)
 \end{pmatrix}\,.
\end{equation}
Solving the leading order RGEs, the evolution matrices in the up-type sector for $q_1 q_2 = c u$ and in the down-type sector for $q_1 q_2 \in \{ds,db,sb\}$ are given by
\begin{align}
U^{cu}(2\,\text{GeV},M_Z)
&=
\setlength{\arraycolsep}{3pt}
\begin{pmatrix}
 0.776 & 0 & 0 & 0 \\
 0 & 0.776 & 0 & 0 \\
 0 & 0 & 0.899 & 0 \\
 0 & 0 &-1.161 & 2.641 \\
\end{pmatrix}\,,
\\
U^{ds}(2\,\text{GeV},M_Z)
&=
\setlength{\arraycolsep}{3pt}
\begin{pmatrix}
 0.785 & 0 & 0 & 0 \\
 0 & 0.785 & 0 & 0 \\
 0 & 0 & 0.891 & 0 \\
 0 & 0 &-1.143 & 2.606 \\
\end{pmatrix}\,,
\\
U^{db}(4.2\,\text{GeV},M_Z)=U^{sb}(4.2\,\text{GeV},M_Z)
&=
\setlength{\arraycolsep}{3pt}
\begin{pmatrix}
 0.843 & 0 & 0 & 0 \\
 0 & 0.843 & 0 & 0 \\
 0 & 0 & 0.922 & 0 \\
 0 & 0 &-0.696 & 1.966 \\
\end{pmatrix}\,,
\end{align}
where we have set $\mu$ to the scale at which the matrix elements are evaluated, $\mu=2\,\text{GeV}$ for $q_1 q_2 \in \{uc,ds\}$ and $\mu=4.2\,\text{GeV}$ for $q_1 q_2 \in \{db,sb\}$.

Combining the results of the SMEFT matching, Eq.~\eqref{eq:SMEFT_matching_4q}, the SMEFT RG evolution, Eq.~\eqref{eq:SMEFT_RG_dominant}, the WET matching, Eq.~\eqref{eq:SMEFT_to_WET_matching}, and the WET RG evolution, Eq.~\eqref{eq:WET_RG_evolution}, we find the following expressions for the WET WCs at the hadronic scale $\mu$:

 \paragraph{\boldmath$q_1 q_2=uc$\unboldmath}
 \begin{equation}\label{eq:C_low_uc}
 \begin{aligned}
  C_{VLL}^{uc}(2\,\text{GeV})
  &\approx
    -0.338
  \
  \frac{\big(\hat\Delta^{uc\,*}_L\big)^2}{M_{Z'}^2}
  \
  \left[1+
     3.26\times10^{-2}
    \,
    \log\left(M_Z/M_{Z'}\right)
    \right]\,,
  \\
  C_{VRR}^{uc}(2\,\text{GeV})
  &\approx
    -0.338
  \
  \frac{\big(\hat\Delta^{uc\,*}_R\big)^2}{M_{Z'}^2}
  \
  \left[1+
     2.92\times10^{-2}
    \,
    \log\left(M_Z/M_{Z'}\right)
    \right]\,,
  \\
  C_{VLR}^{uc}(2\,\text{GeV})
  &\approx
    -0.899
  \
  \frac{\hat\Delta^{uc\,*}_L\,\hat\Delta^{uc\,*}_R}{M_{Z'}^2}
  \
  \left[1+
     1.12\times10^{-2}
    \,
    \log\left(M_Z/M_{Z'}\right)
    \right]\,,
  \\
  C_{SLR}^{uc}(2\,\text{GeV})
  &\approx
    1.161
  \
  \frac{\hat\Delta^{uc\,*}_L\,\hat\Delta^{uc\,*}_R}{M_{Z'}^2}
  \
  \left[1-
     15.71\times10^{-2}
    \,
    \log\left(M_Z/M_{Z'}\right)
    \right]\,.
 \end{aligned}
 \end{equation}

 \paragraph{\boldmath$q_1 q_2=ds$\unboldmath}
 \begin{equation}\label{eq:C_low_ds}
 \begin{aligned}
  C_{VLL}^{ds}(2\,\text{GeV})
  &\approx
    -0.393
  \
  \frac{\big(\Delta^{ds}_L\big)^2}{M_{Z'}^2}
  \
  \left[1+
     3.26\times10^{-2}
    \,
    \log\left(M_Z/M_{Z'}\right)
    \right]\,,
  \\
  C_{VRR}^{ds}(2\,\text{GeV})
  &\approx
    -0.393
  \
  \frac{\big(\Delta^{ds}_R\big)^2}{M_{Z'}^2}
  \
  \left[1+
     2.58\times10^{-2}
    \,
    \log\left(M_Z/M_{Z'}\right)
    \right]\,,
  \\
  C_{VLR}^{ds}(2\,\text{GeV})
  &\approx
    -0.891
  \
  \frac{\Delta^{ds}_L\,\Delta^{ds}_R}{M_{Z'}^2}
  \
  \left[1+
     1.29\times10^{-2}
    \,
    \log\left(M_Z/M_{Z'}\right)
    \right]\,,
  \\
  C_{SLR}^{ds}(2\,\text{GeV})
  &\approx
    1.143
  \
  \frac{\Delta^{ds}_L\,\Delta^{ds}_R}{M_{Z'}^2}
  \
  \left[1-
     15.58\times10^{-2}
    \,
    \log\left(M_Z/M_{Z'}\right)
    \right]\,.
 \end{aligned}
 \end{equation}

 \paragraph{\boldmath$q_1 q_2=d_i b$, $i\in\{1,2\}$\unboldmath}
 \begin{equation}\label{eq:C_low_dib}
 \begin{aligned}
  C_{VLL}^{d_i b}(4.2\,\text{GeV})
  &\approx
    -0.422
  \
  \frac{\big(\Delta^{d_i b}_L\big)^2}{M_{Z'}^2}
  \
  \left[1+
     3.67\times10^{-2}
    \,
    \log\left(M_Z/M_{Z'}\right)
    \right]\,,
  \\
  C_{VRR}^{d_i b}(4.2\,\text{GeV})
  &\approx
    -0.422
  \
  \frac{\big(\Delta^{d_i b}_R\big)^2}{M_{Z'}^2}
  \
  \left[1+
     2.58\times10^{-2}
    \,
    \log\left(M_Z/M_{Z'}\right)
    \right]\,,
  \\
  C_{VLR}^{d_i b}(4.2\,\text{GeV})
  &\approx
    -0.922
  \
  \frac{\Delta^{d_i b}_L\,\Delta^{d_i b}_R}{M_{Z'}^2}
  \
  \left[1+
     1.50\times10^{-2}
    \,
    \log\left(M_Z/M_{Z'}\right)
    \right]\,,
  \\
  C_{SLR}^{d_i b}(4.2\,\text{GeV})
  &\approx
    0.696
  \
  \frac{\Delta^{d_i b}_L\,\Delta^{d_i b}_R}{M_{Z'}^2}
  \
  \left[1-
     19.41\times10^{-2}
    \,
    \log\left(M_Z/M_{Z'}\right)
    \right]\,.
 \end{aligned}
 \end{equation}

\boldmath
\subsection{Meson Mixing Amplitude}\label{step4}
\unboldmath

The meson mixing observables of a given meson $\mathcal M$ depend on the dispersive part $M_{12}^\mathcal{M}$ and the absorptive part $\Gamma_{12}^\mathcal{M}$ of the mixing amplitude.
We consider new physics contributions to $M_{12}^\mathcal{M}$, which is defined as
\begin{equation}
 M_{12}^\mathcal{M} = \frac{\langle \mathcal{M} | \mathcal H_{\mathrm{eff}}^{\Delta F=2} |\bar{\mathcal{M}}\rangle}{2M_\mathcal{M}}\,,
\end{equation}
where $M_\mathcal{M}$ is the mass of the meson $\mathcal{M}$.
Using Eq.~\eqref{eq:Heff_DF2_SM_NP}, we can separate the SM and NP contributions,
\begin{equation}
 M_{12}^\mathcal{M} = {M_{12}^\mathcal{M}}_{\rm SM} + {M_{12}^\mathcal{M}}_{\rm NP}\,,
\end{equation}
and we can express ${M_{12}^\mathcal{M}}_{\rm NP}$ in terms of the WCs and operators in Eq.~\eqref{eq:Heff_q1q2},
\begin{equation}
 {M_{12}^\mathcal{M}}_{\rm NP} =
-  \left(C_{VLL}^{q_1 q_2} + C_{VRR}^{q_1 q_2}\right) \frac{\langle O_{VLL}^{q_1 q_2}\rangle}{2M_\mathcal{M}}
-  C_{VLR}^{q_1 q_2}\, \frac{\langle O_{VLR}^{q_1 q_2} \rangle}{2M_\mathcal{M}}
-  C_{SLR}^{q_1 q_2}\, \frac{\langle O_{SLR}^{q_1 q_2} \rangle}{2M_\mathcal{M}}\,,
\end{equation}
where we used that $\langle O_{VRR}^{q_1 q_2}\rangle = \langle O_{VLL}^{q_1 q_2}\rangle$.
In this equation, $q_1 q_2$ is $cu$ for $\mathcal{M}=D_0$, $sd$ for $\mathcal{M}=K_0$, $db$ for $\mathcal{M}=B_d$, and $sb$ for $\mathcal{M}=B_s$.
The matrix elements $\langle O_i^{q_1 q_2} \rangle$ are defined in Appendix~\ref{app:matrixelements}.

Using our results for the WCs at the scale where the corresponding matrix elements are evaluated, Eqs.~\eqref{eq:C_low_uc},~\eqref{eq:C_low_ds},~\eqref{eq:C_low_dib}, and assuming the left-handed couplings of $Z^\prime$ to quarks to be non-vanishing, we can express ${M_{12}^\mathcal{M}}_{\rm NP}$ as
\begin{equation}\label{eq:M12_CVLL}
 {M_{12}^\mathcal{M}}_{\rm NP} =
- C_{VLL}^{q_1 q_2}\, \frac{\langle O_{VLL}^{q_1 q_2}\rangle}{2M_\mathcal{M}}\,
z_{q_1 q_2}\,,
\end{equation}
where
\begin{equation}\label{eq:z_q1q2}
z_{q_1 q_2} =
 \Big[
 1 + (1+\eta_{q_1 q_2})\, r_{q_1 q_2}^2 + 2\, \kappa_{q_1 q_2}\, r_{q_1 q_2}
\Big]
\end{equation}
parameterizes contributions due to non-vanishing right-handed couplings of $Z^\prime$ to quarks, with
\begin{equation}
 r_{q_1 q_2} = \frac{\Delta_R^{q_1 q_2}}{\Delta_L^{q_1 q_2}}
\end{equation}
defined as in Eq.~\eqref{eq:r_q1q2}.
The quantities $\kappa_{q_1 q_2}$ and $\eta_{q_1 q_2}$ are given by
\begin{equation}
\begin{aligned}
\kappa_{q_1 q_2} &= \left(
\frac{C_{VLR}^{q_1 q_2}\,\langle O_{VLR}^{q_1 q_2}\rangle}{C_{VLL}^{q_1 q_2}\,\langle O_{VLL}^{q_1 q_2}\rangle}
+
\frac{C_{SLR}^{q_1 q_2}\,\langle O_{SLR}^{q_1 q_2}\rangle}{C_{VLL}^{q_1 q_2}\,\langle O_{VLL}^{q_1 q_2}\rangle}
\right)\frac{1}{2\,r_{q_1 q_2}}\,,
\\
 \eta_{q_1 q_2} &= \frac{C_{VRR}^{q_1 q_2}}{C_{VLL}^{q_1 q_2}}\frac{1}{r_{q_1 q_2}^2}-1\,.
\end{aligned}
\end{equation}
The quantity $\eta_{q_1 q_2}$ accounts for the small differences in the SMEFT RG evolution of left- and right-handed WCs $C_{VLL}^{q_1 q_2}$ and $C_{VRR}^{q_1 q_2}$, and is of order $1\%$ in all four meson mixing sectors. It only depends on the SMEFT RGEs and is independent of the hadronic matrix elements.
The quantity $\kappa_{q_1 q_2}$ captures the contributions from the left-right Wilson coefficients $C_{VLR}^{q_1 q_2}$ and $C_{SLR}^{q_1 q_2}$ relative to those from $C_{VLL}^{q_1 q_2}$. In addition to the SMEFT and WET RG effects, it depends crucially on the hadronic matrix elements and therefore has different characteristic values in the four different meson mixing sectors. These values are shown in the left panel of Fig.~\ref{fig:r0} as functions of the $Z^\prime$ mass $M_{Z^\prime}$. We include uncertainty bands stemming from the uncertainties of the hadronic matrix elements $\langle O_i^{q_1 q_2} \rangle$. The $M_{Z^\prime}$ dependence is obtained from the numerical solution of the leading-order (LO) RGEs in the SMEFT and the WET, summing large logarithms.
Note that $\kappa_{sb}$ and $\kappa_{db}$ are essentially equal and cannot be clearly distinguished in Fig.~\ref{fig:r0}.

Using Eqs.~\eqref{eq:C_low_uc}, \eqref{eq:C_low_ds}, \eqref{eq:C_low_dib}, we find the following approximate expressions for $\eta_{q_1 q_2}$ and $\kappa_{q_1 q_2}$ in the four different meson mixing sectors. They are given by

 \paragraph{\boldmath$q_1 q_2=uc$\unboldmath}
 \begin{equation}
 \begin{aligned}
  \kappa_{uc} &\approx
\left(
 1.33\,
 \tfrac{\langle O_{VLR}^{uc}\rangle}{\langle O_{VLL}^{uc}\rangle}
 -
1.72\,
 \tfrac{\langle O_{SLR}^{uc}\rangle}{\langle O_{VLL}^{uc}\rangle}
 \right)
 +
 \left(
 2.85\,
 \tfrac{\langle O_{VLR}^{uc}\rangle}{\langle O_{VLL}^{uc}\rangle}
 -
  32.6\,
 \tfrac{\langle O_{SLR}^{uc}\rangle}{\langle O_{VLL}^{uc}\rangle}
 \right)
 \times
 10^{-2}\, \log\!\left(\tfrac{M_{Z'}}{M_Z}\right)\,,
\\
  \eta_{uc} &\approx
    0.34 \times 10^{-2}\, \log\!\left(\tfrac{M_{Z'}}{M_Z}\right)\,,
 \end{aligned}
 \end{equation}

 \paragraph{\boldmath$q_1 q_2=ds$\unboldmath}
 \begin{equation}
 \begin{aligned}
  \kappa_{ds} &\approx
\left(
 1.13\,
 \tfrac{\langle O_{VLR}^{ds}\rangle}{\langle O_{VLL}^{ds}\rangle}
 -
1.45\,
 \tfrac{\langle O_{SLR}^{ds}\rangle}{\langle O_{VLL}^{ds}\rangle}
 \right)
 +
 \left(
 2.23\,
 \tfrac{\langle O_{VLR}^{ds}\rangle}{\langle O_{VLL}^{ds}\rangle}
 -
  27.4\,
 \tfrac{\langle O_{SLR}^{ds}\rangle}{\langle O_{VLL}^{ds}\rangle}
 \right)
 \times
 10^{-2}\, \log\!\left(\tfrac{M_{Z'}}{M_Z}\right)\,,
\\
  \eta_{ds} &\approx
    0.68 \times 10^{-2}\, \log\!\left(\tfrac{M_{Z'}}{M_Z}\right)\,,
 \end{aligned}
 \end{equation}

 \paragraph{\boldmath$q_1 q_2=d_i b$, $i\in\{1,2\}$\unboldmath}
 \begin{equation}
 \begin{aligned}
  \kappa_{d_i b} &\approx
\left(
 1.09\,
 \tfrac{\langle O_{VLR}^{d_i b}\rangle}{\langle O_{VLL}^{d_i b}\rangle}
 -
0.82\,
 \tfrac{\langle O_{SLR}^{d_i b}\rangle}{\langle O_{VLL}^{d_i b}\rangle}
 \right)
 +
 \left(
 2.37\,
 \tfrac{\langle O_{VLR}^{d_i b}\rangle}{\langle O_{VLL}^{d_i b}\rangle}
 -
  19.0\,
 \tfrac{\langle O_{SLR}^{d_i b}\rangle}{\langle O_{VLL}^{d_i b}\rangle}
 \right)
 \times
 10^{-2}\, \log\!\left(\tfrac{M_{Z'}}{M_Z}\right)\,,
\\
  \eta_{d_i b} &\approx
    1.09 \times 10^{-2}\, \log\!\left(\tfrac{M_{Z'}}{M_Z}\right)\,.
 \end{aligned}
 \end{equation}

\boldmath
\subsection{Suppression of $Z^\prime$ Contributions to Meson Mixing}\label{sec:meson_mixing_suppression}
\unboldmath

%
\begin{figure}
\centering
\includegraphics[height=0.45\textwidth]{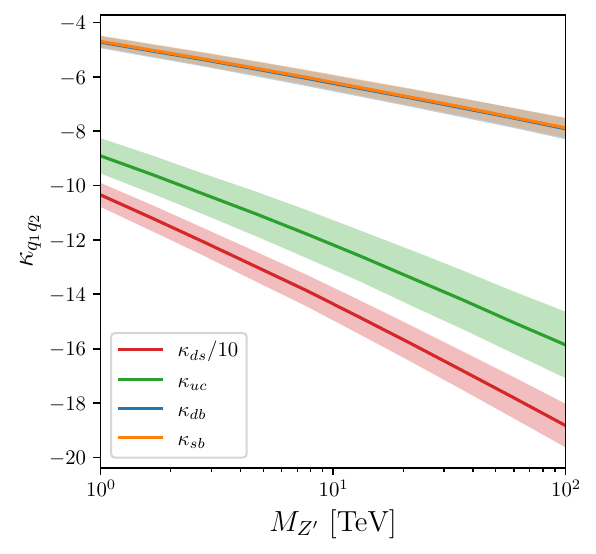}
\hfill
\includegraphics[height=0.45\textwidth]{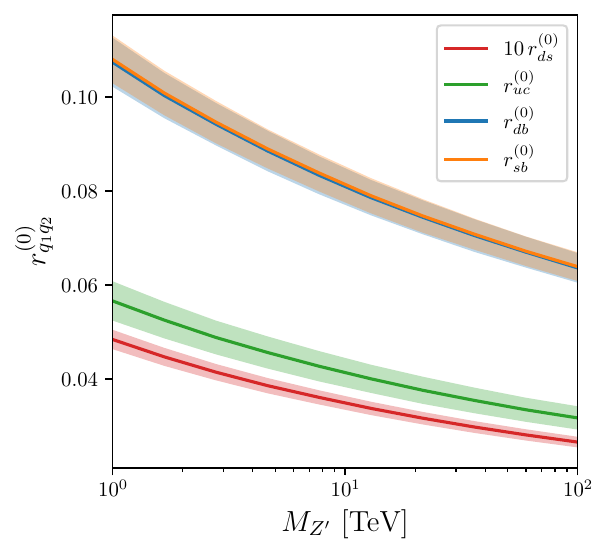}
\caption{Values of $\kappa_{q_1 q_2}$ (left panel) and $r_{q_1 q_2}^{(0)}$ (right panel) for $q_1 q_2\in\{uc, ds, db, sb\}$ as functions of the $Z^\prime$ mass $M_{Z^\prime}$.}
\label{fig:r0}
\end{figure}
%

The NP contribution to the dispersive part of the meson mixing amplitude, ${M_{12}^\mathcal{M}}_{\rm NP}$, is proportional to the quantity $z_{q_1 q_2}$ defined in Eq.~\eqref{eq:z_q1q2}.
While $z_{q_1 q_2}$ simply reduces to $z_{q_1 q_2}=1$ in the case of vanishing right-handed couplings of $Z^\prime$ to quarks, the ratio of left-handed and right-handed couplings $r_{q_1 q_2}$ can take values that result in a vanishing $z_{q_1 q_2}$ and therefore no contribution to ${M_{12}^\mathcal{M}}_{\rm NP}$~(cf.~\cite{Buras:2014zga}).
Solving $z_{q_1 q_2}=0$ for $r_{q_1 q_2}$, we find\footnote{We select the solution for which $|\Delta_R^{q_1 q_2}|<|\Delta_L^{q_1 q_2}|$ for $\kappa_{q_1 q_2}{\ll-1}$.}%
{
\begin{equation}\label{eq:z_q1q1_zero}
\begin{aligned}
r_{q_1 q_2}^{(0)} = r_{q_1 q_2}\Big|_{z_{q_1 q_2}=0}
&=
\frac{-\kappa_{q_1 q_2} -\sqrt{\kappa_{q_1 q_2}^2-1-\eta_{q_1 q_2}}}{1+\eta_{q_1 q_2}}
\\
&\approx
-\kappa_{q_1 q_2} -\sqrt{\kappa_{q_1 q_2}^2-1}
\\
&\approx
-\frac{1}{2\kappa_{q_1 q_2}}\,,
\end{aligned}
\end{equation}
}
where in the second line we used that $\eta_{q_1 q_2}\ll1$ and in the third line that $\kappa_{q_1 q_2}{\ll-1}$.

It follows that  if the relation in Eq.~\eqref{eq:z_q1q1_zero} is approximately satisfied, $Z^\prime$ contributions to the corresponding meson mixing observables
will be strongly suppressed.
Simultaneously, the presence of the right-handed
couplings will have some impact on rare FCNC decays.
The values of $r_{q_1 q_2}^{(0)}$ with $q_1 q_2\in\{uc,ds,db,sb\}$ for which the first line of Eq.~\eqref{eq:z_q1q1_zero} is satisfied are shown in
the right panel of Fig.~\ref{fig:r0}.
As for $\kappa_{q_1 q_2}$, the $M_{Z^\prime}$ dependence is obtained from the full numerical solution of the RGEs at LO and the uncertainty bands correspond to the uncertainties of the hadronic matrix elements. $r_{sb}^{(0)}$ and $r_{db}^{(0)}$ are essentially equal and cannot be clearly distinguished in Fig.~\ref{fig:r0}.
We find that in
the case of
$K^0-\bar K^0$ mixing the corresponding condition reads
\be\label{eq:r0_ds}
\Delta_R^{ds}\approx 0.004\,\Delta_L^{ds},
\ee
{%
implying a large fine tuning between right- and left-handed couplings, but then also negligible impact of right-handed currents on rare $K$ decays.
NP contributions to $\varepsilon_K$ can be suppressed with less fine tuning in a scenario where the relevant $sd$ coupling is nearly imaginary~\cite{Aebischer:2023mbz}, with interesting implications for rare Kaon decays and $\epe$.
}

On the other hand, a cancellation of right- and left-handed NP contributions to $D^0-\bar D^0$, $B_d-\bar B_d$, and $B_s-\bar B_s$ mixing is possible with less fine tuning, under the conditions
\begin{equation}\label{eq:r0_uc_dib}
\hat \Delta_R^{uc}\approx 0.05\,\hat\Delta_L^{uc}
\qquad\text{and}\qquad
\Delta_R^{d_i b}\approx 0.1\,\Delta_L^{d_i b}\,.
\end{equation}
In these cases, the left-handed $Z^\prime$ couplings $\Delta_L^{q_1 q_2}$ can take on large values that are unconstrained by their direct contribution to the corresponding meson mixing amplitude, Eq.~\eqref{eq:M12_CVLL}, as $z_{q_1 q_2} \approx 0$.
However, indirect contributions to other meson mixing sectors that we discuss in the following Sections \ref{sec:SU2_corr} and \ref{sec:SMEFTRGeffects} are still induced and can provide relevant constraints.

\boldmath
\subsection{Correlations in Meson Mixing from $\text{SU(2)}_L$ Gauge Invariance}\label{sec:SU2_corr}
\unboldmath

In the SM, the left-handed up- and down-type quarks are unified into doublets of the $\text{SU(2)}_L$ gauge group.
This means that $\text{SU(2)}_L$ gauge invariance implies relations between the interactions of up- and down-type quarks.
In particular, as discussed in section~\ref{sec:ZpSMEFTmatching}, the $Z^\prime$ couplings of left-handed up-type quarks are related to those of the left-handed down-type quarks by Eq.~\eqref{eq:LH_Zp_relation}.
This allows us to express the $Z^\prime\text{-}u\text{-}c$ coupling $\hat\Delta^{uc}_L$ that contributes to the $D$ meson mixing amplitude $M_{12}^D$ through the Wilson coefficient $C_{VLL}^{uc}$~(cf.~Eq.~\eqref{eq:C_low_uc}) in terms of the couplings of down-type quarks,
\begin{equation}\label{eq:LH_relation}
\begin{aligned}
  \hat\Delta^{uc}_L
  =&\
  \Delta^{ds}_L\,\left(V_{cs}^*\,V_{ud}+e^{-2i\phi_L^{ds}}\,V_{cd}^*\,V_{us}\right)
  \\+&\
 \Delta^{db}_L\,\left(V_{cb}^*\,V_{ud}+e^{-2 i \phi_L^{db}}\,V_{cd}^*\,V_{ub}\right)
 \\+&\
 \Delta^{sb}_L\,\left(V_{cb}^*\,V_{us}+e^{-2 i \phi_L^{sb}}\,V_{cs}^*\,V_{ub}\right)
 \\+&\
 \left(\Delta^{dd}_L - \Delta^{bb}_L\right)\,V_{cd}^*\,V_{ud}
 +
 \left(\Delta^{ss}_L - \Delta^{bb}_L\right)\,V_{cs}^*\,V_{us}\,,
\end{aligned}
\end{equation}
where $\phi_L^{d_i d_j}$ denote the complex phases of the couplings $\Delta^{d_i d_j}_L$.
In the absence of right-handed $Z^\prime$ couplings, the squares of $\Delta^{ds}_L$, $\Delta^{db}_L$, and $\Delta^{sb}_L$ are directly proportional to the NP contribution to the meson mixing amplitude $M_{12}^\mathcal{M}$ with $\mathcal{M}\in \{K,B_d,B_s\}$ and are therefore strongly constrained by experimental data.
In the presence of a right-handed coupling, on the other hand, as discussed in section~\ref{sec:meson_mixing_suppression}, the contribution to $M_{12}^\mathcal{M}$ can become negligibly small, allowing for a potentially large left-handed $Z^\prime$ coupling.
However, Eq.~\eqref{eq:LH_relation} implies that even in this case, a single left-handed coupling cannot be arbitrarily large without the presence of at least one other large left-handed coupling.
Furthermore, if the flavour-conserving left-handed couplings to down-type quarks are approximately equal, their contribution to the relation in Eq.~\eqref{eq:LH_relation} vanishes as a consequence of CKM unitarity. In this case, we find a direct relation between the left-handed contributions to all four meson mixing amplitudes,
\begin{equation}\label{eq:Delta_uc_SU2}
 \hat\Delta^{uc}_L\Big|_{\Delta^{dd}_L \approx \Delta^{ss}_L \approx \Delta^{bb}_L}
 \approx
 \Delta^{ds}_L
 +
 \Delta^{db}_L\,V_{cb}^*
 +
 \Delta^{sb}_L\,\left(V_{cb}^*\,V_{us}+e^{-2 i \phi_L^{sb}}\,V_{ub}\right)\,,
\end{equation}
where we used $V_{ud}\approx V_{cs}\approx 1$ and neglected numerically small terms.

{
This result is particularly important in the presence of a single dominant flavour-changing $Z^\prime$ coupling to only $b$ and $s$ quarks. In this case, a real $\Delta^{sb}_L$ necessarily leads to a \emph{complex} $\hat\Delta^{uc}_L$, contributing to the imaginary part of the dispersive mixing amplitude in the $D^0-\bar D^0$ system.
An observable particularly sensitive to this is
\begin{equation}
 x_{12}^{\text{Im},D} = |x_{12}^D|\,\sin(\phi_{12}^D),
\end{equation}
with $x_{12}^D$ and $\phi_{12}^D$ defined by~\cite{Kagan:2009gb}
\begin{equation}
 x_{12}^D = 2\,\tau_{D}\, |\mathcal{M}_{12}^D|
 \qquad
 \text{and}
 \qquad
 \phi_{12}^D = \arg(\mathcal{M}_{12}^D/\Gamma_{12}^D),
\end{equation}
where $\tau_{D}$ is the average $D^0$ lifetime.

The phenomenological consequences are demonstrated in the examples we present in Sections~\ref{sec:fit_df2} and \ref{sec:DeltaB=DeltaS}.
}

\boldmath
\subsection{SMEFT Renormalization Group Contributions to Meson Mixing}\label{sec:SMEFTRGeffects}
\unboldmath

In addition to the flavour-conserving contributions to the SMEFT RG running and mixing given in Eq.~\eqref{eq:SMEFT_RG_dominant}, the SMEFT RGEs also generate small flavour-changing contributions.
Due to these contributions, in principle any $\Delta F=2$ WC at the scale $\Lambda_\text{NP}$ generates effects in all four meson mixing sectors.
Usually these effects are very small and phenomenologically irrelevant, as the $\Delta F=2$ WC are severely constrained from their tree-level contributions to meson mixing.
However, if these tree-level contributions are suppressed as described in Section~\ref{sec:meson_mixing_suppression}, the corresponding $\Delta F=2$ WC can be large and the RG induced effects can become relevant.

As an example, we consider the contribution to the Kaon mixing observable $\varepsilon_K$ generated from the WC $[C_{qq}^{(1)}]_{2323}(\Lambda_\text{NP})$.
Since $\varepsilon_K$ is particularly sensitive to even very small NP contributions, this effect can in principle become phenomenologically relevant as we will see in Section~\ref{sec:DeltaB=DeltaS}.
The contribution to $\varepsilon_K$ is generated in two steps:

 \begin{itemize}
  \item In the first step, the SM couplings and dimension-six WCs are run from the matching scale $\Lambda_\text{NP}$ down to the electroweak scale $M_Z$.
  This leads to off-diagonal entries in the initially diagonal Yukawa matrix $Y_d$, and $\tilde{C}$ denotes the WC in the corresponding non-canonical flavour basis.

  The WC $[C_{qq}^{(1)}]_{2323}(\Lambda_\text{NP})$ mixes into $[\tilde{C}_{qq}^{(1)}]_{1232}(M_Z)$, which in the first leading-log approximation is given by \cite{Jenkins:2013wua}
  \begin{equation}
  \begin{aligned}
    {[}\tilde{C}_{qq}^{(1)}]_{1232}(M_Z)
    \approx&\
    2\ [C_{qq}^{(1)}]_{2323}(\Lambda_\text{NP})
    \,
    \times
    \,
    [\gamma^{(Y)}_{q}]_{13}\,\frac{\log\left(M_Z/\Lambda_\text{NP}\right)}{16 \pi^2}\,
    \\
    \stackrel{\Lambda_\text{NP}=5\,\text{TeV}}{\approx}&\
    [C_{qq}^{(1)}]_{2323}(5\,\text{TeV})
    \times
    (-1.44-0.58\,i)\times 10^{-4}
  \end{aligned}
  \end{equation}
  where $\gamma^{(Y)}_{q}=\frac{1}{2}[Y_u Y_u^\dagger + Y_d Y_d^\dagger]$ with $Y_d=\text{diag}(y_d,y_s,y_b)$ and $Y_u=V^\dagger_\text{CKM}\,\text{diag}(y_u,y_c,y_t)$ such that
   $[\gamma^{(Y)}_{q}]_{13} \approx \frac{1}{2}\, y_t^2\, V_{tb}\, V_{td}^*\,$.

   The self-mixing of $[C_{qq}^{(1)}]_{2323}$ in the first leading-log approximation results in
    \begin{equation}
  \begin{aligned}
    {[}\tilde{C}_{qq}^{(1)}]_{2323}(M_Z)
    \approx&\
    [C_{qq}^{(1)}]_{2323}(\Lambda_\text{NP})
    \,
    \times
    \\
    &\left[1+
    \Big(
      g_s^2 +\tfrac{1}{3}\,g^{\prime 2} + 2\,[\gamma^{(Y)}_{q}]_{22}+2\,[\gamma^{(Y)}_{q}]_{33}
    \Big)
    \,\frac{\log\left(M_Z/\Lambda_\text{NP}\right)}{16 \pi^2}
    \right]
    \\
    \stackrel{\Lambda_\text{NP}=5\,\text{TeV}}{\approx}&\
    [C_{qq}^{(1)}]_{2323}(5\,\text{TeV})
    \times
    0.96\,.
  \end{aligned}
  \end{equation}
   \item In the second step, the WCs are transformed into the canonical flavour basis, in which $Y_d$ is diagonal.
   To this end, one has to re-diagonalize the Yukawa matrices using flavour rotations that also rotate the flavour indices of the WCs {(for more details and explicit rotation matrices see Appendix~\ref{app:rediag})}. Applying these flavour rotations, we get
   \begin{equation}
   \begin{aligned}
    {[}C_{qq}^{(1)}]_{1212}(M_Z)
    =
    [&\tilde{C}_{qq}^{(1)}]_{1232}(M_Z) \ (U_q^\dagger)_{11}\,(U_q)_{22}\,(U_q^\dagger)_{13}\,(U_q)_{22}
    \\
    +
    [&\tilde{C}_{qq}^{(1)}]_{2323}(M_Z) \ (U_q^\dagger)_{22}\,(U_q)_{13}\,(U_q^\dagger)_{22}\,(U_q)_{13}
    \\
    +\phantom{[}&\dots
    \\
    \stackrel{\Lambda_\text{NP}=5\,\text{TeV}}{\approx}
    [&\tilde{C}_{qq}^{(1)}]_{1232}(M_Z)\times (-1.81 - 0.72\, i)\times 10^{-4}
    \\
    +
    [&\tilde{C}_{qq}^{(1)}]_{2323}(M_Z)\times (2.74 + 2.61\, i)\times 10^{-8}\,,
   \end{aligned}
\end{equation}
where the ellipsis corresponds to numerically irrelevant strongly suppressed contributions.
 \end{itemize}
Combining the above two effects, we find for $\Lambda_\text{NP}=5\,\text{TeV}$ in the first leading-log approximation
\begin{equation}\label{eq:Cqq1212}
 [C_{qq}^{(1)}]_{1212}(M_Z)
 \approx [C_{qq}^{(1)}]_{2323}(5\,\text{TeV})\times (4.8+4.6\, i)\times 10^{-8}\,.
\end{equation}
While this contribution seems to be very small, it can have a relevant impact on $\varepsilon_K$,
as this observable is highly sensitive to the imaginary part of $[C_{qq}^{(1)}]_{1212}(M_Z)$.
To be specific, one can obtain an approximate semi-analytic expression for the dependence of $\varepsilon_K$ on $[C_{qq}^{(1)}]_{1212}(M_Z)$,
\begin{equation}
 \varepsilon_K
 \approx
 \varepsilon_K^\text{SM}
 \times
 \Big(1-1.55\times 10^{8}\,\text{TeV}^2\times\text{Im}\!\left([C_{qq}^{(1)}]_{1212}(M_Z)\right)\Big)\,.
\end{equation}
If we combine this with Eq.~\eqref{eq:Cqq1212} and assume $[C_{qq}^{(1)}]_{2323}(5\,\text{TeV})$ to be real, we find
\begin{equation}
 \varepsilon_K
 \approx
 \varepsilon_K^\text{SM}
 \times
 \Big(1-7.1\,\text{TeV}^2\times[C_{qq}^{(1)}]_{2323}(5\,\text{TeV})\Big)\,.
\end{equation}
Consequently, a shift of $\varepsilon_K$ by around 7-8\%, which corresponds to the theoretical uncertainty of its SM prediction, can be generated by $[C_{qq}^{(1)}]_{2323}(5\,\text{TeV})$ of order $0.01\,\text{TeV}^{-2}$.
Note that if the $Z^\prime$ coupling $\Delta^{bs}_L$ is real, the WC $[C_{qq}^{(1)}]_{2323}(5\,\text{TeV})$ is always real and negative (cf. Eq.~\eqref{eq:SMEFT_matching_4q}), such that the generated shift in $\varepsilon_K$ is {\em positive.}

    This  upward shift in  $\varepsilon_K$ through RG
  effects analyzed here turns out to be welcome, but eventually not crucial. Indeed, as demonstrated
  in Appendix~\ref{app:epsK}, the SM estimate of $\varepsilon_K$ with the
  values of the parameter $\hat B_K$ from
  either
  Dual QCD \cite{Buras:2014maa} or the most recent NLO analysis of the RBC/UKQCD collaboration \cite{Boyle:2024gge},
  and the other input parameters as described in Appendix~\ref{app:epsK},
  is
  around {$5\%$ below the experimental value.} However, the very recent NNLO analysis
    of \cite{Gorbahn:2024qpe} reduces this difference significantly, although
    this depends on the values of $\gamma$ and $\vcb$, as illustrated in
    Table~\ref{tab:eps_K_SM}. Therefore, eventually, in accordance with
    the strategy in \cite{Buras:2021nns,Buras:2022wpw}, NP contributions to
    $\varepsilon_K$ are not required to reproduce the experimental data.

\boldmath
\subsection{Numerical Analysis: Suppression of $Z^\prime$ Contributions to $B_s-\bar B_s$ mixing}\label{sec:fit_df2}
\unboldmath

In this section, we demonstrate the effects discussed in the previous sections in a numerical analysis, using the example of $B_s-\bar B_s$ mixing.
To this end, we perform fits of the $Z^\prime$-$b$-$s$ couplings using the open source python package \texttt{flavio}~\cite{Straub:2018kue}.
The theoretical uncertainties of the observables considered in our analysis depend strongly on the size of the NP WCs and can be much larger than the SM uncertainties.
In particular, $\Delta M_s$ can have a significantly enhanced theory uncertainty in the presence of large NP WCs, even if its central value is SM-like due to the cancellation described in Section~\ref{sec:meson_mixing_suppression}.
It is therefore crucial to account for the new physics dependence of the theory uncertainties in our fits, which we do using the method of~\cite{Altmannshofer:2021qrr}.
We consider constraints from various relevant $\Delta F=2$ observables:
\begin{itemize}
 \item $\Delta M_s$, the mass difference in the $B_s-\bar B_s$ system.
 \item $S_{\psi\phi}$, the mixing induced $C\!P$ asymmetry in $B_s\to J/\psi \phi$.
 \item $x_{12}^{\text{Im},D}$, the normalized imaginary part of the dispersive mixing amplitude in the $D^0-\bar D^0$ system.
 \item $\varepsilon_K$, the indirect $C\!P$ violation parameter in the $K^0-\bar K^0$ system.
 \item $\Delta M_d$, the mass difference in the $B_d-\bar B_d$ system.
\end{itemize}
All of these observables receive considerable contributions either directly or through RG effects, except for $\Delta M_d$. However, the theoretical uncertainties of $\varepsilon_K$ and $\Delta M_d$ are correlated and this correlation slightly affects the global fit, even in the absence of NP contributions to $\Delta M_d$.
Note that in our fit we do not include $S_{\psi K_S}$, the mixing induced $C\!P$ asymmetry in $B_d\to J/\psi K_S$, as this observable is used as input observable to determine the angle $\beta$ of the CKM unitarity triangle.

The $Z^\prime$ couplings $\Delta_{L,R}^{ij}$
  always enter the matching relations in the form of a ratio involving the $Z^\prime$ mass, $\Delta_{L,R}^{ij}/M_{Z^\prime}$, so that the WCs are not individually sensitive to the couplings or the mass.
Consequently, in our numerical analysis we vary the ratios $\Delta_{L,R}^{ij}/M_{Z^\prime}$.
However, the $Z^\prime$ mass enters our results indirectly in terms of the matching scale $\Lambda_\text{NP}\approx M_{Z^\prime}$, which we use as the renormalization scale at which the numerical values of the SMEFT WCs are defined. For our numerical analysis, we set $\Lambda_\text{NP}=5\,\text{TeV}$.

\begin{figure}
\centering
\includegraphics[width=0.48\textwidth]{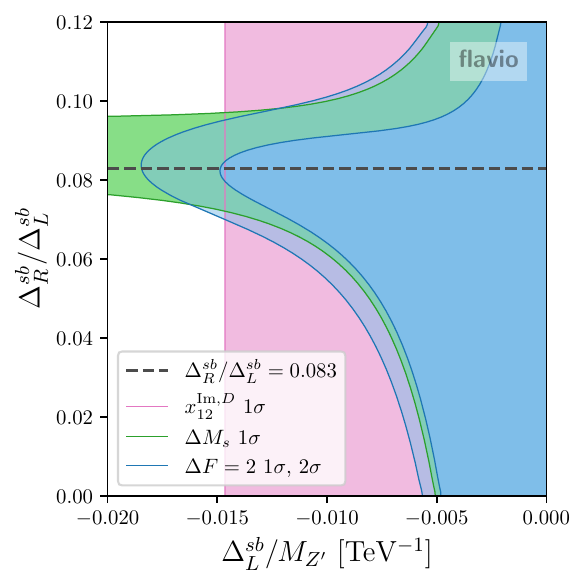}
\includegraphics[width=0.48\textwidth]{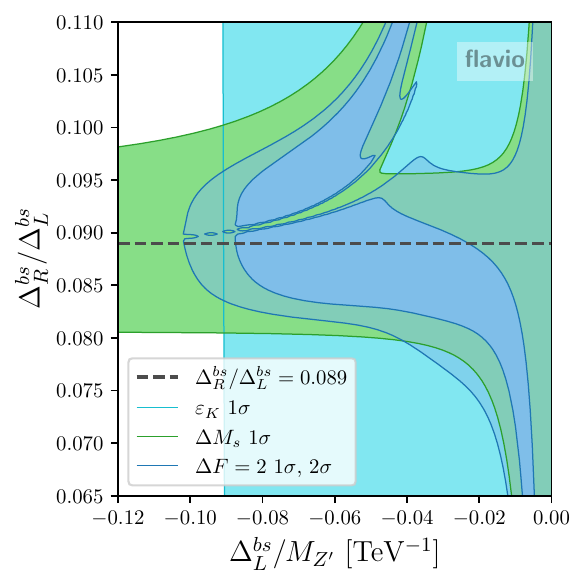}
\caption{Constraints on $Z'$-$b$-$s$ couplings from $\Delta F=2$ observables. The left panel shows the regions allowed by $\Delta M_s$ (in green) and $x_{12}^{\text{Im},D}$ (in pink)  at the $1\sigma$ level, as well as the $1\sigma$ and $2\sigma$ contours of the combined $\Delta F=2$ likelihood.
The right panel shows the regions allowed by $\Delta M_s$ (in green) and $\varepsilon_K$ correlated with $\Delta M_d$ (in cyan) at the $1\sigma$ level, as well as the $1\sigma$ and $2\sigma$ contours of the combined $\Delta F=2$ likelihood.
}
\label{fig:df2_ratio}
\end{figure}

To show how the $B_s-\bar B_s$ constraints on $\Delta^{bs}_L$ are suppressed in the presence of $\Delta^{bs}_R$ satisfying Eq.~\eqref{eq:r0_uc_dib}, we present likelihood contours in the 2D plane $\Delta_{L}^{bs}/M_{Z^\prime}$ vs.\ $\Delta^{bs}_R/\Delta^{bs}_L$ in Fig.~\ref{fig:df2_ratio}.

In the left panel of Fig.~\ref{fig:df2_ratio}, we see that the region allowed by $\Delta M_s$ at the $1\sigma$ level (in green) includes large values of $\Delta_{L}^{bs}/M_{Z^\prime}$ if the ratio $\Delta^{bs}_R/\Delta^{bs}_L$ is roughly between 0.08 and 0.10, demonstrating the suppression of ${M_{12}^{B_s}}_{\rm NP}$ discussed in Section~\ref{sec:meson_mixing_suppression}.
The pink contour shows the constraint from the $D^0-\bar D^0$ mixing observable $x_{12}^{\text{Im},D}$, which is induced by the correlation due to $\text{SU(2)}_L$ gauge invariance discussed in Section~\ref{sec:SU2_corr}, and which places a limit on the magnitude of $\Delta_{L}^{bs}/M_{Z^\prime}$.
We show the combined constraint from $\Delta F=2$ observables in blue, demonstrating that for $\Delta^{bs}_R/\Delta^{bs}_L\approx0.08$, the left-handed $Z^\prime$-$b$-$s$ coupling $\Delta_{L}^{bs}$ is allowed to be roughly two times larger than for vanishing right-handed $Z^\prime$-$b$-$s$ coupling.

In the right panel of Fig.~\ref{fig:df2_ratio}, we show a scenario in which the contributions to $D^0-\bar D^0$ mixing are also suppressed, which is e.g.\ possible in the presence of right-handed $Z^\prime$-$c$-$u$ couplings that compensate the effect of Eq.~\eqref{eq:Delta_uc_SU2}.
Consequently, the magnitude of $\Delta_{L}^{bs}/M_{Z^\prime}$ is not limited by $x_{12}^{\text{Im},D}$ and can be considerably larger than in the left panel of Fig.~\ref{fig:df2_ratio}.
However, in this case RG effects, and in particular those described in Section~\ref{sec:SMEFTRGeffects}, become important.
The cyan band shown in the right panel of Fig.~\ref{fig:df2_ratio} corresponds to the region allowed by $\varepsilon_K$ at the $1\sigma$ level, clearly limiting the magnitude of $\Delta_{L}^{bs}/M_{Z^\prime}$.
In green and blue, we again show the regions allowed by $\Delta M_s$ and by the combined $\Delta F=2$ observables, respectively.
For $\Delta^{bs}_R/\Delta^{bs}_L\approx0.085$, the left-handed $Z^\prime$-$b$-$s$ coupling $\Delta_{L}^{bs}$ is allowed to be about ten times larger than for vanishing right-handed $Z^\prime$-$b$-$s$ coupling.

The cut through the blue region visible in the right panel of Fig.~\ref{fig:df2_ratio} corresponds to a constraint from $S_{\psi\phi}$.
In this narrow region, the NP contribution to $M_{12}^{B_s}$ has a similar magnitude and opposite sign to the real part of the SM contribution, ${M_{12}^{B_s}}_{\rm NP} \approx - \re({M_{12}^{B_s}}_{\rm SM})$.
This leads to a vanishing $\re(M_{12}^{B_s})$, and consequently to $|M_{12}^{B_s}| \approx |\im(M_{12}^{B_s})|$.
Since $S_{\psi\phi}$ is approximately proportional to $\im(M_{12}^{B_s})/|M_{12}^{B_s}|$, it is considerably enhanced in this case, leading to a strong experimental constraint.
While $|M_{12}^{B_s}| \approx |\im(M_{12}^{B_s})|$ also implies a significant suppression of the central value of $\Delta M_s$, the additional dependence on the NP bag parameters increases the theoretical uncertainty of $\Delta M_s$ to about 50\% of its SM central value, making even the suppressed central value compatible with experimental data.
On the other hand, the dependence of $S_{\psi\phi}$ on the NP bag parameters is partially cancelled when the ratio $|\im(M_{12}^{B_s})|/|M_{12}^{B_s}|$ is close to unity, so that in this case $S_{\psi\phi}$ can provide a strong constraint even in the presence of large NP WCs.

In the left and right panels of Fig.~\ref{fig:df2_ratio}, the black dashed lines show the values of $\Delta^{bs}_R/\Delta^{bs}_L$ that allow for the largest magnitude of $\Delta_{L}^{bs}/M_{Z^\prime}$, $\Delta^{bs}_R/\Delta^{bs}_L=0.083$ in the presence of $D^0-\bar D^0$ mixing constraints, and $\Delta^{bs}_R/\Delta^{bs}_L=0.089$ in their absence.
We use these values as benchmark scenarios in Section~\ref{sec:DeltaB=DeltaS}.

\boldmath
\section{The Impact on  $\Delta B=\Delta S=1$ Transitions}\label{sec:DeltaB=DeltaS}
\unboldmath
In the previous section we have studied the WCs of the four-quark operators including RG effects both in the SMEFT and WET under the conditions
  of suppressed NP contributions to $\Delta F=2$ processes. At the NP scale
  $\Lambda_\text{NP}$ these conditions are summarized in (\ref{eq:r0_ds}) and
  (\ref{eq:r0_uc_dib}). In the present section we will investigate what
  is the impact of the relation in (\ref{eq:r0_uc_dib}) for the $B_s-\bar B_s$
  mixing on the transitions $b\to s\nu\bar\nu$ and $b\to s\mu^+\mu^-$.
  This involves not only $B\to K(K^*)\nu\bar\nu$ and  $B\to K(K^*)\mu^+\mu^-$  decays but also $B_s\to \mu^+\mu^-$.

  Our goal is to find out the correlations between these five decays
  and in particular the implications for $B\to K(K^*)\nu\bar\nu$ and $B_s\to \mu^+\mu^-$ taking into account the suppressions of $B\to K(K^*)\mu^+\mu^-$
  relative to the SM predictions observed by the LHCb experiment.

  To this end,
  in Sections~\ref{sec:bs_Zp_SMEFT_matching} and~\ref{sec:bs_Zp_WET_matching}  we will
  perform
  the steps done for the four-quark operators in the previous section, this time
  for semi-leptonic operators relevant for the decays considered here.
  This includes the matching between the $Z^\prime$ model to the SMEFT, RG
  running within the SMEFT, the matching of the SMEFT onto the WET and
  RG running within WET.

  Subsequently,
  in Section~\ref{sec:correlations_bsll_bsnunu}
  we will study correlations between the WCs for $b\to s\nu\bar\nu$ and $b\to s\mu^+\bar\mu^-$ transitions within the WET and their dependence
  on the $Z^\prime$ couplings to $\mu^+\mu^-$ concentrating on vector and left-handed couplings. Here the requirement of the suppression of NP contributions
  to for $B_s-\bar B_s$ mixing has an important impact on these correlations.

  In Section~\ref{sec:bs_Zp_observables} we define a number of
    observables for all decays analysed by us. We summarize their experimental
    status and discuss the relevant formfactors.
    Subsequently in Section~\ref{sec:numerical_analysis} we perform a numerical
    analysis. This includes the global fit of $Z^\prime$ couplings and in particular the correlations between various observables that are the most important
    phenomenological results of our paper.

\subsection{Matching to the SMEFT and RG Evolution in the SMEFT}\label{sec:bs_Zp_SMEFT_matching}
\unboldmath

When we match the $Z^\prime$ model defined by Eq.~\eqref{eq:Zp_couplings} to the SMEFT, the matching relations relevant for the $B\to K(K^*)\nu\bar\nu$,  $B\to K(K^*)\mu^+\mu^-$, and $B_s\to \mu^+\mu^-$ decays at tree-level are those of the quark flavour changing semi-leptonic operators (see Appendix~\ref{app:SMEFT_matching} for the full tree-level matching results),

\begin{equation}\label{eq:SMEFT_matching_bsll}
\begin{aligned}{}
 [C_{qe}]_{23ii} &= -\frac{\Delta^{e_i}_R\,\Delta^{sb}_L}{M_{Z'}^2}\,,
 \qquad
 [C_{lq}^{(1)}]_{ii23} &= -\frac{\Delta^{l_i}_L\,\Delta^{sb}_L}{M_{Z'}^2}\,,
 \\
 [C_{ed}]_{ii23} &= -\frac{\Delta^{e_i}_R\,\Delta^{sb}_R}{M_{Z'}^2}\,,
 \qquad
 [C_{ld}]_{ii23} &= -\frac{\Delta^{l_i}_L\,\Delta^{sb}_R}{M_{Z'}^2}\,.
\end{aligned}
\end{equation}
The relations in Eq.~\eqref{eq:SMEFT_matching_bsll} hold at the UV matching scale $\Lambda_{NP}\approx M_{Z^\prime}$. In order to obtain the predictions for $B\to K(K^*)\nu\bar\nu$, $B\to K(K^*)\mu^+\mu^-$, and  $B_s\to \mu^+\mu^-$ observables, we use the SMEFT RGEs to evolve the WCs down to the electroweak scale, where we match the SMEFT to the WET, which we then evolve further down using the WET RGEs to the scale at which the $B\to K(K^*)$ matrix elements are evaluated.
The  dominant contributions to the SMEFT RG running and mixing of the Wilson coefficients in Eq.~\eqref{eq:SMEFT_matching_bsll} in the first leading-log approximation are given by~\cite{Jenkins:2013zja,Jenkins:2013wua,Alonso:2013hga}
\begin{equation}\label{eq:SMEFT_RG_bsll_1}\small
 \begin{aligned}
      {[}{C}_{qe}]_{23ii}(M_Z)
    \approx&\
    [C_{qe}]_{23ii}(\Lambda_\text{NP})
    \left[
    1
    +
    [\beta_{qe}^{qe}]_{ii}\,
    \tfrac{\log\left(\frac{M_Z}{\Lambda_\text{NP}}\right)}{16 \pi^2}
    \right]
    +
    [C_{lq}^{(1)}]_{ii23}(\Lambda_\text{NP})
    \left[
     [\beta^{lq^{(1)}}_{qe}]_{ii}\,
    \tfrac{\log\left(\frac{M_Z}{\Lambda_\text{NP}}\right)}{16 \pi^2}
    \right]
    ,
    \\
      {[}{C}_{ed}]_{ii23}(M_Z)
    \approx&\
    [C_{ed}]_{ii23}(\Lambda_\text{NP})
    \left[
    1
    +
    [\beta_{ed}^{ed}]_{ii}\,
    \tfrac{\log\left(\frac{M_Z}{\Lambda_\text{NP}}\right)}{16 \pi^2}
    \right]
    +
    [C_{ld}]_{ii23}(\Lambda_\text{NP})
    \left[
     [\beta^{ld}_{ed}]_{ii}\,
    \tfrac{\log\left(\frac{M_Z}{\Lambda_\text{NP}}\right)}{16 \pi^2}
    \right]
    ,
    \\
      {[}{C}_{lq}^{(1)}]_{ii23}(M_Z)
    \approx&\
    [C_{lq}^{(1)}]_{ii23}(\Lambda_\text{NP})
    \left[
    1
    +
     [\beta_{lq^{(1)}}^{lq^{(1)}}]_{ii}\,
    \tfrac{\log\left(\frac{M_Z}{\Lambda_\text{NP}}\right)}{16 \pi^2}
    \right]
    +
    [C_{qe}]_{23ii}(\Lambda_\text{NP})
    \left[
     [\beta^{qe}_{lq^{(1)}}]_{ii}\,
    \tfrac{\log\left(\frac{M_Z}{\Lambda_\text{NP}}\right)}{16 \pi^2}
    \right]
    ,
    \\
      {[}{C}_{ld}]_{ii23}(M_Z)
    \approx&\
    [C_{ld}]_{ii23}(\Lambda_\text{NP})
    \left[
    1
    +
     [\beta_{ld}^{ld}]_{ii}\,
    \tfrac{\log\left(\frac{M_Z}{\Lambda_\text{NP}}\right)}{16 \pi^2}
    \right]
    +
    [C_{ed}]_{ii23}(\Lambda_\text{NP})
    \left[
     [\beta^{ed}_{ld}]_{ii}\,
    \tfrac{\log\left(\frac{M_Z}{\Lambda_\text{NP}}\right)}{16 \pi^2}
    \right]
    ,
 \end{aligned}
\end{equation}
where we have defined
\begin{equation}\small
 \begin{aligned}
{[}\beta_{qe}^{qe}]_{ii} &=
      \tfrac{10}{3}\,g^{\prime 2} + [\gamma^{(Y)}_{q}]_{22} + [\gamma^{(Y)}_{q}]_{33} + 2\,[\gamma^{(Y)}_{e}]_{ii}
    \,,
    \qquad
    &
    [\beta^{lq^{(1)}}_{qe}]_{ii} &=
      \tfrac{4}{3}\,g^{\prime 2} - 2\left|[Y_e]_{ii}\right|^2
    \,,
    \\
    [\beta^{ed}_{ed}]_{ii} &=
      \tfrac{16}{3}\,g^{\prime 2} + [\gamma^{(Y)}_{d}]_{22} + [\gamma^{(Y)}_{d}]_{33} + 2\,[\gamma^{(Y)}_{e}]_{ii}
    \,,
    &
    [\beta^{ld}_{ed}]_{ii} &=
      \tfrac{4}{3}\,g^{\prime 2} - 2\left|[Y_e]_{ii}\right|^2
    \,,
    \\
    [\beta^{lq^{(1)}}_{lq^{(1)}}]_{ii} &=
      -\tfrac{1}{3}\,g^{\prime 2} + [\gamma^{(Y)}_{q}]_{22} + [\gamma^{(Y)}_{q}]_{33} + 2\,[\gamma^{(Y)}_{e}]_{ii}
    \,,
    &
    [\beta^{qe}_{lq^{(1)}}]_{ii} &=
      \tfrac{2}{3}\,g^{\prime 2} - \left|[Y_e]_{ii}\right|^2
    \,,
    \\
    [\beta^{ld}_{ld}]_{ii} &=
      -\tfrac{4}{3}\,g^{\prime 2} + [\gamma^{(Y)}_{d}]_{22} + [\gamma^{(Y)}_{d}]_{33} + 2\,[\gamma^{(Y)}_{l}]_{ii}
    \,,
    &
    [\beta^{ed}_{ld}]_{ii} &=
      \tfrac{2}{3}\,g^{\prime 2} - \left|[Y_e]_{ii}\right|^2
    \,.
    \\
 \end{aligned}
\end{equation}
The quantities $\gamma^{(Y)}_{q}$ and $\gamma^{(Y)}_{d}$ are defined in Eq.~\eqref{eq:gammaY} and
\begin{equation}
  \gamma^{(Y)}_{l}  = \frac{1}{2}[Y_e Y_e^\dagger]\,,
  \qquad
  \gamma^{(Y)}_{e}  = [Y_e^\dagger Y_e]\,.
\end{equation}
Apart from the WCs in Eq.~\eqref{eq:SMEFT_matching_bsll}, which are generated by the tree-level matching, additional WCs relevant for $B\to K(K^*)\nu\bar\nu$,  $B\to K(K^*)\mu^+\mu^-$, and  $B_s\to \mu^+\mu^-$ decays are generated through SMEFT RG effects.
In particular, $[C_{lq}^{(1)}]_{ii23}(\Lambda_\text{NP})$ generates a contribution to $[C_{lq}^{(3)}]_{ii23}(M_Z)$, and both the semi-leptonic WCs in Eq.~\eqref{eq:SMEFT_matching_bsll} and the four-quark WCs in Eq.~\eqref{eq:SMEFT_matching_4q} generate contributions to the coefficients $[C_{\phi q}^{(1)}]_{23}(M_Z)$,  $[C_{\phi q}^{(3)}]_{23}(M_Z)$, and $[C_{\phi d}]_{23}(M_Z)$, which correspond to effective $Z$-$b$-$s$ couplings.
In the first leading-log approximation, these contributions are given by~\cite{Jenkins:2013zja,Jenkins:2013wua,Alonso:2013hga}
\begin{equation}\label{eq:SMEFT_RG_bsll_2}\small
 \begin{aligned}
      {[}{C}_{lq}^{(3)}]_{ii23}(M_Z)
    \approx&\
    \tfrac{\log\left(\frac{M_Z}{\Lambda_\text{NP}}\right)}{16 \pi^2}
    \,
     [\beta^{lq^{(1)}}_{lq^{(3)}}]\,
    [C_{lq}^{(1)}]_{ii23}(\Lambda_\text{NP})
    \,,
    \\
      {[}{C}_{\phi q}^{(3)}]_{23}(M_Z)
    \approx&\
    \tfrac{\log\left(\frac{M_Z}{\Lambda_\text{NP}}\right)}{16 \pi^2}
     [\beta^{qq^{(1)}}_{\phi q^{(3)}}]\,
    [C_{qq}^{(1)}]_{2323}(\Lambda_\text{NP})
    ,
    \\
      {[}{C}_{\phi q}^{(1)}]_{23}(M_Z)
    \approx&\
    \tfrac{\log\left(\frac{M_Z}{\Lambda_\text{NP}}\right)}{16 \pi^2}
    \,
    \Big[
     [\beta^{lq^{(1)}}_{\phi q^{(1)}}]_{ii}\,
    [C_{lq}^{(1)}]_{ii23}(\Lambda_\text{NP})
    +
     [\beta^{qe}_{\phi q^{(1)}}]_{ii}\,
    [C_{qe}]_{23ii}(\Lambda_\text{NP})
    \\
      &\qquad\qquad
      +
     [\beta^{qq^{(1)}}_{\phi q^{(1)}}]_{2323}\,
    [C_{qq}^{(1)}]_{2323}(\Lambda_\text{NP})
    +
     [\beta^{qq^{(1)}}_{\phi q^{(1)}}]_{2332}\,
    [C_{qq}^{(1)}]_{2332}(\Lambda_\text{NP})
     \Big]
    \,,
    \\
      {[}{C}_{\phi d}]_{23}(M_Z)
    \approx&\
    \tfrac{\log\left(\frac{M_Z}{\Lambda_\text{NP}}\right)}{16 \pi^2}
    \,
    \Big[
     [\beta^{ld}_{\phi d}]_{ii}\,
    [C_{ld}]_{ii23}(\Lambda_\text{NP})
    +
     [\beta^{ed}_{\phi d}]_{ii}\,
    [C_{ed}]_{ii23}(\Lambda_\text{NP})
    \\
      &\qquad\qquad
      +
     [\beta^{qd^{(1)}}_{\phi d}]_{2323}\,
    [C_{qd}^{(1)}]_{2323}(\Lambda_\text{NP})
    +
     [\beta^{qd^{(1)}}_{\phi d}]_{2332}\,
    [C_{qd}^{(1)}]_{2332}^*(\Lambda_\text{NP})
     \Big]
    \,,
 \end{aligned}
\end{equation}
where we have defined
\begin{equation}\small
 \begin{aligned}
{[}\beta^{lq^{(1)}}_{lq^{(3)}}] &=
      3\,g^{2}
      \,,
    &
    [\beta^{qq^{(1)}}_{\phi q^{(3)}}] &=
      -2\,[Y_u Y_u^\dagger]_{32}
\,,
\\
    [\beta^{lq^{(1)}}_{\phi q^{(1)}}]_{ii} &=
      -\tfrac{2}{3}\,g^{\prime 2} - 2 \left|[Y_e]_{ii}\right|^2
    \,,
    \qquad
    &
    [\beta^{qq^{(1)}}_{\phi q^{(1)}}]_{2323} &=
      14\,[Y_u Y_u^\dagger]_{32}
\,,
    \\
    [\beta^{qe}_{\phi q^{(1)}}]_{ii} &=
      -\tfrac{2}{3}\,g^{\prime 2} + 2 \left|[Y_e]_{ii}\right|^2
    \,,
    &
    [\beta^{qq^{(1)}}_{\phi q^{(1)}}]_{2332} &=
      6\,[Y_u Y_u^\dagger]_{23}
\,,
    \\
    [\beta^{ld}_{\phi d}]_{ii} &=
      -\tfrac{2}{3}\,g^{\prime 2} - 2 \left|[Y_e]_{ii}\right|^2
    \,,
    &
    [\beta^{qd^{(1)}}_{\phi d}]_{2323} &=
      6\,[Y_u Y_u^\dagger]_{32}
\,,
    \\
    [\beta^{ed}_{\phi d}]_{ii} &=
      -\tfrac{2}{3}\,g^{\prime 2} + 2 \left|[Y_e]_{ii}\right|^2
    \,,
&
    [\beta^{qd^{(1)}}_{\phi d}]_{2332} &=
      6\,[Y_u Y_u^\dagger]_{23}
\,.
 \end{aligned}
\end{equation}
In $Z^\prime$ models with real $\Delta^{bs}_L$, we have $[C_{qq}^{(1)}]_{2323}(\Lambda_\text{NP})={\frac{1}{2}} [C_{qq}^{(1)}]_{2332}(\Lambda_\text{NP})<0$.
In this case, the contribution to $[C_{\phi q}^{(1)}]_{23}(M_Z)$ from left-handed four-quark operators, which is usually the dominant one in the scenario given by Eq.~\eqref{eq:r0_uc_dib}, is always {\em negative.}

\boldmath
\subsection{Matching to the WET and RG Evolution in the WET}\label{sec:bs_Zp_WET_matching}
\unboldmath

For the low-energy phenomenology of rare semi-leptonic decays, we work  in the  WET and define the effective Hamiltonian at the $b$-quark scale $\mu_b=4.8\,\text{GeV}$,
\begin{equation}
 {\cal H}_{\rm eff} = {\cal H}_{\rm eff, SM} + {\cal H}_{\rm eff, NP}\,,
\end{equation}
where the first and second term contains the SM and NP contributions, respectively.
For the NP part, we consider
\begin{equation}
 {\cal H}_{\rm eff, NP} = {\cal H}_{\rm eff, NP}^{bs\nu\nu} + {\cal H}_{\rm eff, NP}^{bs\ell\ell}\,,
\end{equation}
where ${\cal H}_{\rm eff, NP}^{bs\nu\nu}$ and ${\cal H}_{\rm eff, NP}^{bs\ell\ell}$ parameterise the $b\to s \nu\bar\nu$ and $b\to s \ell^+\ell^-$ transitions, respectively.
The contributions relevant for $Z^\prime$ models with couplings to left- and right-handed quark currents are
\begin{equation} \label{eq:Heffnunu}
{\cal H}_{\rm eff, NP}^{bs\nu\nu} = -\mathcal{N}  \left(C^{bs\nu\nu}_L O^{bs\nu\nu}_L +C^{bs\nu\nu}_R O^{bs\nu\nu}_R  \right) ~+~ {\rm h.c.} ~,
\end{equation}
and
\begin{equation}\label{eq:HeffqllZprime}
  \mathcal{H}_{\rm eff, NP}^{bs\ell\ell}
=
-  \mathcal{N} \sum_{i = 9,10} \left(C_i^{bs\ell\ell} O_i^{bs\ell\ell}+C_i^{\prime,bs\ell\ell} O_i^{\prime,bs\ell\ell}\right) ~+~ {\rm h.c.} ~,
\end{equation}
with the normalization factor
\begin{equation}\label{eq:Heff_normalization}
 \mathcal{N}=\frac{4\,G_F}{\sqrt{2}}\frac{\alpha}{4\pi}V_{ts}^* V_{tb}
 \,,
\end{equation}
which renders all the WCs in Eqs.~\eqref{eq:Heffnunu} and \eqref{eq:HeffqllZprime} dimensionless.
Note that we define the WCs $C_i$ to correspond to NP contributions only, while we explicitly denote the SM contributions as $C_{i,\text{SM}}$.
The $b\to s \nu\bar\nu$ operators are defined as
\begin{equation}
{O^{bs\nu\nu}_{L} =
(\bar{s}  \gamma_{\mu} P_L b)(\bar{\nu} \gamma^\mu(1- \gamma_5) \nu)~,\qquad
O^{bs\nu\nu}_{R} =(\bar{s}\gamma_{\mu} P_R b)(  \bar{\nu} \gamma^\mu(1- \gamma_5) \nu)~,}
\end{equation}
with only the first one present in the SM,
and the $b\to s \ell^+\ell^-$ operators are given by
\begin{equation}\label{QAQVL}
O_9^{bs\ell\ell}  = (\bar s\gamma_\mu P_L b)(\bar \ell\gamma^\mu\ell),\qquad
O_{10}^{bs\ell\ell}  = (\bar s\gamma_\mu P_L b)(\bar \ell\gamma^\mu\gamma_5\ell),
\end{equation}
\begin{equation}\label{QAQVR}
O_9^{\prime\,bs\ell\ell}  = (\bar s\gamma_\mu P_R b)(\bar \ell\gamma^\mu\ell), \qquad
O_{10}^{\prime\,bs\ell\ell}  = (\bar s\gamma_\mu P_R b)(\bar \ell\gamma^\mu\gamma_5\ell)\,,
\end{equation}
with only the first two present in the SM. The coefficients $C_R^{bs\nu\nu}$, $C_9^{\prime\,bs\ell\ell}$ and $ C_{10}^{\prime\,bs\ell\ell}$, which are absent in the SM, signal the
presence of flavour violating right-handed quark currents.


In order to connect the SMEFT Wilson coefficient in Section~\ref{sec:bs_Zp_SMEFT_matching} to the WET Wilson coefficients defined in Eqs.~\eqref{eq:Heffnunu} and \eqref{eq:HeffqllZprime}, we match the SMEFT to the WET at the electroweak scale and find the relations

  \begin{equation}\label{eq:SMEFTtoWETmatching}
  \begin{aligned}
   2\,\mathcal{N}\,C_{9}^{bs\ell_i\ell_i} &= [C_{qe}]_{23ii} + [C_{lq}^{(1)}]_{ii23} + [C_{lq}^{(3)}]_{ii23} - \zeta\,c_Z\,,
   \\
   2\,\mathcal{N}\,C_{10}^{bs\ell_i\ell_i} &= [C_{qe}]_{23ii} - [C_{lq}^{(1)}]_{ii23} - [C_{lq}^{(3)}]_{ii23} + c_Z\,,
   \\
   2\,\mathcal{N}\,C^{bs\nu_i\nu_i}_L &= [C_{lq}^{(1)}]_{ii23} - [C_{lq}^{(3)}]_{ii23} + c_Z\,,
   \\
   2\,\mathcal{N}\,C_{9}^{\prime\,bs\ell_i\ell_i} &= [C_{ed}]_{ii23}  + [C_{ld}]_{ii23} -\zeta\, c_Z^\prime\,,
   \\
   2\,\mathcal{N}\,C_{10}^{\prime\,bs\ell_i\ell_i} &= [C_{ed}]_{ii23} -[C_{ld}]_{ii23} + c_Z^\prime\,,
   \\
   2\,\mathcal{N}\,C^{bs\nu_i\nu_i}_R &= [C_{ld}]_{ii23} + c_Z^\prime\,,
  \end{aligned}
  \end{equation}
  where $\mathcal{N}$ is the normalization factor defined in~\eqref{eq:Heff_normalization}, $c_Z$ and $c_Z^\prime$ denote the contribution from modified $Z$ couplings,
  \begin{equation}
   c_Z = [C_{\phi q}^{(1)}]_{23} + [C_{\phi q}^{(3)}]_{23}\,,
   \qquad
   c_Z^\prime = [C_{\phi d}]_{23}\,,
  \end{equation}
  and $\zeta = 1- 4 s_w^2\approx 0.08$ is the accidentally small vector coupling of the $Z$ to the charged leptons.
  While the WET WCs are dimensionless due to the normalization factor $\mathcal{N}$, the SMEFT WCs are dimensionful and proportional to $1/\Lambda_\text{NP}^{2}$ with $\Lambda_\text{NP}$ being the NP scale.

  It should be emphasized that in $Z^\prime$ models
    $[C_{lq}^{(3)}]_{ii23},$ $ c_Z$ and $ c_Z^\prime$ vanish at the NP scale
    $\Lambda_\text{NP}$.
However, as discussed in Section~\ref{sec:bs_Zp_SMEFT_matching}, they all can be generated at the electroweak scale through RG
running from $\Lambda_\text{NP}$ down to the electroweak scale.
At the scale $M_Z$, the RG induced contribution to $c_Z$, using the expressions for $[C_{\phi q}^{(1)}]_{23}$
and $[C_{\phi q}^{(3)}]_{23}$ from Section~\ref{sec:bs_Zp_SMEFT_matching}, can be expressed in the first leading-log approximation  as
\begin{equation}\label{eq:cZ_RG}
\begin{aligned}
 c_Z
 &\approx
 \tfrac{\log\left(\frac{\Lambda_\text{NP}}{M_Z}\right)}{16 \pi^2}
 \,\frac{\Delta^{bs}_L}{M_{Z^\prime}^2} \,
 \Bigg(
 2 \left|[Y_e]_{ii}\right|^2 \left( \Delta_R^{e_i} - \Delta_L^{l_i} \right)
 -\tfrac{2}{3}\,g^{\prime 2}  \left( \Delta_R^{e_i} + \Delta_L^{l_i} \right)
 +12\, \re\Big( [Y_u Y_u^\dagger]_{32}\,\Delta^{bs}_L  \Big)
 \Bigg)
 \\
 &\approx
 \tfrac{\log\left(\frac{\Lambda_\text{NP}}{M_Z}\right)}{16 \pi^2}
 \,\frac{\Delta^{bs}_L}{M_{Z^\prime}^2} \,
 \Bigg(
 -0.33\,\re\Big( \Delta^{bs}_L  \Big)
 \Bigg)\,,
\end{aligned}
\end{equation}
and $ c_Z^\prime$  is related to  $c_Z$ by
\begin{equation}
 c_Z^\prime = \frac{\Delta^{bs}_R}{\Delta^{bs}_L}\, c_Z\,.
\end{equation}
In the second line of Eq.~\eqref{eq:cZ_RG} we have inserted the SM parameters at the scale $\Lambda_\text{NP}$, using $\Lambda_\text{NP}=5\,\text{TeV}$ as our reference scale, for which in particular the top Yukawa coupling is suppressed compared to the electroweak scale, $y_t(5\,\text{TeV})\approx0.81$.
We observe that for sizable $\Delta^{bs}_L$, for which the terms proportional to $\Delta_R^{e_i}$ and $\Delta_L^{l_i}$ can be safely neglected, the RG induced contribution to $\re(c_Z)$ is always {\em negative}.
Since the normalization factor $\mathcal{N}$ is approximately real and negative, this leads to a contribution to $\re(C_{10}^{bs\ell\ell})$ that is always positive, which in particular {\em suppresses} the $B_s\to\mu^+\mu^-$ branching ratio \footnote{Recall that the SM contribution to $C_{10}^{bs\ell\ell}$ is negative.}.

Moreover,
in the process of electroweak symmetry breaking $ c_Z$ and $ c_Z^\prime$ can receive
contributions from $Z^\prime-Z$ mixing. This mixing is clearly model dependent
and we will not include it in our analysis. It has been investigated in 331 models in \cite{Buras:2014yna}.

As the hadronic matrix elements entering $b\to s\nu\bar\nu$ and $b\to s\ell^+\ell^-$ processes are evaluated by the Lattice QCD collaborations at the scale $\mu_b=4.8\, \text{GeV}$, the coefficients at the scale $M_Z$ in Eq.~\eqref{eq:SMEFTtoWETmatching} have to be evolved down to $\mu_b$ using the WET RGEs.
For WCs involving charged leptons, we express those at the scale $\mu_b$ in terms of those at $M_Z$ and the RG evolution matrix $U^{bs\ell\ell}(\mu_b,M_Z)$,
\begin{equation}\label{eq:WET_RG_evolution_bsll}
 \begin{pmatrix}
  C_{9}^{bs\ell\ell}(\mu_b)\\
  C_{10}^{bs\ell\ell}(\mu_b)\\
  C_{9}^{\prime\,bs\ell\ell}(\mu_b)\\
  C_{10}^{\prime\,bs\ell\ell}(\mu_b)
 \end{pmatrix}
 =
 U^{bs\ell\ell}(\mu_b,M_Z)
 \begin{pmatrix}
  C_{9}^{bs\ell\ell}(M_Z)\\
  C_{10}^{bs\ell\ell}(M_Z)\\
  C_{9}^{\prime\,bs\ell\ell}(M_Z)\\
  C_{10}^{\prime\,bs\ell\ell}(M_Z)
 \end{pmatrix}\,.
\end{equation}
Solving the leading order RGEs, the evolution matrix is given by
\begin{align}
U^{bs\ell\ell}(\mu_b,M_Z)
&=
\setlength{\arraycolsep}{3pt}
\begin{pmatrix}
 0.995 & 0.008 & 0 & 0 \\
 0.008 & 1.000 & 0 & 0 \\
 0 & 0 & \phantom{-}0.995 & -0.008 \\
 0 & 0 &-0.008 & \phantom{-}1.000 \\
\end{pmatrix}\,.
\end{align}
The WCs involving neutrinos are essentially invariant under the RG evolution and we simply use
\begin{equation}
 C^{bs\nu\nu}_L(\mu_b) = C^{bs\nu\nu}_L(M_Z)\,,
 \qquad
 C^{bs\nu\nu}_R(\mu_b) = C^{bs\nu\nu}_R(M_Z)\,.
\end{equation}
{%
Combining the results of the SMEFT matching, Eq.~\eqref{eq:SMEFT_matching_bsll}, the SMEFT RG evolution, Eqs.~\eqref{eq:SMEFT_RG_bsll_1} and~\eqref{eq:SMEFT_RG_bsll_2}, the WET matching, Eq.~\eqref{eq:SMEFTtoWETmatching}, and the WET RG evolution, Eq.~\eqref{eq:WET_RG_evolution_bsll}, we can express the WET WCs at the scale $\mu_b$ in terms of the $Z^\prime$ couplings.
For the WCs involving charged leptons, we find
\begin{align}\label{eq:C_low_bsll}
C_{9}^{bs\ell_i\ell_i}(\mu_b)
&\approx
\frac{
    \Delta_L^{bs}\, \Delta_L^{l_i}
}{
    -2\,\mathcal{N}\,M_{Z^\prime}^2
}
&&\!\!\!\!\!\times
\Bigg[
    0.987
    +1.003\,\frac{\Delta_R^{e_i}}{\Delta_L^{l_i}}
    \nonumber\\
    &&&\!\!\!\!\!+
  \Bigg(
- 1.04
- 0.55\, \frac{\Delta_R^{e_i}}{\Delta_L^{l_i}}
- 0.02\, \frac{ \re(\Delta_L^{bs}) }{\Delta_L^{l_i}}
\Bigg)
\times
10^{-2}\,\log\!\left(\tfrac{M_{Z^\prime}}{M_Z}\right)
\Bigg]\,,
\displaybreak[1]\nonumber\\
C_{10}^{bs\ell_i\ell_i}(\mu_b)
&\approx
\frac{
    \Delta_L^{bs}\, \Delta_L^{l_i}
}{
    -2\,\mathcal{N}\,M_{Z^\prime}^2
}
&&\!\!\!\!\!\times
\Bigg[
    -0.992
    +1.008\,\frac{\Delta_R^{e_i}}{\Delta_L^{l_i}}
    \nonumber\\
    &&&\!\!\!\!\!+
  \Bigg(
0.87
- 0.38\, \frac{\Delta_R^{e_i}}{\Delta_L^{l_i}}
+ 0.21\, \frac{ \re(\Delta_L^{bs}) }{\Delta_L^{l_i}}
\Bigg)
\times
10^{-2}\,\log\!\left(\tfrac{M_{Z^\prime}}{M_Z}\right)
\Bigg]\,,
\displaybreak[1]\nonumber\\
C_{9}^{\prime\,bs\ell_i\ell_i}(\mu_b)
&\approx
\frac{
    \Delta_R^{bs}\, \Delta_L^{l_i}
}{
    -2\,\mathcal{N}\,M_{Z^\prime}^2
}
&&\!\!\!\!\!\times
\Bigg[
    1.003
    +0.987\,\frac{\Delta_R^{e_i}}{\Delta_L^{l_i}}
    \nonumber\\
    &&&\!\!\!\!\!+
  \Bigg(
-0.51\, \frac{\Delta_R^{e_i}}{\Delta_L^{l_i}}
- 0.02\, \frac{ \re(\Delta_L^{bs}) }{\Delta_L^{l_i}}
\Bigg)
\times
10^{-2}\,\log\!\left(\tfrac{M_{Z^\prime}}{M_Z}\right)
\Bigg]\,,
\displaybreak[1]\nonumber\\
C_{10}^{\prime\,bs\ell_i\ell_i}(\mu_b)
&\approx
\frac{
    \Delta_R^{bs}\, \Delta_L^{l_i}
}{
    -2\,\mathcal{N}\,M_{Z^\prime}^2
}
&&\!\!\!\!\!\times
\Bigg[
    -1.008
    +0.992\,\frac{\Delta_R^{e_i}}{\Delta_L^{l_i}}
    \nonumber\\
    &&&\!\!\!\!\!+
  \Bigg(
  -0.17
- 0.33\, \frac{\Delta_R^{e_i}}{\Delta_L^{l_i}}
+ 0.21\, \frac{ \re(\Delta_L^{bs}) }{\Delta_L^{l_i}}
\Bigg)
\times
10^{-2}\,\log\!\left(\tfrac{M_{Z^\prime}}{M_Z}\right)
\Bigg]\,.
\end{align}
All these expressions depend on the ratio of left-handed to right-handed lepton couplings, $\Delta_R^{e_i}/\Delta_L^{l_i}$, and are therefore different for vector and purely left-handed $Z^\prime$ lepton couplings.
On the other hand, the WCs involving neutrinos are independent of $\Delta_R^{e_i}$, and are given by
\begin{equation}\label{eq:C_low_bsnunu}
\begin{aligned}
C_{L}^{bs\nu_i\nu_i}(\mu_b)
&\approx
\frac{
    \Delta_L^{bs}\,\Delta_L^{l_i}
}{
    -2\,\mathcal{N}\,M_{Z^\prime}^2
}
&&\!\!\!\!\!\times
\Bigg[
    1
+\Bigg(
0.63
+ 0.21\, \frac{ \re(\Delta_L^{bs}) }{\Delta_L^{l_i}}
\Bigg)
\times
10^{-2}\,\log\!\left(\tfrac{M_{Z^\prime}}{M_Z}\right)
\Bigg]\,,
\\
C_{R}^{bs\nu_i\nu_i}(\mu_b)
&\approx
\frac{
    \Delta_R^{bs}\, \Delta_L^{l_i}
}{
    -2\,\mathcal{N}\,M_{Z^\prime}^2
}
&&\!\!\!\!\!\times
\Bigg[
    1 +
  \Bigg(
  0.17
+ 0.21\, \frac{ \re(\Delta_L^{bs}) }{\Delta_L^{l_i}}
\Bigg)
\times
10^{-2}\,\log\!\left(\tfrac{M_{Z^\prime}}{M_Z}\right)
\Bigg]\,.
\end{aligned}
\end{equation}
For small $Z^\prime$-lepton couplings and large $Z^\prime$-b-s couplings, for which $\left|\frac{ \re(\Delta_L^{bs}) }{\Delta_L^{l_i}}\right|=\mathcal{O}(100)$, the contributions proportional to this ratio become very relevant, since such a large ratio compensates for the loop suppression.
However, as we will see in Section~\ref{sec:fit_global}, ${b\to s\ell^+\ell^-}$ data combined with the bounds from $D^0-\bar D^0$ and $K^0-\bar K^0$ mixing discussed in Section~\ref{sec:fit_df2} require
$\left|\frac{ \re(\Delta_L^{bs}) }{\Delta_L^{l_i}}\right|\lesssim 10$, and thus the RG effects in $C_{L}^{bs\nu_i\nu_i}(\mu_b)$ and $C_{R}^{bs\nu_i\nu_i}(\mu_b)$ are very small.

The special cases of vector and left-handed $Z^\prime$-lepton couplings correspond to $\Delta_R^{e_i}/\Delta_L^{l_i}=1$ and $\Delta_R^{e_i}/\Delta_L^{l_i}=0$, respectively. In these two cases, the expressions for WCs involving charged leptons simplify considerably.

For vector $Z^\prime$-lepton couplings we find
\begin{align}\label{eq:C_low_bsll_vector}
C_{9}^{bs\ell_i\ell_i}(\mu_b)
&\approx
\frac{
    \Delta_L^{bs}\, \Delta_L^{l_i}
}{
    -2\,\mathcal{N}\,M_{Z^\prime}^2
}
\times
\Bigg[
    1.99+
  \Bigg(
- 1.59
- 0.02\, \frac{ \re(\Delta_L^{bs}) }{\Delta_L^{l_i}}
\Bigg)
\times
10^{-2}\,\log\!\left(\tfrac{M_{Z^\prime}}{M_Z}\right)
\Bigg]\,,
\displaybreak[1]\nonumber\\
C_{10}^{bs\ell_i\ell_i}(\mu_b)
&\approx
\frac{
    \Delta_L^{bs}\, \Delta_L^{l_i}
}{
    -2\,\mathcal{N}\,M_{Z^\prime}^2
}
\times
\Bigg[
    0.016
    +
  \Bigg(
0.49
+ 0.21\, \frac{ \re(\Delta_L^{bs}) }{\Delta_L^{l_i}}
\Bigg)
\times
10^{-2}\,\log\!\left(\tfrac{M_{Z^\prime}}{M_Z}\right)
\Bigg]\,,
\displaybreak[1]\nonumber\\
C_{9}^{\prime\,bs\ell_i\ell_i}(\mu_b)
&\approx
\frac{
    \Delta_R^{bs}\, \Delta_L^{l_i}
}{
    -2\,\mathcal{N}\,M_{Z^\prime}^2
}
\times
\Bigg[
    1.99
    +
  \Bigg(
-0.51
- 0.02\, \frac{ \re(\Delta_L^{bs}) }{\Delta_L^{l_i}}
\Bigg)
\times
10^{-2}\,\log\!\left(\tfrac{M_{Z^\prime}}{M_Z}\right)
\Bigg]\,,
\displaybreak[1]\nonumber\\
C_{10}^{\prime\,bs\ell_i\ell_i}(\mu_b)
&\approx
\frac{
    \Delta_R^{bs}\, \Delta_L^{l_i}
}{
    -2\,\mathcal{N}\,M_{Z^\prime}^2
}
\times
\Bigg[
    -0.016
    +
  \Bigg(
  -0.50
+ 0.21\, \frac{ \re(\Delta_L^{bs}) }{\Delta_L^{l_i}}
\Bigg)
\times
10^{-2}\,\log\!\left(\tfrac{M_{Z^\prime}}{M_Z}\right)
\Bigg]\,.
\end{align}
As mentioned above, we observe that $C_{10}^{bs\ell_i\ell_i}(\mu_b)$ and $C_{10}^{\prime\,bs\ell_i\ell_i}(\mu_b)$ are entirely generated from RG effects.
We find the following correlations between $C_{9}^{(\prime)\,bs\ell_i\ell_i}(\mu_b)$ and $C_{10}^{(\prime)\,bs\ell_i\ell_i}(\mu_b)$:
\begin{equation}
\begin{aligned}
 C_{10}^{bs\ell_i\ell_i}(\mu_b)
 &\approx
 0.01\,C_{9}^{bs\ell_i\ell_i}(\mu_b)
 \times
 \left[
 0.80 + \left(0.25 + 0.11\, \frac{ \re(\Delta_L^{bs}) }{\Delta_L^{l_i}} \right) \log\!\left(\tfrac{M_{Z^\prime}}{M_Z}\right)
 \right]\,,
 \\
 C_{10}^{\prime\,bs\ell_i\ell_i}(\mu_b)
 &\approx
 0.01\,C_{9}^{\prime\,bs\ell_i\ell_i}(\mu_b)
 \times
 \left[
 -0.80 + \left(-0.25 + 0.11\, \frac{ \re(\Delta_L^{bs}) }{\Delta_L^{l_i}} \right) \log\!\left(\tfrac{M_{Z^\prime}}{M_Z}\right)
 \right]\,,
\end{aligned}
\end{equation}
which for $\left|\frac{ \re(\Delta_L^{bs}) }{\Delta_L^{l_i}}\right|\ll1$ and $M_{Z^\prime}=5\,\text{TeV}$ results in
\begin{equation}
 C_{10}^{bs\ell_i\ell_i}(\mu_b)
 \approx
 0.02\,C_{9}^{bs\ell_i\ell_i}(\mu_b)\,,
 \qquad
 C_{10}^{\prime\,bs\ell_i\ell_i}(\mu_b)
 \approx
 -0.02\,C_{9}^{\prime\,bs\ell_i\ell_i}(\mu_b)\,.
\end{equation}
Since $b\to s\ell^+\ell^-$ data requires $\Delta_L^{bs}/\Delta_L^{l_i}<0$, the correlation between $C_{9}^{bs\ell_i\ell_i}(\mu_b)$ and $C_{10}^{bs\ell_i\ell_i}(\mu_b)$ can turn into an anti-correlation for large values of $\left|\frac{ \re(\Delta_L^{bs}) }{\Delta_L^{l_i}}\right|\gtrsim 5$, while $C_{9}^{\prime\,bs\ell_i\ell_i}(\mu_b)$ and $C_{10}^{\prime\,bs\ell_i\ell_i}(\mu_b)$ are always anti-correlated.
But in any case, since $D^0-\bar D^0$ and $K^0-\bar K^0$ mixing constraints require
$\left|\frac{ \re(\Delta_L^{bs}) }{\Delta_L^{l_i}}\right|\lesssim 10$, the values of $C_{10}^{(\prime)\,bs\ell_i\ell_i}(\mu_b)$ are usually only few percent of $C_{9}^{(\prime)\,bs\ell_i\ell_i}(\mu_b)$.

For left-handed $Z^\prime$-lepton couplings we find
\begin{align}\label{eq:C_low_bsll_left}
C_{9}^{bs\ell_i\ell_i}(\mu_b)
&\approx
\frac{
    \Delta_L^{bs}\, \Delta_L^{l_i}
}{
    -2\,\mathcal{N}\,M_{Z^\prime}^2
}
\times
\Bigg[
    0.987+
  \Bigg(
- 1.04
- 0.02\, \frac{ \re(\Delta_L^{bs}) }{\Delta_L^{l_i}}
\Bigg)
\times
10^{-2}\,\log\!\left(\tfrac{M_{Z^\prime}}{M_Z}\right)
\Bigg]\,,
\displaybreak[1]\nonumber\\
C_{10}^{bs\ell_i\ell_i}(\mu_b)
&\approx
\frac{
    \Delta_L^{bs}\, \Delta_L^{l_i}
}{
    -2\,\mathcal{N}\,M_{Z^\prime}^2
}
\times
\Bigg[
    -0.992
    +
  \Bigg(
0.87
+ 0.21\, \frac{ \re(\Delta_L^{bs}) }{\Delta_L^{l_i}}
\Bigg)
\times
10^{-2}\,\log\!\left(\tfrac{M_{Z^\prime}}{M_Z}\right)
\Bigg]\,,
\displaybreak[1]\nonumber\\
C_{9}^{\prime\,bs\ell_i\ell_i}(\mu_b)
&\approx
\frac{
    \Delta_R^{bs}\, \Delta_L^{l_i}
}{
    -2\,\mathcal{N}\,M_{Z^\prime}^2
}
\times
\Bigg[
    1.003
    +\Bigg(
- 0.02\, \frac{ \re(\Delta_L^{bs}) }{\Delta_L^{l_i}}
\Bigg)
\times
10^{-2}\,\log\!\left(\tfrac{M_{Z^\prime}}{M_Z}\right)
\Bigg]\,,
\displaybreak[1]\nonumber\\
C_{10}^{\prime\,bs\ell_i\ell_i}(\mu_b)
&\approx
\frac{
    \Delta_R^{bs}\, \Delta_L^{l_i}
}{
    -2\,\mathcal{N}\,M_{Z^\prime}^2
}
\times
\Bigg[
    -1.008
    +
  \Bigg(
  -0.17
+ 0.21\, \frac{ \re(\Delta_L^{bs}) }{\Delta_L^{l_i}}
\Bigg)
\times
10^{-2}\,\log\!\left(\tfrac{M_{Z^\prime}}{M_Z}\right)
\Bigg]\,.
\end{align}
In this case, $C_{10}^{bs\ell_i\ell_i}(\mu_b)$ and $C_{10}^{\prime\,bs\ell_i\ell_i}(\mu_b)$ are already generated from tree-level matching.
We find the following correlations between $C_{9}^{(\prime)\,bs\ell_i\ell_i}(\mu_b)$ and $C_{10}^{(\prime)\,bs\ell_i\ell_i}(\mu_b)$:
\begin{equation}
\begin{aligned}
 C_{10}^{bs\ell_i\ell_i}(\mu_b)
 &\approx
 -C_{9}^{bs\ell_i\ell_i}(\mu_b)
 \times
 \left[
 1.005 + \left(0.18 - 0.19\, \frac{ \re(\Delta_L^{bs}) }{\Delta_L^{l_i}} \right)\times 10^{-2}\,\log\!\left(\tfrac{M_{Z^\prime}}{M_Z}\right)
 \right]\,,
 \\
 C_{10}^{\prime\,bs\ell_i\ell_i}(\mu_b)
 &\approx
 -C_{9}^{\prime\,bs\ell_i\ell_i}(\mu_b)
 \times
 \left[
 1.005 + \left(0.17 - 0.19\, \frac{ \re(\Delta_L^{bs}) }{\Delta_L^{l_i}} \right)\times 10^{-2}\,\log\!\left(\tfrac{M_{Z^\prime}}{M_Z}\right)
 \right]\,,
\end{aligned}
\end{equation}
which for $\left|\frac{ \re(\Delta_L^{bs}) }{\Delta_L^{l_i}}\right|\ll1$ and $M_{Z^\prime}=5\,\text{TeV}$ results in
\begin{equation}
 C_{10}^{(\prime)\,bs\ell_i\ell_i}(\mu_b)
 \approx
 -1.01\,C_{9}^{(\prime)\,bs\ell_i\ell_i}(\mu_b)\,.
\end{equation}
Once again, the $D^0-\bar D^0$ and $K^0-\bar K^0$ mixing constraints requiring
$\left|\frac{ \re(\Delta_L^{bs}) }{\Delta_L^{l_i}}\right|\lesssim 10$ keep the RG effects at the level of only few percent.

}

\boldmath
\subsection{Correlations Between WET Wilson Coefficients}\label{sec:correlations_bsll_bsnunu}
\unboldmath

  The matching relations in (\ref{eq:SMEFTtoWETmatching}) imply
   \begin{equation}\label{eq:CL}
  \begin{aligned}
    C^{bs\nu_i\nu_i}_L &=\frac{C_{9}^{bs\ell_i\ell_i}-C_{10}^{bs\ell_i\ell_i}}{2}+\frac{(3+\zeta)}{4\mathcal{N}}c_Z -\frac{1}{\mathcal{N}}[C_{lq}^{(3)}]_{ii23}\,,
    \\
    C^{bs\nu_i\nu_i}_R &=\frac{C_{9}^{\prime,bs\ell_i\ell_i}-C_{10}^{\prime,bs\ell_i\ell_i}}{2}+\frac{(3+\zeta)}{4\mathcal{N}}c^\prime_Z \,
    \end{aligned}
  \end{equation}
    which have been presented already in \cite{Buras:2014fpa} neglecting the
    contribution from $[C_{lq}^{(3)}]_{ii23}$.

  For a {\bf vector $Z^\prime$-lepton coupling}, i.e.\ $\Delta_R^{e_i}=\Delta_L^{l_i}$ at the NP scale we have  $[C_{qe}]_{23ii} = [C_{lq}^{(1)}]_{ii23}$
  at this scale but through RG effects this relation is violated at the electroweak scale:
  \be
\Delta_{23ii}=[C_{qe}]_{23ii}- [C_{lq}^{(1)}]_{ii23}\not=0.
\ee
From Eq.~\eqref{eq:SMEFTtoWETmatching} we obtain then  the following relation between the WET WCs at the scale $M_Z$
\begin{equation}\label{CORR1}
  C_{L}^{bs\nu_i\nu_i}= \frac{1}{2}C_{9}^{bs\ell_i\ell_i}
  +\frac{(2+\zeta)}{2} C_{10}^{bs\ell_i\ell_i}\,
  -\frac{(1-\zeta)}{4\mathcal{N}}[C_{lq}^{(3)}]_{ii23}-\frac{1}{4\mathcal{N}}(3+\zeta)\Delta_{23ii}.
\end{equation}

On the other hand, for a purely {\bf left-handed $Z^\prime$-lepton coupling}, i.e.\ $\Delta_R^{e_i}=0$ at the NP scale we have $[C_{qe}]_{23ii}=0$ at this scale.
However, again through RG evolution $[C_{qe}]_{23ii}\not=0$ at the electroweak scale. We obtain then
\begin{equation}\label{CORR2}
  C_{L}^{bs\nu_i\nu_i}= \frac{2}{(1-\zeta)}C_{9}^{bs\ell_i\ell_i}
  +\frac{(1+\zeta)}{(1-\zeta)} C_{10}^{bs\ell_i\ell_i}\,
  -\frac{1}{\mathcal{N}}[C_{lq}^{(3)}]_{ii23}-\frac{1}{2\mathcal{N}}\frac{(3+\zeta)}{(1-\zeta)}[C_{qe}]_{23ii}\,.
\end{equation}

 In fact the relations in (\ref{CORR1}) and
  (\ref{CORR2}) are at the basis of the pattern of correlations between the
  $B\to K(K^*)\nu\bar\nu$,  $B\to K(K^*)\mu^+\mu^-$ and  $B_s\to\mu^+\mu^-$  decay rates that we will find in Section~\ref{sec:numerical_analysis}.

 Of interest are also the RG effects that lead to the violation of the
   following NP-scale relations for vector and left-handed $Z^\prime$
   couplings to leptons respectively:
   \be
   C_{10}^{bs\ell_i\ell_i}=0,\qquad  C_{9}^{bs\ell_i\ell_i}=- C_{10}^{bs\ell_i\ell_i}\,,
   \qquad (\mu=M_{Z^\prime}).
   \ee
   At the electroweak scale, the inclusion of RG effects implies respectively
   \be\label{RG1}
   2\,\mathcal{N}\,C_{10}^{bs\ell_i\ell_i} = \Delta_{23ii} - [C_{lq}^{(3)}]_{ii23} +
   c_Z\,,\qquad (\mu=M_Z)\,,
   \ee
   \be\label{RG2}
    C_{9}^{bs\ell_i\ell_i}=- C_{10}^{bs\ell_i\ell_i} +
    \frac{1}{\mathcal{N}}[C_{qe}]_{23ii}+\frac{1}{2\mathcal{N}}(1-\zeta)c_Z\,,
    \qquad (\mu=M_Z).
\ee
In particular, as mentioned in Section~\ref{sec:bs_Zp_WET_matching}, the RG effect contributing to $c_Z$ always increases
{%
$\re(C_{10}^{bs\ell_i\ell_i})$.
}%
In models with {vector} $Z^\prime$ couplings, which fulfil Eq.~\eqref{RG1}, this could be the dominant contribution to $C_{10}^{bs\ell_i\ell_i}$, slightly suppressing the
$B_s\to\mu^+\mu^-$ branching ratio below its  SM prediction and improving the agreement
with experimental data.
However, as discussed in the previous section, this would require $\re(\Delta_L^{bs}) \gg |\Delta_L^{l_i}|$, which is strongly disfavoured by the $b\to s\ell^+\ell^-$ data combined with the bounds from $D^0-\bar D^0$ and $K^0-\bar K^0$ mixing on $\Delta_L^{bs}$ discussed in Section~\ref{sec:fit_df2}.

The relation (\ref{eq:z_q1q1_zero}) necessary for the suppression of $Z^\prime$ contributions to $B_s^0-\bar B_s^0$  mixing implies that all four WCs of the operators in \eqref{QAQVL} and \eqref{QAQVR} are affected by $Z^\prime$ contributions but in a correlated
manner.
{%
Up to RG effects, we have
}%
\be\label{LHRH}
C_9^{\prime,bs\ell\ell}= r_{bs}\, C_9^{bs\ell\ell},\qquad C_{10}^{\prime,bs\ell\ell}= r_{bs}\, C_{10}^{bs\ell\ell}.
\ee
Therefore, only two of them are independent.
In addition, with (\ref{eq:CL}) and (\ref{LHRH})
we also have up to RG effects
\be\label{eq:CRCL}
C_{R}^{bs\nu_i\nu_i}=r_{bs}\,C_{L}^{bs\nu_i\nu_i}\,.
\ee
Consequently, determining the NP contributions to
$C_9^{bs\ell\ell}$ and $C_{10}^{bs\ell\ell}$ from the $b\to s\mu^+\mu^-$ data
will automatically determine their right-handed counterparts, as well as the ratio of $C_{R}^{bs\nu_i\nu_i}$ and $C_{L}^{bs\nu_i\nu_i}$.

In this manner the $b\to s \nu\bar\nu$ and $b\to s \mu^+\mu^-$ transitions
are correlated and this correlation is governed by the parameter $r_{bs}$ and
the $\text{SU(2)}_L$ gauge symmetry relation
\be\label{SU(2)}
\Delta_L^{\nu\nu}=\Delta_L^{\mu\mu}
\ee
that is already taken into account in all relations above.

{
\boldmath
\subsection{$\Delta B=\Delta S=1$ Observables}\label{sec:bs_Zp_observables}
\unboldmath
}

\boldmath
\subsubsection{$B\to K(K^*)\nu\bar\nu$}
\unboldmath

The effect of right-handed currents can be tested in $b\to s\nu\bar\nu$
transitions. Defining
\begin{equation}\label{RKRK*}
\mathcal{R}_{K\nu\nu}=\frac{\mathcal{B}(B^+ \to K^+ \nu \bar\nu)}{\mathcal{B}_{\rm SM}(B^+ \to K^+ \nu\bar\nu)} \, , \quad \mathcal{R}_{K^*\nu\nu}=\frac{\mathcal{B}(B^0 \to K^{0*} \nu\bar\nu)}{\mathcal{B}_{\rm SM}(B^0 \to K^{0*} \nu\bar\nu)} \,,
\end{equation}
we have (cf.~\cite{Grossman:1995gt,Melikhov:1998ug,Altmannshofer:2009ma,Buras:2014fpa})
\begin{equation}\label{RKRK*new}
\begin{aligned}
 \mathcal{R}_{K\nu\nu}
 &=
 \epsilon^2+ 2\, \tilde{\eta}\,,
 \\
 \mathcal{R}_{K^*\nu\nu}
 &=
 \epsilon^2-\kappa_\eta\,\tilde{\eta}\,,
\end{aligned}
\end{equation}
where $\kappa_\eta=1.33\pm0.05$  and $\epsilon^2$ and $\tilde{\eta}$ are given by\footnote{%
We define and use $\tilde{\eta} \equiv -\epsilon^2\eta$, while $\eta$ is used in~\cite{Grossman:1995gt,Melikhov:1998ug,Altmannshofer:2009ma,Buras:2014fpa}.}
\begin{equation}\label{epsilon}
 \epsilon^2 =
 \frac{
 |C^{bs\nu\nu}_{L,\text{SM}} + C^{bs\nu\nu}_L|^2 + |C^{bs\nu\nu}_R|^2
 }{|C^{bs\nu\nu}_{L,\text{SM}}|^2}
\end{equation}
and
\begin{equation}\label{tildeeta}
\tilde{\eta} =
\text{Re}\left( C_R^{bs\nu\nu}\right)
\,
\frac{
    \text{Re}\left(C^{bs\nu\nu}_{L,\text{SM}} + C^{bs\nu\nu}_L\right)
}{|C^{bs\nu\nu}_{L,\text{SM}}|^2}
+
\text{Im}\left( C_R^{bs\nu\nu}\right)
\,
 \frac{
    \text{Im}\left(C^{bs\nu\nu}_{L,\text{SM}} + C^{bs\nu\nu}_L\right)
 }{|C^{bs\nu\nu}_{L,\text{SM}}|^2}\,,
\end{equation}
such that the difference of $\mathcal{R}_{K\nu\nu}$ and $\mathcal{R}_{K^*\nu\nu}$,
\begin{equation}\label{DIFF}
 \mathcal{R}_{K\nu\nu} - \mathcal{R}_{K^*\nu\nu} = (2+\kappa_\eta)\, \tilde{\eta}
\end{equation}
is proportional to $\tilde\eta$ and thus depends linearly on the real and imaginary parts of the right-handed coefficient $C_R^{bs\nu\nu}$, which  in turn, given Eq.~\eqref{eq:CRCL}, is proportional to $r_{bs}$.
A non-zero difference between $\mathcal{R}_{K\nu\nu}$ and $\mathcal{R}_{K^*\nu\nu}$ directly signals the presence of right-handed currents.

The analytic formulae for the branching ratios $\mathcal{B}(B \to K \nu \bar\nu)$ and $\mathcal{B}(B \to K^* \nu \bar\nu)$ can be found in \cite{Buras:2014fpa} and in Section 9.6 of \cite{Buras:2020xsm}. They all are incorporated
  in the {open source python package \texttt{flavio}~\cite{Straub:2018kue}} that we will be using in our numerical analysis.
  The branching ratios in the SM depend quadratically on $\vcb$ which is subject to   known tensions between its inclusive and exclusive determinations \cite{Bordone:2021oof,FlavourLatticeAveragingGroupFLAG:2021npn}. Moreover, they
  depend on the chosen {$B\to K^{(*)}$} form factors. As in the recent papers
  \cite{Bause:2021cna,He:2021yoz,Bause:2022rrs,Becirevic:2023aov,Bause:2023mfe,Allwicher:2023xba,Dreiner:2023cms} different choices of $\vcb$ and of form factors have been made,  we
  illustrate this dependence in Table~\ref{tab:nunubar1} by presenting SM results which
  correspond  to  two choices of $\vcb$ and two choices of form factors.

 For $\vcb$ we use
  \be\label{eq:Vcb_input}
  \vcb= 42.6(4)\times 10^{-3} \quad \text{and} \quad \vcb_{\rm incl}={41.97(48)}\times 10^{-3}\,.
  \ee
  The first value follows from the  strategies of \cite{Buras:2021nns,Buras:2022wpw,Buras:2022qip} that  allow to avoid the $\vcb$ tensions in question by
  determining CKM parameters solely from $\Delta F=2$ observables. The
  second value follows from inclusive decays  \cite{Finauri:2023kte}.
  {
  In our numerical analysis presented in section~\ref{sec:numerical_analysis}, we use $\vcb_{\rm incl}$.
  }

For the $B\to K$ form factors we use either the HPQCD 2022~\cite{Parrott:2022zte,Parrott:2022rgu,Parrott:2022smq} form factors or the average
of HPQCD~2013~\cite{Bouchard:2013eph}, FNAL+MILC~2015~\cite{Bailey:2015dka}, and HPQCD~2022 form factors as presented in GRvDV~2023~\cite{Gubernari:2023puw}.
For the $B\to K^*$ form factors we use the combination of LQCD and LCSR results presented in BSZ~2015~\cite{Bharucha:2015bzk}.
  In our numerical analysis presented in section~\ref{sec:numerical_analysis}, we use the GRvDV 2023 $B\to K$ form factors and the BSZ 2015 $B\to K^*$ form factors.

Now in \cite{Parrott:2022zte,Parrott:2022rgu,Parrott:2022smq} that use the first value of $\vcb$ in Eq.~\eqref{eq:Vcb_input} and the HPQCD~2022
form factors, the SM prediction for $B^+\to K^+\nu\bar\nu$
includes a $10\%$ upward shift from a tree-level long distance contribution pointed out in \cite{Kamenik:2009kc}. This results in
\begin{align}\label{BVNEW}
{\mathcal{B}}(B^+\to K^+\nu\bar\nu)_{\rm SM}^{\rm SD+LD} &={(5.53\pm 0.30)\times 10^{-6}},\\
{\mathcal{B}}(B^0\to K^{0*}\nu\bar\nu)_{\rm SM} &= {(10.11\pm 0.96)\times 10^{-6}}.
\end{align}
Otherwise, as seen in Table~\ref{tab:nunubar1}%
, the first branching ratio would be {${(4.92\pm 0.30)\times 10^{-6}}$}. In fact
the latter result should be compared  with the experimental  result
given below and in what follows we will leave out this tree level {long distance}
contribution from our analysis. {We observe that the results for
 the SD+LD and the purely SD branching ratios differ  roughly by 2$\sigma$
  taking uncertainties in  $\vcb$ and the form factors into account.}
  They  are typical  by $10\%$ higher than {those} used in most of the recent analyses   \cite{Bause:2021cna,He:2021yoz,Bause:2022rrs,Becirevic:2023aov,Bause:2023mfe,Allwicher:2023xba,Dreiner:2023cms}. This difference is presently immaterial in view of large experimental errors
but could turn out to be important when the experimental errors will be significantly reduced.

\begin{table}
\centering
\renewcommand{\arraystretch}{1.4}
\resizebox{\columnwidth}{!}{
\begin{tabular}{|l|l|l|l|}
\hline
Observable& Form factors & $\vcb = 42.6(4)\times 10^{-3}$ & {$\vcb = {41.97(48)}\times 10^{-3}$}
\\
\hline
\hline
 \multirow{2}{*}{${\mathcal{B}}(B^+\to K^+\nu\bar\nu)_{\rm SM}^{\rm SD}$}&HPQCD 2022~\cite{Parrott:2022rgu} & {$(4.92 \pm 0.30)\times 10^{-6}$}  & {${(4.78 \pm 0.30)}\times 10^{-6}$}
 \\
 &GRvDV~2023~\cite{Gubernari:2023puw} & {$(4.85 \pm 0.23)\times 10^{-6}$} & {${(4.71 \pm 0.23)}\times 10^{-6}$}
\\
\hline
\hline
 ${\mathcal{B}}(B^0\to K^{0*}\nu\bar\nu)_{\rm SM}$&BSZ 2015~\cite{Bharucha:2015bzk} & {$(10.11 \pm 0.96)\times 10^{-6}$}  & {${(9.81 \pm 0.96)}\times 10^{-6}$}
 \\
\hline

\end{tabular}
}
\renewcommand{\arraystretch}{1.0}
\caption{\label{tab:nunubar1}
  \small
  SM predictions for $B^+\to K^+\nu\bar\nu$ and $B^0\to K^{0*}\nu\bar\nu$ for different $\vcb$ and form factor values.}
\end{table}

On the other hand the best current experimental bounds~\cite{Olive:2016xmw,Grygier:2017tzo} including the BaBar results \cite{BaBar:2010oqg,BaBar:2013npw} and
most recent results from Belle II \cite{Belle-II:2023esi} read\footnote{%
Note that Ref.~\cite{Grygier:2017tzo} reports ${\mathcal{B}}(B^0\to K_S\nu\bar\nu) \leq 13\times 10^{-6}\ \text{@ 90\% CL}$, which has been converted to a bound on ${\mathcal{B}}(B^0\to K^0\nu\bar\nu)$ by assuming ${\mathcal{B}}(B^0\to K^0\nu\bar\nu) = 2\,{\mathcal{B}}(B^0\to K_S\nu\bar\nu)$.
}
\begin{align}
{\mathcal{B}}(B^+\to K^+\nu\bar\nu) &={(13\pm4)}\times 10^{-6},\\
{\mathcal{B}}(B^0\to K^0\nu\bar\nu) &\leq 26\times 10^{-6}\quad \text{@ 90\% CL}~,\\
{\mathcal{B}}(B^+\to  K^{+*} \nu\bar\nu) &\leq 40\times 10^{-6}\quad \text{@ 90\% CL}~,\\
{\mathcal{B}}(B^0\to K^{0*}\nu\bar\nu) &\leq 18\times 10^{-6}\quad \text{@ 90\% CL}\,.
\end{align}

\boldmath
\subsubsection{$B\to K(K^*)\mu^+\mu^-$ and $B_s\to \mu^+\mu^-$}
\unboldmath

For $B\to K(K^*)\mu^+\mu^-$ decays we define
\be\label{RKRK*mumu}
\mathcal{R}_{K\mu\mu}=\frac{\mathcal{B}(B^+ \to K^+ \mu \bar\mu)^{{[1.1,6.0]}}}{\mathcal{B}_{\rm SM}(B^+ \to K^+ \mu\bar\mu)^{{[1.1,6.0]}}} \, , \quad \mathcal{R}_{K^{0*}\mu\mu}=\frac{\mathcal{B}(B^0 \to K^{*0} \mu\bar\mu)^{{[1.1,6.0]}}}{\mathcal{B}_{\rm SM}(B^0 \to K^{0*} \mu\bar\mu)^{{[1.1,6.0]}}} \,.
\ee
We will also consider $B_s\to \mu^+\mu^-$ decay for which we define
  \be\label{mumu}
  \mathcal{R}_{\mu\mu}=\frac{\overline{\mathcal{B}}(B_s \to \mu \bar\mu)}{\overline{\mathcal{B}}_{\rm SM}(B_s \to \mu\bar\mu)},
  \ee
  with the {\em overline} indicating the inclusion of $\Delta\Gamma_s$ effects
\cite{deBruyn:2012wk,Fleischer:2012fy,Buras:2013uqa}.

The analytic formulae for the branching ratios $\mathcal{B}(B \to K \mu^+ \mu^-)$, $\mathcal{B}(B \to K^* \mu^+ \mu^-)$ and $\overline{\mathcal{B}}(B \to \mu^+ \mu^-)$in terms of the WCs are well known in the
  context of $b\to s\mu^+\mu^-$ anomalies and can also be found in  \cite{Buras:2020xsm}. They all are incorporated
  in the open source python package \texttt{flavio}~\cite{Straub:2018kue} that we will be using in our numerical analysis.
  {
  For the $B\to K$ form factors we use the average
  of HPQCD~2013~\cite{Bouchard:2013eph}, FNAL+MILC~2015~\cite{Bailey:2015dka}, and HPQCD~2022 form factors as presented in GRvDV~2023~\cite{Gubernari:2023puw}, and for the $B\to K^*$ form factors we use the combination of LQCD and LCSR results presented in BSZ~2015~\cite{Bharucha:2015bzk}.
  }

\subsection{Numerical Analysis}\label{sec:numerical_analysis}
We perform a numerical analysis of the observables described in section~\ref{sec:bs_Zp_observables} to study the implications of the $Z^\prime$ parameter space with suppressed NP contributions to $B_s-\bar B_s$, as identified in section~\ref{sec:fit_df2}.
Following the discussion in section~\ref{sec:fit_df2}, we vary the ratios $\Delta_{L,R}^{ij}/M_{Z^\prime}$ and set $\Lambda_\text{NP}=M_{Z^\prime}=5\,\text{TeV}$.
We consider two benchmark scenarios:
\begin{enumerate}
 \item[\textbf{Benchmark 1}] We consider the bound of the \boldmath\textbf{$D^0-\bar D^0$ mixing observable $x_{12}^{\text{Im},D}$}\unboldmath\ on the magnitude of $\Delta_{L}^{bs}/M_{Z^\prime}$. We choose the benchmark value
 \begin{equation}
\Delta^{bs}_R/\Delta^{bs}_L=0.083\,,
 \end{equation}
which maximizes the allowed magnitude of $\Delta_{L}^{bs}/M_{Z^\prime}$ for $M_{Z^\prime}=5\,\text{TeV}$ in the presence of the $D^0-\bar D^0$ bound.
 \item[\textbf{Benchmark 2}] We consider the scenario in which the contributions to $D^0-\bar D^0$ mixing are suppressed and the \boldmath\textbf{$K^0-\bar K^0$ mixing observable $\varepsilon_K$}\unboldmath\ provides the strongest bound on the magnitude of $\Delta_{L}^{bs}/M_{Z^\prime}$.  We choose the benchmark value
 \begin{equation}
  \Delta^{bs}_R/\Delta^{bs}_L=0.089\,,
 \end{equation}
 which maximizes the allowed magnitude of $\Delta_{L}^{bs}/M_{Z^\prime}$ for $M_{Z^\prime}=5\,\text{TeV}$ in this scenario.
\end{enumerate}

\subsubsection{The Global $b\to s\ell^+\ell^-$ Fit}\label{sec:fit_bsll}

\begin{figure}
\centering
\includegraphics[width=0.48\textwidth]{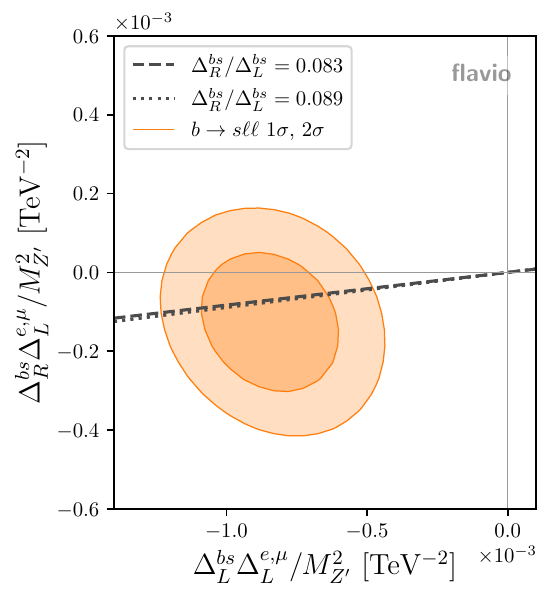}
\includegraphics[width=0.48\textwidth]{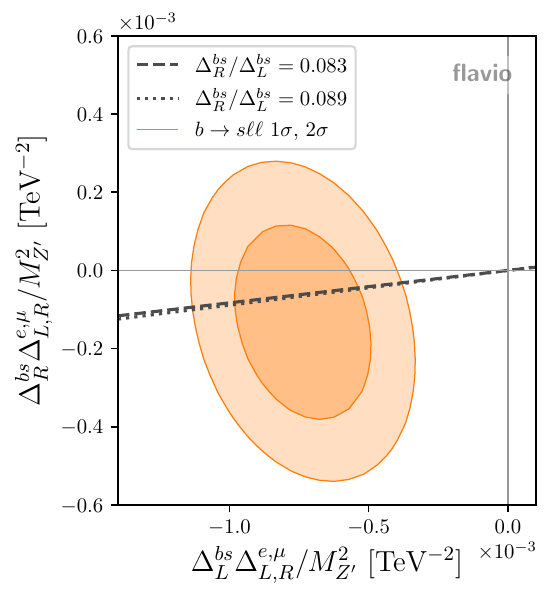}
\caption{%
{
Constraints on product of quark and lepton $Z'$ couplings from $b\to s\ell^+\ell^-$ observables.
The left and right panels show the scenarios with left-handed and vector $Z^\prime$-lepton couplings, respectively.
}
}
\label{fig:bsll}
\end{figure}

The products of $Z^\prime$-lepton and $Z'$-$b$-$s$ couplings enter the predictions of semi-leptonic rare $B$ decays based on the $b\to s\ell^+\ell^-$ transition.
Since SM predictions of ${b\to s\ell^+\ell^-}$ observables show tensions with experimental data, we investigate whether these tensions can be reduced by $Z'$ contributions that are compatible with Benchmark~1 and Benchmark~2.
To this end, we perform a fit to experimental data in the 2D-plane spanned by the products of quark and lepton couplings.
We consider two scenarios of flavour universal lepton couplings:
\begin{itemize}
\item \textbf{Left-handed $Z^\prime$ couplings}, for which we define
\begin{equation}
\Delta^{e,\mu}_{L} \equiv \Delta^{e}_{L} = \Delta^{\mu}_{L}\,.
\end{equation}
 \item \textbf{Vector $Z^\prime$ couplings}, for which we define
\begin{equation}
\Delta^{e,\mu}_{L,R} \equiv \Delta^{e}_{L} = \Delta^{\mu}_{L} = \Delta^{e}_{R} = \Delta^{\mu}_{R}\,.
\end{equation}
\end{itemize}
The results of the fits in these two scenarios are shown in Fig.~\ref{fig:bsll} with left-handed and vector $Z^\prime$ couplings shown in the left and right panels, respectively.
We find a clear preference for a negative $\Delta^{bs}_L\Delta^{e,\mu}_{L}$ (left panel) or $\Delta^{bs}_L\Delta^{e,\mu}_{L,R}$ (right panel), while $\Delta^{bs}_R\Delta^{e,\mu}_{L,R}$ (left panel) and $\Delta^{bs}_R\Delta^{e,\mu}_{L}$ (right panel) are compatible with zero, but show a preference for positive $\Delta^{bs}_R/\Delta^{bs}_L$.
This result is fully compatible with the positive $\Delta^{bs}_R/\Delta^{bs}_L$ values of Benchmark~1 and Benchmark~2, which are shown as
(nearly overlapping)
dashed and dotted lines in both plots.

In contrast to a previous numerical study~\cite{DiLuzio:2019jyq} that considered the effects of left- and right-handed $Z^\prime$-$b$-$s$ couplings on $b\to s\ell^+\ell^-$ observables and $B_s^0-\bar B_s^0$ mixing, the picture has significantly changed due to the recent measurement of $R_{K^{(*)}}$ by LHCb~\cite{LHCb:2022qnv,LHCb:2022zom}.
Previously, the slightly larger value of $R_K$ compared to $R_{K^*}$ indicated negative $\Delta^{bs}_R/\Delta^{bs}_L$ and was therefore incompatible with the positive values required for the suppression of $B_s^0-\bar B_s^0$ mixing discussed in section~\ref{sec:meson_mixing}.
An explanation of the $b\to s\ell^+\ell^-$ anomalies by a $Z^\prime$ then required very small $Z^\prime$-$b$-$s$ couplings to be compatible with $B_s^0-\bar B_s^0$ mixing.
The fact that $R_{K^{(*)}}$ are now in agreement with LFU means that they do not affect an LFU fit to $b\to s\ell^+\ell^-$ data anymore, which can now comfortably accommodate the positive values of $\Delta^{bs}_R/\Delta^{bs}_L$ shown by the black dashed and dotted lines in Fig.~\ref{fig:bsll} that can hardly be distinguished from each other.

\subsubsection{Global Fit of $Z^\prime$ Couplings}\label{sec:fit_global}

\begin{figure}
\centering
\includegraphics[width=0.49\textwidth]{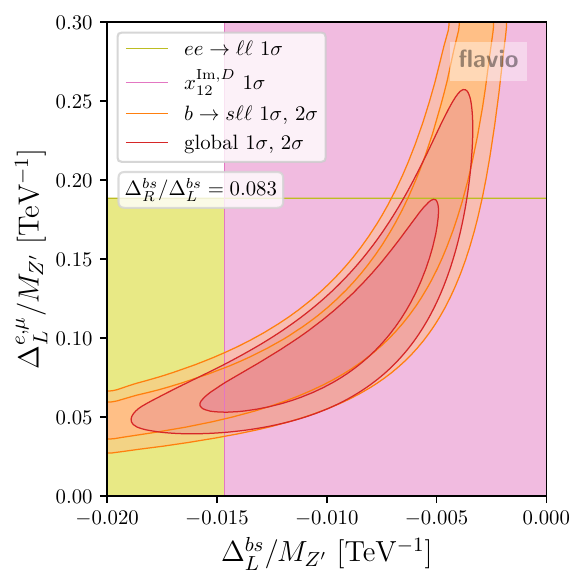}
\includegraphics[width=0.49\textwidth]{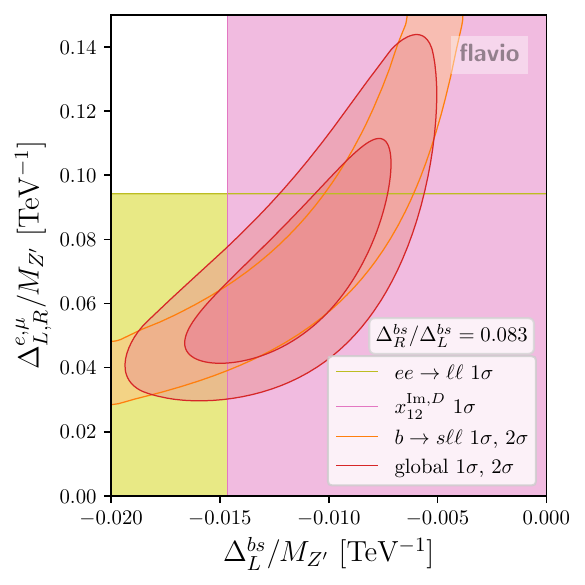}
\includegraphics[width=0.48\textwidth]{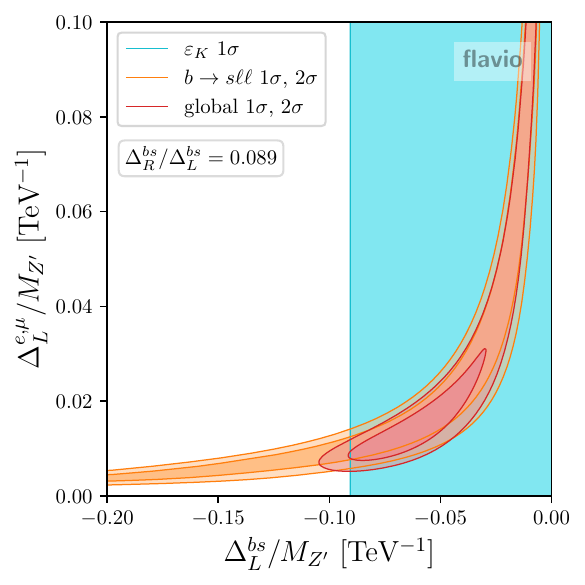}
\includegraphics[width=0.48\textwidth]{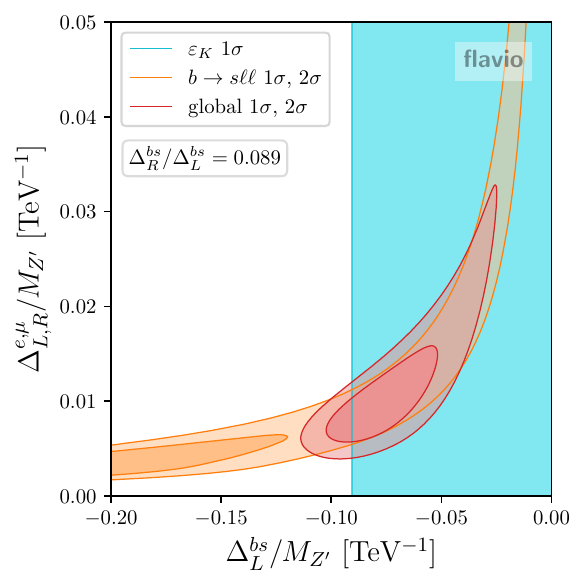}
\caption{%
{%
Global fit of $Z^\prime$ couplings $\Delta^{bs}_L$ and $\Delta^{e,\mu}_{L}$ or $\Delta^{e,\mu}_{L,R}$.
The two upper panels show Benchmark~1, while the two lower panels show Benchmark~2.
The two left-panels show scenarios with left-handed $Z^\prime$-lepton couplings (Scenario~1 for Benchmark~1 and Scenario~2 for Benchmark~2), while the two right panels show scenarios with vector $Z^\prime$-lepton couplings (Scenario~3 for Benchmark~1 and Scenario~4 for Benchmark~2).
}
}
\label{fig:bsll_df2}
\end{figure}

Having demonstrated the compatibility of Benchmark~1 and Benchmark~2 with the $b\to s\ell^+\ell^-$ observables, we perform global fits to all experimental data constraining the $Z^\prime$ quark and lepton couplings. We consider the following scenarios:
\begin{enumerate}
 \item[\textbf{Scenario 1}] Benchmark~1 with left-handed $Z^\prime$-lepton couplings: (1,1).
 \item[\textbf{Scenario 2}] Benchmark~2 with left-handed $Z^\prime$-lepton couplings: (2,1).
 \item[\textbf{Scenario 3}] Benchmark~1 with vector $Z^\prime$-lepton couplings: (1,2).
 \item[\textbf{Scenario 4}] Benchmark~2 with vector $Z^\prime$-lepton couplings: (2.2).
\end{enumerate}
The results of these fits are shown as a $2\times 2$ matrix in Fig.~\ref{fig:bsll_df2} with the different entries allocated as indicated above. The left and right panels show the results for left-handed and vector lepton-$Z^\prime$ couplings, respectively.
The top and bottom panels show the results for Benchmark~1 and Benchmark~2, respectively.
We find the following results for the four scenarios in question:
\begin{itemize}
\item
  The plots for Scenarios 1 and 3 (top panels) show that values of the $Z^\prime$ couplings that can explain the $b\to s\ell^+\ell^-$ data are allowed by the combined constraints from $D^0-\bar D^0$ mixing and LEP2 $e^+ e^-\to\ell^+\ell^-$ data. These three data sets select compact best-fit regions shown as red contours.
\item
  The plots for Scenarios 2 and 4 (bottom panels) show that in these cases the bound from $\varepsilon_K$ allows considerably larger magnitudes of left-handed $Z^\prime$-quark couplings $\Delta^{bs}_L$, and even slightly prefers non-zero values.
  Consequently, the $b\to s\ell^+\ell^-$ data can be explained with relatively small $Z^\prime$-lepton couplings, and the bound from LEP2 $e^+ e^-\to\ell^+\ell^-$ data plays no important role.

\item
In Scenarios 3 and 4   with vector $Z^\prime$-lepton couplings (right panels), no contribution to $C_{10}^{bs\ell_i\ell_i}(\mu_b)$ is generated from tree-level matching of the $Z^\prime$ model. Since the $b\to s\ell^+\ell^-$ data prefers a slightly positive $C_{10}^{bs\ell_i\ell_i}(\mu_b)$ (see e.g.~\cite{Greljo:2022jac}), which can be generated from four-quark WCs through the RG effects described in sections~\ref{sec:bs_Zp_SMEFT_matching} and~\ref{sec:bs_Zp_WET_matching}, a sizeable $\Delta^{bs}_L\approx -0.2$ gives the best fit to $b\to s\ell^+\ell^-$ data.
While such a large magnitude of $\Delta^{bs}_L$ is disfavoured by the bounds from $D^0-\bar D^0$ and $K^0-\bar K^0$ mixing, the preference for non-zero $C_{10}^{bs\ell_i\ell_i}(\mu_b)$ disfavours small magnitudes of $\Delta^{bs}_L$, which can be clearly seen in Scenario~4 (lower right panel). This effect is less pronounced in Scenario~3 (upper right panel), as the $D^0-\bar D^0$ mixing bounds restrict the magnitude of $\Delta^{bs}_L$ to much smaller values.
\item
  In Scenarios 1 and 2 with left-handed $Z^\prime$-lepton couplings (left panels), a contribution to $C_{10}^{bs\ell_i\ell_i}(\mu_b)$ is already generated from tree-level matching, which slightly overshoots the value preferred by $b\to s\ell^+\ell^-$ data (see e.g.~\cite{Greljo:2022jac}). Consequently, the opposite effect as in the scenarios 3 and 4 with vector $Z^\prime$-lepton couplings can be observed: smaller magnitudes of $\Delta^{bs}_L$ are preferred by  $b\to s\ell^+\ell^-$ data. However, in Scenario 2 (lower left panel), this effect is partially compensated in the global fit by the preference of $\varepsilon_K$ for non-zero $\Delta^{bs}_L$. In Scenario 1 (upper left panel), the effect of $C_{10}^{bs\ell_i\ell_i}(\mu_b)$ is again less pronounced due to the stringent bound from $D^0-\bar D^0$ mixing on the magnitude of $\Delta^{bs}_L$.
  \end{itemize}

\subsubsection{Predictions for $b\to s\nu\bar\nu$ Observables}

\begin{figure}
\centering
\includegraphics[width=0.49\textwidth]{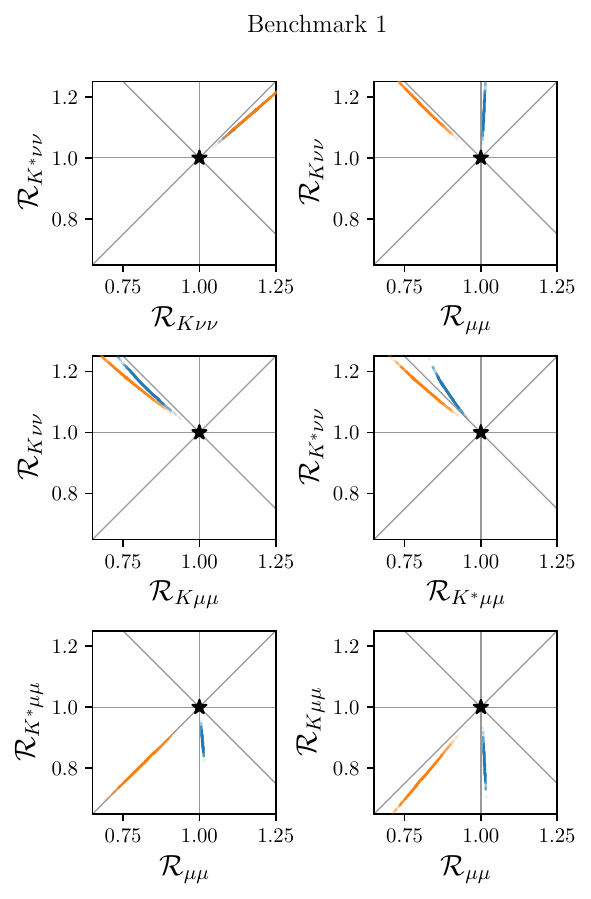}
\includegraphics[width=0.49\textwidth]{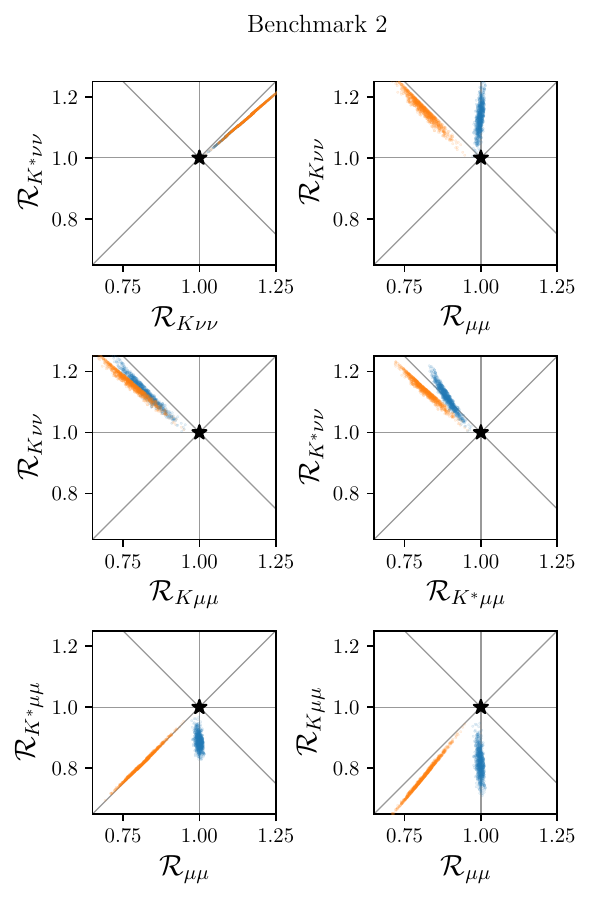}
\caption{%
Predictions for various observables distributed according to the global fits in section~\ref{sec:fit_global}. The two left panels show the results for Benchmark~1, while the two right panels show the results Benchmark~2. The blue and orange points correspond to scenarios with vector and left-handed $Z^\prime$ lepton couplings, respectively.
}
\label{fig:predictions_R}
\end{figure}

The global fits presented in section~\ref{sec:fit_global} select compact regions in the parameter space of $Z^\prime$ couplings.
This in turn implies correlations between the theory predictions of various observables.
In order to study these correlations, we generate samples of $Z^\prime$ couplings that are distributed according to the global likelihood,
while assuming LFU in all three lepton generations.
From these samples we make predictions for $b\to s\mu^+\mu^-$ and $b\to s\nu\bar\nu$ observables.
The results are shown in Fig.~\ref{fig:predictions_R} for the four scenarios discussed in section~\ref{sec:fit_global}.
%
%
%
%
%
We observe the following:
\begin{itemize}
\item As expected from the data entering the fits, $\mathcal{R}_{K\mu\mu}$ and $\mathcal{R}_{K^*\mu\mu}$ are always suppressed below unity.
 The NP effects in both of these ratios are slightly larger in the case of purely left-handed leptonic couplings (orange dots) than in the case of vector leptonic couplings (blue dots).
  The main reason for this is that the size of NP effects in the vector couplings is constrained by angular observables like $P_{5^\prime}$, which are included in the fit. The additional axial-vector component present in the left-handed case enhances the NP effect in branching fractions without being significantly constrained by the angular observables.

\item $\mathcal{R}_{K\nu\nu}$ and $\mathcal{R}_{K^*\nu\nu}$ are always {\em anti-correlated} with $\mathcal{R}_{K\mu\mu}$ and $\mathcal{R}_{K^*\mu\mu}$.
The suppression of $\mathcal{R}_{K\mu\mu}$ and $\mathcal{R}_{K^*\mu\mu}$ below
  unity as observed by the LHCb experiment implies enhancements of the
  $\mathcal{R}_{K\nu\nu}$ and $\mathcal{R}_{K^*\nu\nu}$ ratios by up to
  $20\%$.
  {
  The ratios of the anti-correlated quantities,
  \begin{equation}\label{eq:Rnumu}
   \mathcal{R}_{\nu/\mu}(K) = \frac{\mathcal{R}_{K\nu\nu}}{\mathcal{R}_{K\mu\mu}}
   \qquad
   \text{and}
   \qquad
   \mathcal{R}_{\nu/\mu}(K^*) = \frac{\mathcal{R}_{K^*\nu\nu}}{\mathcal{R}_{K^*\mu\mu}}\,,
  \end{equation}
  are particularly sensitive to the NP effects and are enhanced by up to 60\%.}

\item $R_{\mu\mu}$ clearly distinguishes between scenarios with left-handed (orange dots) and vector (blue dots) $Z^\prime$ lepton couplings. Only in the former case can a contribution to $C_{10}^{bs\ell_i\ell_i}(\mu_b)$, and thus to $R_{\mu\mu}$, be sizeable, since it is generated from the tree-level matching of the $Z^\prime$ model.
In the latter case, contributions to $R_{\mu\mu}$ are purely RG generated and small.
These two different kind of contributions lead to different correlations.
In the case of left-handed couplings (orange dots), $R_{\mu\mu}$ and $\mathcal{R}_{K^{(*)}\mu\mu}$ are correlated and equally sensitive to the NP effects, being nearly directly proportional.
In the case of vector lepton couplings (blue dots), $R_{\mu\mu}$ and $\mathcal{R}_{K^{(*)}\mu\mu}$ are anti-correlated, but the sensitivity of $\mathcal{R}_{\mu\mu}$ to the NP effects is very weak compared to $\mathcal{R}_{K^{(*)}\mu\mu}$.
Even if $\mathcal{R}_{K^{(*)}\mu\mu}$ are significantly suppressed, $R_{\mu\mu}$ remains practically SM-like as opposed to the left-handed case.
A similar behaviour, but with correlations and anti-correlations exchanged, can be observed in the relationships between $R_{\mu\mu}$ and $\mathcal{R}_{K^{(*)}\nu\nu}$.
This is of course expected from the anti-correlation between $\mathcal{R}_{K^{(*)}\mu\mu}$ and $\mathcal{R}_{K^{(*)}\nu\nu}$.
Evidently, the constraint on the chirality of leptonic couplings will improve with the experimental precision of $\mathcal{R}_{\mu\mu}$ and it is possible
  that a scenario with a linear combination of left-handed and vector leptonic couplings will fit the data best.

\item
  The correlation between $\mathcal{R}_{K\nu\nu}$ and $\mathcal{R}_{K^*\nu\nu}$
  reflects the ratio of left-handed over right-handed quark currents fixed by Benchmark 1 and 2.
As seen in~\eqref{DIFF}, both ratios would be equal in the absence of right-handed currents, and the violation of this equality corresponds to the benchmark values $\Delta^{bs}_R/\Delta^{bs}_L\approx10\%$.

\item The size of all effects is very similar in Benchmark~1 and Benchmark~2. The main difference between the two is that RG effects generating $C_{10}^{bs\ell_i\ell_i}(\mu_b)$ are allowed to be considerably larger in Benchmark~2, given that $Z^\prime$-quark couplings entering these RG effects are allowed to be larger.
This is in particular reflected by the slightly less strict (anti-)correlations between $R_{\mu\mu}$ and the other ratios.
The difference between Benchmark~1 and Benchmark~2 is more pronounced in the case of vector lepton couplings (blue dots), where $C_{10}^{bs\ell_i\ell_i}(\mu_b)$ is generated exclusively from RG effects.
\end{itemize}

\section{Summary}\label{sec:summary}

{
In the present paper, we have provided a comprehensive discussion of meson mixing constraints in $Z^\prime$ models, in particular taking into account effects from the SMEFT and WET RG evolution and the implications of $\text{SU(2)}_L$ gauge invariance.

We have reviewed how NP contributions to the meson mixing amplitude can be suppressed in the presence of both left- and right handed $Z^\prime$-quark couplings.
This suppression depends both on the ratio of left- and right-handed couplings,
\begin{equation}
 r_{q_1 q_2} = \frac{\Delta_R^{q_1 q_2}}{\Delta_L^{q_1 q_2}},
\end{equation}
and on the renormalization scale $\mu=\Lambda_{\rm NP}$ at which this ratio is defined and the $Z^\prime$ model is matched to the SMEFT.
For a reference scale $\mu=5\,\text{TeV}$, we find suppressions of the four different meson mixing amplitudes as follows:
\begin{itemize}
 \item The  $B^0_d-\bar B_d^0$ and $B^0_s-\bar B_s^0$ mixing amplitudes are suppressed for
 \begin{equation}\label{eq:summary_r_db}
  r_{d_i b}\approx 0.1\,,\qquad i\in\{1,2\}.
 \end{equation}
 \item The $K^0-\bar K^0$ mixing amplitude is suppressed for
 \begin{equation}\label{eq:summary_r_ds}
  r_{ds} \approx 0.004\,.
 \end{equation}
 \item The $D^0-\bar D^0$ mixing amplitude is suppressed for
 \begin{equation}\label{eq:summary_r_uc}
  r_{uc} \approx 0.05\,.
 \end{equation}
\end{itemize}

If the NP contribution to one of the four meson mixing amplitudes is suppressed due to Eqs.~\eqref{eq:summary_r_db}, \eqref{eq:summary_r_ds}, or \eqref{eq:summary_r_uc}, $\text{SU(2)}_L$ gauge invariance implies a contribution to the other meson mixing amplitudes.
In addition, RG effects in the SMEFT lead to correlations between all four meson mixing sectors. Focusing on the example of $Z^\prime$-$b$-$s$ couplings, we have made the following observations:

\begin{itemize}
 \item If NP contributions to $B^0_s-\bar B_s^0$ mixing are suppressed due to Eq.~\eqref{eq:summary_r_db}, constraints on $C\!P$ violation in $D^0-\bar D^0$ mixing provide stringent bounds on the $Z^\prime$-$b$-$s$ couplings. These bounds are due to  $\text{SU(2)}_L$ gauge invariance and are only by a factor three weaker than the unsuppressed bounds from $B^0_s-\bar B_s^0$ mixing.
 The contribution to $C\!P$ violation in $D^0-\bar D^0$ mixing even from real $Z^\prime$-$b$-$s$ couplings stems from the fact that the $\text{SU(2)}_L$ relation between left-handed $Z^\prime$-$b$-$s$ and $Z^\prime$-$u$-$c$ couplings involves the CKM phase.

 \item If in addition, the $D^0-\bar D^0$ mixing contributions are also suppressed due to Eq.~\eqref{eq:summary_r_uc}, constraints on $C\!P$ violation in $K^0-\bar K^0$ mixing provide bounds on the $Z^\prime$-$b$-$s$ couplings. These bounds are due to RG effects in the SMEFT and are by around a factor six weaker than the unsuppressed bounds from $C\!P$ violation in $D^0-\bar D^0$ mixing mentioned above.
 The contribution to $C\!P$ violation in $K^0-\bar K^0$ mixing even from real $Z^\prime$-$b$-$s$ couplings stems from the fact that the RG mixing through the SM Yukawa couplings, as well as the re-diagonalization of the running SM Yukawa matrices both involve the CKM phase.
\end{itemize}

The suppression of the NP contributions to $B^0_s-\bar B_s^0$ mixing due to Eq.~\eqref{eq:summary_r_db} has important implications for rare semi-leptonic $b\to s$ decays.
We have found that in this case it is possible to explain the
present anomalies in $b\to s \mu^+\mu^-$ with the help of a $Z^\prime$ while satisfying all existing constraints (see Figs.~\ref{fig:df2_ratio}-\ref{fig:bsll_df2}).
}%
The determination of the $\Delta^{bs}_{L,R}$ couplings {from a global fit including $b\to s \mu^+\mu^-$ data,} combined
with the  $\text{SU(2)}_L$ gauge symmetry and RG effects within the SMEFT,
imply
\begin{itemize}
\item
  {\em enhancements} of  $B\to K(K^*)\nu\bar\nu$ branching ratios by up     to $20\%$ relatively to the SM predictions that are correlated with the observed
  {\em suppressions}  of  $B\to K(K^*)\mu^+\mu^-$ branching ratios as
  seen in Fig.~\ref{fig:predictions_R}.
The larger the suppression
  of $b\to s\mu^+\mu^-$ branching ratios the larger the enhancement
  of $b\to s \nu\bar\nu$ branching ratios. Therefore the ratios in Eq.~\eqref{eq:Rnumu} can be enhanced by up to $60\%$.
\item
{
  significant suppression of the $B_s\to \mu^+\mu^-$ branching ratio in a scenario with purely left-handed $Z^\prime$-lepton couplings, and practically no effect on $B_s\to \mu^+\mu^-$ branching ratios in a scenario with vector $Z^\prime$-lepton couplings.
  }
  \end{itemize}

We are looking forward to improved  Belle II data  on $B\to K(K^*)\nu\bar\nu$,
on  $B\to K(K^*)\mu^+\mu^-$ decays from Belle II and LHCb and for
$B_s\to \mu^+\mu^-$ from LHCb, CMS and ATLAS.
{%
The correlations we have found between these decays in $Z^\prime$ models will allow us to further test the viability of a $Z^\prime$ explanation of the $b\to s \mu^+\mu^-$ anomalies.
Interestingly, through the suppression of NP contributions to $B^0_s-\bar B_s^0$ mixing, the scenarios we have studied also predict additional $C\!P$ violation in $D^0-\bar D^0$ and $K^0-\bar K^0$ mixing.
However, to identify NP in meson mixing will require significant improvement on the precision of input parameters, as we show for $\varepsilon_K$ in Appendix~\ref{app:epsK}.
Particularly important will be the precise determination of the CKM elements that currently dominate the theory uncertainties of many meson mixing observables.
}%

\vskip1em

{\bf Acknowledgements}
We thank  Jason Aebischer for a discussion about the non-redundant basis in SMEFT and  Aleks Smolkovic for contributing to the numerical code used in our analysis.  Financial support of AJB by the Excellence Cluster ORIGINS,
funded by the Deutsche Forschungsgemeinschaft (DFG, German Research
Foundation), Excellence Strategy, EXC-2094, 390783311 is acknowledged.

\appendix

\section{SM Predictions for $\varepsilon_K$}\label{app:epsK}

The formula for $\varepsilon_K$ used in our numerical analysis is by now well
known and we refer to \cite{Brod:2019rzc} for details. As stressed in particular
in \cite{Buras:2021nns} it depends strongly on the value of $\vcb$, which is used as an input parameter for the CKM elements.
For $\vcb$ we use either the value determined from inclusive $B\to X_c\ell\nu$ decays ($\vcb=41.97(48)\times 10^{-3}$)~\cite{Finauri:2023kte} or from $\Delta F=2$ observables ($\vcb=42.6(4)\times 10^{-3}$)~\cite{Buras:2021nns,Buras:2022wpw,Buras:2022qip}.
In this appendix, we compare the results obtained with each of these values, while we use the first value in our numerical analysis.

Here we want to show that also the dependence on the non-perturbative parameter
$\hat B_K$ is still sizable.
In fact, for $\hat B_K$ one can use not only the FLAG~\cite{FlavourLatticeAveragingGroupFLAG:2024oxs} averages for $N_f=2+1+1$~\cite{Carrasco:2015pra} or $N_f=2+1$~\cite{BMW:2011zrh,Laiho:2011np,RBC:2014ntl,SWME:2015oos,Boyle:2024gge}, but might want to use the latest and most precise single lattice determination obtained by the RBC/UKQCD collaboration~\cite{Boyle:2024gge}.
This latest lattice determination agrees in an impressive manner with $\hat B_K=0.73(2)$ from the Dual QCD approach \cite{Buras:2014maa} obtained already ten years ago.
Recently, also next-to-next-to-leading-order (NNLO) results for $\hat B_K$ have become available~\cite{Gorbahn:2024qpe}, which increase the previous results by 1--4\%.
We collect various values of $\hat B_K$ in Table~\ref{tab:eps_K_SM}.
In our numerical analysis we use the NNLO version~\cite{Gorbahn:2024qpe} of the latest and most precise single lattice determination by RBC/UKQCD~\cite{Boyle:2024gge}, rather than using a combination including older lattice data that are in slight tension with the more recent results.

Finally, in view of the experimental progress on the determination of
the angle $\gamma$ in the UT, it is of interest to compare the value
for  $\varepsilon_K$ obtained with
the 2024 HFLAV average ($\gamma = 66.4(30)^\circ$)~\cite{HFLAV:2022esi} with that obtained using the latest LHCb determination ($\gamma = 64.6(28)^\circ$)~\cite{LHCb:2024yxi}.
  For the UT angle $\beta$ we use the most recent HFLAV~\cite{HFLAV:2022esi} average $\sin(2\beta)=0.709(11)$ that implies
    $\beta=22.6(4)^\circ$ compared to the previous average of  $\beta=22.2(7)^\circ$.

\begin{table}
  \centering
  \renewcommand{\arraystretch}{1.4}
  \resizebox{\columnwidth}{!}{
  \begin{tabular}{|l|l||c|c|c|c|}
  \hhline{~~|--|--|}
  \multicolumn{2}{c|}{} & \multicolumn{2}{c|}{\boldmath$\vcb = 41.97(48)\times 10^{-3}$\unboldmath\ } & \multicolumn{2}{c|}{$\vcb = 42.6(4)\times 10^{-3}$} \\
  \multicolumn{2}{c|}{} & $\gamma = 64.6(28)^\circ$ & \boldmath$\gamma = 66.4(30)^\circ$\unboldmath\ & $\gamma = 64.6(28)^\circ$ & $\gamma = 66.4(30)^\circ$\\
  \hhline{~~ ====}
  \hhline{|--||~~~~|}
  \multicolumn{2}{|c||}{$\hat B_K$} & \multicolumn{4}{c|}{$\varepsilon_K^{\rm SM}\times 10^{3}$}\\
  \hhline{|--||----|}
   $N_f=2+1+1$, NLO~\cite{FlavourLatticeAveragingGroupFLAG:2024oxs}\tablefootnote{%
   Note that we have converted the four-flavour version of the bag parameter, $\hat{B}_K(N_f=4)$, to the three-flavour version $\hat{B}_K\equiv \hat{B}_K(N_f=3)$, reducing the numerical value by 1.5\%~\cite{Gorbahn:2024qpe}.}%
   &$0.706(18)(16)$ & $1.98 \pm 0.15$ & $2.03 \pm 0.15$ & $2.08 \pm 0.15$  & $2.13 \pm 0.16$
   \\
   $N_f=2+1+1$, NNLO~\cite{Gorbahn:2024qpe} &$0.733(26)$ & $2.06 \pm 0.16$ & $2.10 \pm 0.16$ & $2.16 \pm 0.16$  & $2.21 \pm 0.16$
   \\
   RBC/UKQCD 24, NLO~\cite{Boyle:2024gge} &$0.7436(82)$ & $2.09 \pm 0.15$ & $2.13 \pm 0.15$ & $2.19 \pm 0.15$  & $2.24 \pm 0.15$
    \\
   RBC/UKQCD 24, NNLO~\cite{Gorbahn:2024qpe} &\boldmath$0.7600(53)$\unboldmath\ & $2.13 \pm 0.14$ & \boldmath$2.18 \pm 0.15$\unboldmath\ & $2.24 \pm 0.15$  & $2.29 \pm 0.15$
    \\
   $N_f=2+1$, NLO~\cite{FlavourLatticeAveragingGroupFLAG:2024oxs} &$0.7533(91)$ & $2.07 \pm 0.14$ & $2.11 \pm 0.14$ & $2.17 \pm 0.14$  & $2.22 \pm 0.14$
    \\
   $N_f=2+1$, NNLO~\cite{Gorbahn:2024qpe} &$0.7637(62)$ & $2.14 \pm 0.15$ & $2.19 \pm 0.15$ & $2.25 \pm 0.15$  & $2.30 \pm 0.15$
    \\
   all LQCD, NNLO~\cite{Gorbahn:2024qpe} &$0.7627(60)$ & $2.14 \pm 0.15$ & $2.19 \pm 0.15$ & $2.25 \pm 0.15$  & $2.30 \pm 0.15$
    \\
  \hhline{|--||----|}

  \end{tabular}
  }
  \renewcommand{\arraystretch}{1.0}
  \caption{\label{tab:eps_K_SM}
    \small
    SM predictions for $\varepsilon_K\times 10^3$ for different values of $\hat B_K$, $\vcb$, and $\gamma$.
        The experimental value is $\varepsilon_K^{\rm exp} = (2.228 \pm 0.011)\times 10^{-3}$~\cite{Zyla:2020zbs}.
        The value used in our numerical analysis is shown in bold, which is {in good agreement with} the experimental value.
    }
  \end{table}
SM predictions for $\varepsilon_K\times 10^3$ for different values of $\hat B_K$, $\vcb$ and $\gamma$ are given in Table~\ref{tab:eps_K_SM}. They should
be compared with the experimental value, which is $\varepsilon_K^{\rm exp} = (2.228 \pm 0.011)\times 10^{-3}$~\cite{Zyla:2020zbs}.

We find that for the full range of parameters considered, the SM predictions
are in agreement with the experimental value at the $1\sigma$ level, but in most cases for the inclusive value of $\vcb$ the central SM
values are below the experimental value.
 Interestingly, our analysis in Section~\ref{sec:SMEFTRGeffects}
  provides an upward shift in $\varepsilon_K$ {from NP} through RG
  effects.
Yet, to identify NP contributions
to  $\varepsilon_K$ will require significant improvements in the three
parameters considered, as well as in the parameter $\eta_{tt}$ (the QCD correction factor for the top contribution to $K^0$ mixing).
On the other hand,
  as expected on the basis of \cite{Buras:2021nns,Buras:2022wpw,Buras:2022qip},
  for the higher value of $\vcb$ and lower value of $\gamma$, the central SM values for    $\varepsilon_K$, in particular in the NNLO case, are  in a very good agreement with experiment.

%
%
\begin{figure}
  \centering
  \includegraphics[width=0.6\textwidth]{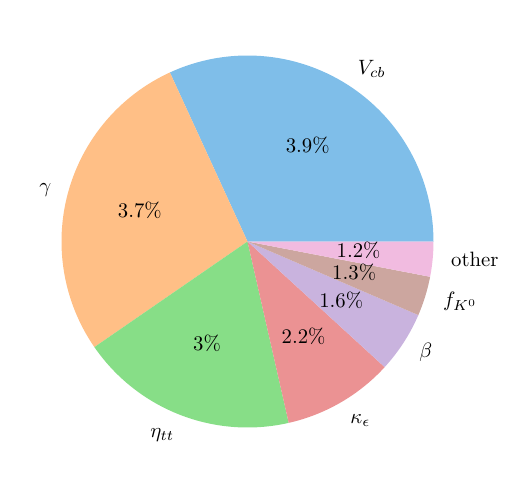}
  \caption{%
  Error budget of $\varepsilon_K$. The total uncertainty is $7.0\%$, which equals the numbers shown in the plot summed in quadrature. The area of each wedge in the pie chart corresponds to the square of the attached number.
}
  \label{fig:epsK_error_budget}
  \end{figure}
%
%
In Fig.~\ref{fig:epsK_error_budget} we display the error budget of $\varepsilon_K$ corresponding to the choice of parameters used in our numerical analysis (shown in bold in Table~\ref{tab:eps_K_SM}).
The overall uncertainty of $7.0\%$ is dominated by the uncertainties of $\vcb$, $\gamma$, and $\eta_{tt}$.
{%
Note that we use $\eta_{tt}=0.550(23)$, which includes the residual theory uncertainty from missing higher-order perturbative corrections estimated in~\cite{Brod:2019rzc}.
While the uncertainty of $\hat B_K$ plays a very minor role (0.7\%, contained in ``other'' in Fig.~\ref{fig:epsK_error_budget}), it should be kept in mind that different values of $\hat B_K$ used in the literature differ significantly from each other and lead to considerably different central values of $\varepsilon_K$, as shown in Table~\ref{tab:eps_K_SM}.
}%

\section{$\Delta F=2$ Matrix Elements}\label{app:matrixelements}

In this appendix, we list the matrix elements of the operators defined in Eq.~\eqref{eq:O_DF2}.
We define
\begin{equation}
 r(\mu) = \left(
 \frac{m_\mathcal{M}}{m_{q_1}(\mu)+m_{q_2}(\mu)}
 \right)^2\,,
\end{equation}
where $M_\mathcal{M}$, $m_{q_1}$, and $m_{q_2}$ are the masses of the meson $\mathcal{M}$ and the quarks $q_1$ and $q_2$ and $\mu$ is a renormalization scale.
We recall that $q_1 q_2$ is $cu$ for $\mathcal{M}=D_0$, $sd$ for $\mathcal{M}=K_0$, $db$ for $\mathcal{M}=B_d$, and $sb$ for $\mathcal{M}=B_s$.
The matrix elements are then given by
\begin{equation}\label{eq:bag_vsa_convention}
 \begin{aligned}
 \frac{\langle O_{VLL}^{q_1 q_2}\rangle(\mu)}{2 M_\mathcal{M}}
 =
 \frac{\langle O_{VRR}^{q_1 q_2}\rangle(\mu)}{2 M_\mathcal{M}}
 &= \frac{1}{3}\, M_\mathcal{M}\, f_\mathcal{M}^2\,B_\mathcal{M}^{(1)}(\mu)\,,
\\
 \frac{\langle O_{SLR}^{q_1 q_2}\rangle(\mu)}{2 M_\mathcal{M}}
 &= \frac{1}{4}\, M_\mathcal{M}\, f_\mathcal{M}^2\,B_\mathcal{M}^{(4)}(\mu)\left(r(\mu)+\frac{1}{6}\right)\,,
\\
 \frac{\langle O_{VLR}^{q_1 q_2}\rangle(\mu)}{2 M_\mathcal{M}}
 &= -\frac{1}{6}\, M_\mathcal{M}\, f_\mathcal{M}^2\,B_\mathcal{M}^{(5)}(\mu)\left(r(\mu)+\frac{3}{2}\right)\,,
 \end{aligned}
\end{equation}
where $f_\mathcal{M}$ is the decay constant of the meson $\mathcal{M}$ and the $B_\mathcal{M}^{(i)}$ are its so-called bag parameters, which are normalized such that in the vacuum saturation approximation~(VSA), they are all given by $B_\mathcal{M}^{(i)}|_{\rm VSA}=1$.

{
Sometimes an alternative convention is used for $B_\mathcal{M}^{(4)}$ and $B_\mathcal{M}^{(5)}$ (see e.g.~\cite{Allton:1998sm}), which we denote in the following with a tilde to distinguish them from those in Eq.~\eqref{eq:bag_vsa_convention}. In the alternative convention, terms that are higher order in the chiral expansion are omitted, which simplifies the scaling properties of the bag parameters. In this convention, $\tilde B_\mathcal{M}^{(4)}$ and $\tilde B_\mathcal{M}^{(5)}$ are given by
\begin{equation}\label{eq:bag_alternative_convention}
 \begin{aligned}
 \frac{\langle O_{SLR}^{q_1 q_2}\rangle(\mu)}{2 M_\mathcal{M}}
 &= \frac{1}{4}\, M_\mathcal{M}\, f_\mathcal{M}^2\,\tilde B_\mathcal{M}^{(4)}(\mu)r(\mu)\,,
\\
 \frac{\langle O_{VLR}^{q_1 q_2}\rangle(\mu)}{2 M_\mathcal{M}}
 &= -\frac{1}{6}\, M_\mathcal{M}\, f_\mathcal{M}^2\,\tilde B_\mathcal{M}^{(5)}(\mu)r(\mu)\,,
 \end{aligned}
\end{equation}
and $\tilde B_\mathcal{M}^{(1)}(\mu) = B_\mathcal{M}^{(1)}(\mu)$.

In our numerical analysis, we use the results of the HPQCD collaboration for the $B_d$ and $B_s$ bag parameters~\cite{Dowdall:2019bea}, which are given in the convention of Eq.~\eqref{eq:bag_vsa_convention}.
For the $D^0$ bag parameters we use the result of the ETM collaboration~\cite{Carrasco:2015pra} given in the alternative convention of Eq.~\eqref{eq:bag_alternative_convention}.
For the $K^0$ bag parameter we use the result of the RBC/UKQCD collaboration~\cite{Boyle:2024gge} given in the alternative convention of  Eq.~\eqref{eq:bag_alternative_convention}.
}

\section{Re-diagonalizing the Running Quark Yukawa Matrices}\label{app:rediag}
Starting at a high scale $\Lambda_{\rm NP}$ with the quark Yukawa matrices $Y_d(\Lambda_{\rm NP})$ and $Y_u(\Lambda_{\rm NP})$  in a canonical flavour basis (e.g. the one where $Y_d(\Lambda_{\rm NP})=\hat{Y}_d(\Lambda_{\rm NP})$ is diagonal) and running them down to the electroweak scale leads to off-diagonal entries in an initially diagonal Yukawa matrix. In order to obtain results in the initial canonical flavour basis, the Yukawa matrices in the non-canonical flavour basis after the running, $\tilde Y_d(M_Z)$ and $\tilde Y_u(M_Z)$, have to be diagonalised,
\begin{equation}
 U_{d_L}^\dagger\ \tilde Y_d(M_Z)\ U_{d_R} = \hat Y_d(M_Z)\,,
 \qquad
 U_{u_L}^\dagger\ \tilde Y_u(M_Z)\ U_{u_R} = \hat Y_u(M_Z)\,,
\end{equation}
and all WCs have to be rotated back to the initial flavour basis.\footnote{Various phenomenological consequences of this so-called ''back-rotation`` have been explored in~\cite{Aebischer:2020lsx}.} E.g., the flavour basis in which $Y_d(M_Z)=\hat{Y}_d(M_Z)$ is diagonal is obtained by choosing $U_q=U_{d_L}$ and rotating all flavour indices associated with $q_L$, $d_R$, and $u_R$ fields using the matrices $U_q$, $U_{d_R}$, and $U_{u_R}$.

For $\Lambda_{\rm NP}=5\,\text{TeV}$ and vanishing $C_{d\phi}$, $C_{u\phi}$, the rotation matrices $U_q$, $U_{d_R}$, and $U_{u_R}$ for rotating to the canonical basis with diagonal $Y_d=\hat Y_d$ after running in the first leading-log approximation to the scale $M_Z$ are given by
\begin{equation}
\arraycolsep=10pt
 U_q - \mathbb{1} =
 \begin{pmatrix}
 \phantom{+}0.00+0.00\,i & -0.08-0.03\,i & \phantom{+}1.81+0.72\,i\\
 \phantom{+}0.08-0.03\,i & \phantom{+}0.00+0.28\,i & -8.88+0.17\,i\\
 -1.81+0.72\,i & \phantom{+}8.88+0.17\,i &  \phantom{+}0.00+0.28\,i\\
 \end{pmatrix}\times 10^{-4}\,,
\end{equation}
\begin{equation}
\arraycolsep=10pt
 U_{d_R} - \mathbb{1} =
 \begin{pmatrix}
 \phantom{+}0.00+0.00\,i & -0.07-0.03\,i & \phantom{+}0.03+0.01\,i\\
 \phantom{+}0.07-0.03\,i & \phantom{+}0.00+2.78\,i & -3.50+0.07\,i\\
 -0.03+0.01\,i & \phantom{+}3.50+0.07\,i &  \phantom{+}0.00+2.75\,i\\
 \end{pmatrix}\times 10^{-5}\,,
\end{equation}
\begin{equation}
\arraycolsep=10pt
 U_{u_R} - \mathbb{1} =
 \begin{pmatrix}
 \phantom{+}0.00+0.00\,i & \phantom{+}0.00+0.00\,i & \phantom{+}0.00+0.00\,i\\
 \phantom{+}0.00+0.00\,i & \phantom{+}0.00+2.78\,i & \phantom{+}0.00+0.00\,i\\
 \phantom{+}0.00+0.00\,i & \phantom{+}0.00+0.00\,i & \phantom{+}0.00+2.75\,i\\
 \end{pmatrix}\times 10^{-5}\,,
\end{equation}
where we used the freedom to choose an overall phase to make $(U_q)_{11}$ real and positive.

If the one-loop RGEs are numerically solved to resum all logs, the resulting rotation matrices are
\begin{equation}
\arraycolsep=10pt
 U_q - \mathbb{1} =
 \begin{pmatrix}
 \phantom{+}0.00+0.00\,i & -0.11-0.04\,i & \phantom{+}2.54+1.01\,i\\
 \phantom{+}0.11-0.05\,i & -0.01+0.39\,i & -12.47+0.23\,i\phantom{0}\\
 -2.54+1.01\,i & \phantom{+}12.47+0.23\,i\phantom{0} &  -0.01+0.39\,i\\
 \end{pmatrix}\times 10^{-4}\,,
\end{equation}
\begin{equation}
\arraycolsep=10pt
 U_{d_R} - \mathbb{1} =
 \begin{pmatrix}
 \phantom{+}0.00+0.00\,i & -0.10-0.04\,i & \phantom{+}0.05+0.02\,i\\
 \phantom{+}0.10-0.04\,i & \phantom{+}0.00+3.89\,i & -4.89+0.09\,i\\
 -0.05+0.02\,i & \phantom{+}4.89+0.09\,i &  \phantom{+}0.00+3.87\,i\\
 \end{pmatrix}\times 10^{-5}\,,
\end{equation}
\begin{equation}
\arraycolsep=10pt
 U_{u_R} - \mathbb{1} =
 \begin{pmatrix}
 \phantom{+}0.00+0.00\,i & \phantom{+}0.00+0.00\,i & \phantom{+}0.00+0.00\,i\\
 \phantom{+}0.00+0.00\,i & \phantom{+}0.00+3.89\,i & \phantom{+}0.00+0.00\,i\\
 \phantom{+}0.00+0.00\,i & \phantom{+}0.00+0.00\,i & \phantom{+}0.00+3.87\,i\\
 \end{pmatrix}\times 10^{-5}\,.
\end{equation}

\section{Matching of Simplified $Z^\prime$ Model to SMEFT}\label{app:SMEFT_matching}
{
\begin{equation}\label{eq:SMEFT_matching}
\begin{aligned}{}
 [C_{qq}^{(1)}]_{ijij} &= -\frac{\big(\Delta^{q_i q_j}_L\big)^2}{2\,M_{Z'}^2}\,,
 \qquad
 [C_{qq}^{(1)}]_{ijji}  = -\frac{\big|\Delta^{q_i q_j}_L\big|^2}{M_{Z'}^2}\,,
 \\
 [C_{dd}]_{ijij} &= -\frac{\big(\Delta^{d_i d_j}_R\big)^2}{2\,M_{Z'}^2}\,,
 \qquad
 [C_{dd}]_{ijji}  = -\frac{\big|\Delta^{d_i d_j}_R\big|^2}{M_{Z'}^2}\,,
 \\
 [C_{uu}]_{ijij} &= -\frac{\big(\Delta^{u_i u_j}_R\big)^2}{2\,M_{Z'}^2}\,,
 \qquad
 [C_{uu}]_{ijji}  = -\frac{\big|\Delta^{u_i u_j}_R\big|^2}{M_{Z'}^2}\,,
 \\
 [C_{qd}^{(1)}]_{ijij} &= -\frac{\Delta^{q_i q_j}_L\,\Delta^{d_i d_j}_R}{M_{Z'}^2}\,,
 \qquad
 [C_{qd}^{(1)}]_{ijji}  = -\frac{\Delta^{q_i q_j}_L\,\big(\Delta^{d_i d_j}_R\big)^*}{M_{Z'}^2}\,,
 \\
 [C_{qu}^{(1)}]_{ijij} &= -\frac{\Delta^{q_i q_j}_L\,\Delta^{u_i u_j}_R}{M_{Z'}^2}\,,
 \qquad
 [C_{qu}^{(1)}]_{ijji}  = -\frac{\Delta^{q_i q_j}_L\,\big(\Delta^{u_i u_j}_R\big)^*}{M_{Z'}^2}\,,
 \\
 [C_{ud}^{(1)}]_{ijij} &= -\frac{\Delta^{u_i u_j}_R\,\Delta^{d_i d_j}_R}{M_{Z'}^2}\,,
 \qquad
 [C_{ud}^{(1)}]_{ijji}  = -\frac{\Delta^{u_i u_j}_R\,\big(\Delta^{d_i d_j}_R\big)^*}{M_{Z'}^2}\,,
 \\
  [C_{ll}]_{kkkk} &= -\frac{\left(\Delta^{e_k}_L\right)^2}{2\,M_{Z'}^2}\,,
 \qquad
 [C_{ll}]_{kkll} = -\frac{\Delta^{e_k}_L\,\Delta^{e_l}_L}{M_{Z'}^2}\,,
 \\
  [C_{ee}]_{kkkk} &= -\frac{\left(\Delta^{e_k}_R\right)^2}{2\,M_{Z'}^2}\,,
 \qquad
 [C_{ee}]_{kkll} = -\frac{2\,\Delta^{e_k}_R\,\Delta^{e_l}_R}{M_{Z'}^2}\,,
 \\
  [C_{le}]_{kkkk} &= -\frac{\Delta^{e_k}_L\,\Delta^{e_k}_R}{M_{Z'}^2}\,,
 \qquad
 [C_{le}]_{kkll} = -\frac{\Delta^{e_k}_L\,\Delta^{e_l}_R}{M_{Z'}^2}\,,
 \qquad
  [C_{le}]_{llkk} &= -\frac{\Delta^{e_l}_L\,\Delta^{e_k}_R}{M_{Z'}^2}\,,
 \\
 [C_{lq}^{(1)}]_{kkij} &= -\frac{\Delta^{e_k}_L\,\Delta^{q_i q_j}_L}{M_{Z'}^2}\,,
 \qquad
 [C_{qe}]_{ijkk} = -\frac{\Delta^{e_k}_R\,\Delta^{q_i q_j}_L}{M_{Z'}^2}\,,
 \\
 [C_{ed}]_{kkij} &= -\frac{\Delta^{e_k}_R\,\Delta^{q_i q_j}_R}{M_{Z'}^2}\,,
 \qquad
 [C_{ld}]_{kkij} = -\frac{\Delta^{e_k}_L\,\Delta^{q_i q_j}_R}{M_{Z'}^2}\,,
\end{aligned}
\end{equation}
where $k<l$ and $i<j$.
}

\section{SMEFT RGE Effects in $\Delta F=2$ Wilson Coefficients}\label{AppC}

\begin{itemize}
 \item Contributions to $[C_{qq}^{(1)}]_{ijij}(M_Z)$:
\begin{equation}
 \begin{aligned}
    {[}{C}_{qq}^{(1)}]_{ijij}(M_Z)
    \approx&\
    [C_{qd}^{(1)}]_{ijij}(\Lambda_\text{NP})
    \,
    \times
    \left[
    - [Y_d]_{ii}\,[Y_d^*]_{jj}
    \,
    \frac{\log\left(M_Z/\Lambda_\text{NP}\right)}{16 \pi^2}
    \right]
    \\
    +&\
    [C_{qd}^{(1)}]_{ijji}(\Lambda_\text{NP})
    \,
    \times
    \left[
    - [Y_d]_{ij}\,[Y_d^*]_{ji}
    \,
    \frac{\log\left(M_Z/\Lambda_\text{NP}\right)}{16 \pi^2}
    \right]
    \\
    +&\
    [C_{qu}^{(1)}]_{ijij}(\Lambda_\text{NP})
    \,
    \times
    \left[
    - [Y_u]_{ii}\,[Y_u^*]_{jj}
    \,
    \frac{\log\left(M_Z/\Lambda_\text{NP}\right)}{16 \pi^2}
    \right]
    \\
    +&\
    [C_{qu}^{(1)}]_{ijji}(\Lambda_\text{NP})
    \,
    \times
    \left[
    - [Y_u]_{ij}\,[Y_u^*]_{ji}
    \,
    \frac{\log\left(M_Z/\Lambda_\text{NP}\right)}{16 \pi^2}
    \right]
    \\
    +&\
    [C_{qq}^{(1)}]_{ijij}(\Lambda_\text{NP})
    \,
    \times
    \left[1+
    {[}\beta_{qq^{(1)}}^{qq^{(1)}}]_{ij}
    \,\frac{\log\left(M_Z/\Lambda_\text{NP}\right)}{16 \pi^2}
    \right]
 \end{aligned}
\end{equation}
\begin{itemize}

 \item $ij=12$, up-aligned flavour basis:
\begin{equation}
 \begin{aligned}
    {[}{\hat C}_{qq}^{(1)}]_{1212}(M_Z)
    \approx&\
    [\hat C_{qd}^{(1)}]_{1212}(\Lambda_\text{NP})
    \,
    \times
    \left[
    - 1.95\times 10^{-11}
    \,
    \log\left(M_Z/\Lambda_\text{NP}\right)
    \right]
    \\
    +&\
    [\hat C_{qd}^{(1)}]_{1221}(\Lambda_\text{NP})
    \,
    \times
    \left[
    1.01\times 10^{-12}
    \,
    \log\left(M_Z/\Lambda_\text{NP}\right)
    \right]
    \\
    +&\
    [\hat C_{qu}^{(1)}]_{1212}(\Lambda_\text{NP})
    \,
    \times
    \left[
    - 1.10\times 10^{-10}
    \,
    \log\left(M_Z/\Lambda_\text{NP}\right)
    \right]
    \\
    +&\
    [\hat C_{qu}^{(1)}]_{1221}(\Lambda_\text{NP})
    \,
    \times
    0
    \\
    +&\
    [\hat C_{qq}^{(1)}]_{1212}(\Lambda_\text{NP})
    \,
    \times
    \left[1+
    6.45\times 10^{-3}
    \,\log\left(M_Z/\Lambda_\text{NP}\right)
    \right]
 \end{aligned}
\end{equation}

 \item $ij=12$, down-aligned flavour basis:
\begin{equation}
 \begin{aligned}
    {[}{C}_{qq}^{(1)}]_{1212}(M_Z)
    \approx&\
    [C_{qd}^{(1)}]_{1212}(\Lambda_\text{NP})
    \,
    \times
    \left[
    - 2.05\times 10^{-11}
    \,
    \log\left(M_Z/\Lambda_\text{NP}\right)
    \right]
    \\
    +&\
    [C_{qd}^{(1)}]_{1221}(\Lambda_\text{NP})
    \,
    \times
    0
    \\
    +&\
    [C_{qu}^{(1)}]_{1212}(\Lambda_\text{NP})
    \,
    \times
    \left[
    - 1.08\times 10^{-10}
    \,
    \log\left(M_Z/\Lambda_\text{NP}\right)
    \right]
    \\
    +&\
    [C_{qu}^{(1)}]_{1221}(\Lambda_\text{NP})
    \,
    \times
    \left[
    2.24\times10^{-12}
    \,
    \log\left(M_Z/\Lambda_\text{NP}\right)
    \right]
    \\
    +&\
    [C_{qq}^{(1)}]_{1212}(\Lambda_\text{NP})
    \,
    \times
    \left[1+
    6.46\times 10^{-3}
    \,\log\left(M_Z/\Lambda_\text{NP}\right)
    \right]
 \end{aligned}
\end{equation}

 \item $ij=13$, down-aligned flavour basis:
\begin{equation}
 \begin{aligned}
    {[}{C}_{qq}^{(1)}]_{1313}(M_Z)
    \approx&\
    [C_{qd}^{(1)}]_{1313}(\Lambda_\text{NP})
    \,
    \times
    \left[
    - 1.03\times 10^{-9}
    \,
    \log\left(M_Z/\Lambda_\text{NP}\right)
    \right]
    \\
    +&\
    [C_{qd}^{(1)}]_{1331}(\Lambda_\text{NP})
    \,
    \times
    0
    \\
    +&\
    [C_{qu}^{(1)}]_{1313}(\Lambda_\text{NP})
    \,
    \times
    \left[
    - 2.94\times 10^{-8}
    \,
    \log\left(M_Z/\Lambda_\text{NP}\right)
    \right]
    \\
    +&\
    [C_{qu}^{(1)}]_{1331}(\Lambda_\text{NP})
    \,
    \times
    \left[
    (-3.22+9.03\,i)\times10^{-11}
    \,
    \log\left(M_Z/\Lambda_\text{NP}\right)
    \right]
    \\
    +&\
    [C_{qq}^{(1)}]_{1313}(\Lambda_\text{NP})
    \,
    \times
    \left[1+
    1.06\times 10^{-2}
    \,\log\left(M_Z/\Lambda_\text{NP}\right)
    \right]
 \end{aligned}
\end{equation}

 \item $ij=23$, down-aligned flavour basis:
\begin{equation}
 \begin{aligned}
    {[}{C}_{qq}^{(1)}]_{2323}(M_Z)
    \approx&\
    [C_{qd}^{(1)}]_{2323}(\Lambda_\text{NP})
    \,
    \times
    \left[
    - 2.10\times 10^{-8}
    \,
    \log\left(M_Z/\Lambda_\text{NP}\right)
    \right]
    \\
    +&\
    [C_{qd}^{(1)}]_{2332}(\Lambda_\text{NP})
    \,
    \times
    0
    \\
    +&\
    [C_{qu}^{(1)}]_{2323}(\Lambda_\text{NP})
    \,
    \times
    \left[
    - 1.47\times 10^{-5}
    \,
    \log\left(M_Z/\Lambda_\text{NP}\right)
    \right]
    \\
    +&\
    [C_{qu}^{(1)}]_{2332}(\Lambda_\text{NP})
    \,
    \times
    \left[
    (2.26-0.05\,i)\times10^{-8}
    \,
    \log\left(M_Z/\Lambda_\text{NP}\right)
    \right]
    \\
    +&\
    [C_{qq}^{(1)}]_{2323}(\Lambda_\text{NP})
    \,
    \times
    \left[1+
    1.06\times 10^{-2}
    \,\log\left(M_Z/\Lambda_\text{NP}\right)
    \right]
 \end{aligned}
\end{equation}

\end{itemize}

 \item Contributions to $[C_{qq}^{(3)}]_{ijij}(M_Z)$:
\begin{equation}
 \begin{aligned}
    {[}{C}_{qq}^{(3)}]_{ijij}(M_Z)
    \approx&\
    [C_{qq}^{(1)}]_{ijij}(\Lambda_\text{NP})
    \,
    \times
    \left[
    [\beta_{qq^{(3)}}^{qq^{(1)}}]
    \,
    \frac{\log\left(M_Z/\Lambda_\text{NP}\right)}{16 \pi^2}
    \right]
 \end{aligned}
\end{equation}

\begin{itemize}

  \item $ij=12$, up-aligned flavour basis:
\begin{equation}
 \begin{aligned}
    {[}{\hat C}_{qq}^{(3)}]_{1212}(M_Z)
    \approx&\
    [\hat C_{qq}^{(1)}]_{1212}(\Lambda_\text{NP})
    \,
    \times
    \left[
    2.61\times10^{-2}
    \,
    \log\left(M_Z/\Lambda_\text{NP}\right)
    \right]
 \end{aligned}
\end{equation}

  \item $ij=12,13,23$, down-aligned flavour basis:
\begin{equation}
 \begin{aligned}
    {[}{C}_{qq}^{(3)}]_{ijij}(M_Z)
    \approx&\
    [C_{qq}^{(1)}]_{ijij}(\Lambda_\text{NP})
    \,
    \times
    \left[
    2.61\times10^{-2}
    \,
    \log\left(M_Z/\Lambda_\text{NP}\right)
    \right]
 \end{aligned}
\end{equation}

\end{itemize}

 \item Contributions to $[C_{dd}]_{ijij}(M_Z)$:
\begin{equation}
 \begin{aligned}
    {[}{C}_{dd}]_{ijij}(M_Z)
    \approx&\
    [C_{qd}^{(1)}]_{ijij}(\Lambda_\text{NP})
    \,
    \times
    \left[
    - 2\, [Y_d]_{jj}\,[Y_d^*]_{ii}
    \,
    \frac{\log\left(M_Z/\Lambda_\text{NP}\right)}{16 \pi^2}
    \right]
    \\
    +&\
    [C_{qd}^{(1)}]_{ijji}(\Lambda_\text{NP})
    \,
    \times
    \left[
    - 2\, [Y_d]_{ij}\,[Y_d^*]_{ji}
    \,
    \frac{\log\left(M_Z/\Lambda_\text{NP}\right)}{16 \pi^2}
    \right]
    \\
    +&\
    [C_{dd}]_{ijij}(\Lambda_\text{NP})        \times
    \left[1+ [\beta_{dd}^{dd}]_{ij}
        \,\frac{\log\left(M_Z/\Lambda_\text{NP}\right)}{16 \pi^2}
    \right]
 \end{aligned}
\end{equation}

\begin{itemize}

   \item $ij=12$, up-aligned flavour basis:
\begin{equation}
 \begin{aligned}
    {[}{\hat C}_{dd}]_{1212}(M_Z)
    \approx&\
    [\hat C_{qd}^{(1)}]_{1212}(\Lambda_\text{NP})
    \,
    \times
    \left[
    - 3.9\times10^{-12}
    \,
    \log\left(M_Z/\Lambda_\text{NP}\right)
    \right]
    \\
    +&\
    [\hat C_{qd}^{(1)}]_{1221}(\Lambda_\text{NP})
    \,
    \times
    \left[
    2.01\times10^{-12}
    \,
    \log\left(M_Z/\Lambda_\text{NP}\right)
    \right]
    \\
    +&\
    [\hat C_{dd}]_{1212}(\Lambda_\text{NP})
    \,
    \times
    \left[1+
      2.58\times10^{-2}
    \,\log\left(M_Z/\Lambda_\text{NP}\right)
    \right]
 \end{aligned}
\end{equation}

   \item $ij=12$, down-aligned flavour basis:
\begin{equation}
 \begin{aligned}
    {[}{C}_{dd}]_{1212}(M_Z)
    \approx&\
    [C_{qd}^{(1)}]_{1212}(\Lambda_\text{NP})
    \,
    \times
    \left[
    - 4.1\times10^{-11}
    \,
    \log\left(M_Z/\Lambda_\text{NP}\right)
    \right]
    \\
    +&\
    [C_{qd}^{(1)}]_{1221}(\Lambda_\text{NP})
    \,
    \times
    0
    \\
    +&\
    [C_{dd}]_{1212}(\Lambda_\text{NP})
    \,
    \times
    \left[1+
      2.58\times10^{-2}
    \,\log\left(M_Z/\Lambda_\text{NP}\right)
    \right]
 \end{aligned}
\end{equation}

   \item $ij=13$, down-aligned flavour basis:
\begin{equation}
 \begin{aligned}
    {[}{C}_{dd}]_{1313}(M_Z)
    \approx&\
    [C_{qd}^{(1)}]_{1313}(\Lambda_\text{NP})
    \,
    \times
    \left[
    - 2.06\times10^{-9}
    \,
    \log\left(M_Z/\Lambda_\text{NP}\right)
    \right]
    \\
    +&\
    [C_{qd}^{(1)}]_{1331}(\Lambda_\text{NP})
    \,
    \times
    0
    \\
    +&\
    [C_{dd}]_{1313}(\Lambda_\text{NP})
    \,
    \times
    \left[1+
      2.58\times10^{-2}
    \,\log\left(M_Z/\Lambda_\text{NP}\right)
    \right]
 \end{aligned}
\end{equation}

   \item $ij=23$, down-aligned flavour basis:
\begin{equation}
 \begin{aligned}
    {[}{C}_{dd}]_{2323}(M_Z)
    \approx&\
    [C_{qd}^{(1)}]_{2323}(\Lambda_\text{NP})
    \,
    \times
    \left[
    - 4.2\times10^{-8}
    \,
    \log\left(M_Z/\Lambda_\text{NP}\right)
    \right]
    \\
    +&\
    [C_{qd}^{(1)}]_{2332}(\Lambda_\text{NP})
    \,
    \times
    0
    \\
    +&\
    [C_{dd}]_{2323}(\Lambda_\text{NP})
    \,
    \times
    \left[1+
      2.58\times10^{-2}
    \,\log\left(M_Z/\Lambda_\text{NP}\right)
    \right]
 \end{aligned}
\end{equation}

\end{itemize}

 \item Contributions to $[C_{uu}]_{ijij}(M_Z)$:
\begin{equation}
 \begin{aligned}
    {[}{C}_{uu}]_{ijij}(M_Z)
    \approx&\
    [C_{qu}^{(1)}]_{ijij}(\Lambda_\text{NP})
    \,
    \times
    \left[
    - 2\, [Y_u]_{jj}\,[Y_u^*]_{ii}
    \,
    \frac{\log\left(M_Z/\Lambda_\text{NP}\right)}{16 \pi^2}
    \right]
    \\
    +&\
    [C_{qu}^{(1)}]_{ijji}(\Lambda_\text{NP})
    \,
    \times
    \left[
    - 2\, [Y_u]_{ij}\,[Y_u^*]_{ji}
    \,
    \frac{\log\left(M_Z/\Lambda_\text{NP}\right)}{16 \pi^2}
    \right]
    \\
    +&\
    [C_{uu}]_{ijij}(\Lambda_\text{NP})
    \,
    \times
    \left[1+
          [\beta_{uu}^{uu}]_{ij}
        \,\frac{\log\left(M_Z/\Lambda_\text{NP}\right)}{16 \pi^2}
    \right]
 \end{aligned}
\end{equation}

\begin{itemize}
   \item $ij=12$, up-aligned flavour basis:
\begin{equation}
 \begin{aligned}
    {[}{\hat C}_{uu}]_{1212}(M_Z)
    \approx&\
    [\hat C_{qu}^{(1)}]_{1212}(\Lambda_\text{NP})
    \,
    \times
    \left[
    -2.21\times10^{-10}
    \,
    \log\left(M_Z/\Lambda_\text{NP}\right)
    \right]
    \\
    +&\
    [\hat C_{qu}^{(1)}]_{1212}(\Lambda_\text{NP})
    \,
    \times
0
    \\
    +&\
    [\hat C_{uu}]_{1212}(\Lambda_\text{NP})
    \,
    \times
    \left[1+
    2.92\times10^{-2}
    \,\log\left(M_Z/\Lambda_\text{NP}\right)
    \right]
 \end{aligned}
\end{equation}

   \item $ij=12$, down-aligned flavour basis:
\begin{equation}
 \begin{aligned}
    {[}{C}_{uu}]_{1212}(M_Z)
    \approx&\
    [C_{qu}^{(1)}]_{1212}(\Lambda_\text{NP})
    \,
    \times
    \left[
    -2.16\times10^{-10}
    \,
    \log\left(M_Z/\Lambda_\text{NP}\right)
    \right]
    \\
    +&\
    [C_{qu}^{(1)}]_{1221}(\Lambda_\text{NP})
    \,
    \times
    \left[
    4.48\times10^{-12}
    \,
    \log\left(M_Z/\Lambda_\text{NP}\right)
    \right]
    \\
    +&\
    [C_{uu}]_{1212}(\Lambda_\text{NP})
    \,
    \times
    \left[1+
    2.92\times10^{-2}
    \,\log\left(M_Z/\Lambda_\text{NP}\right)
    \right]
 \end{aligned}
\end{equation}

   \item $ij=13$, down-aligned flavour basis:
\begin{equation}
 \begin{aligned}
    {[}{C}_{uu}]_{1313}(M_Z)
    \approx&\
    [C_{qu}^{(1)}]_{1313}(\Lambda_\text{NP})
    \,
    \times
    \left[
    -5.88\times10^{-8}
    \,
    \log\left(M_Z/\Lambda_\text{NP}\right)
    \right]
    \\
    +&\
    [C_{qu}^{(1)}]_{1331}(\Lambda_\text{NP})
    \,
    \times
    \left[
    (-0.64+1.81\,i)\times10^{-10}
    \,
    \log\left(M_Z/\Lambda_\text{NP}\right)
    \right]
    \\
    +&\
    [C_{uu}]_{1313}(\Lambda_\text{NP})
    \,
    \times
    \left[1+
    3.75\times10^{-2}
    \,\log\left(M_Z/\Lambda_\text{NP}\right)
    \right]
 \end{aligned}
\end{equation}

   \item $ij=23$, down-aligned flavour basis:
\begin{equation}
 \begin{aligned}
    {[}{C}_{uu}]_{2323}(M_Z)
    \approx&\
    [C_{qu}^{(1)}]_{2323}(\Lambda_\text{NP})
    \,
    \times
    \left[
    -2.95\times10^{-5}
    \,
    \log\left(M_Z/\Lambda_\text{NP}\right)
    \right]
    \\
    +&\
    [C_{qu}^{(1)}]_{2332}(\Lambda_\text{NP})
    \,
    \times
    \left[
    (4.53-0.10\,i)\times10^{-8}
    \,
    \log\left(M_Z/\Lambda_\text{NP}\right)
    \right]
    \\
    +&\
    [C_{uu}]_{2323}(\Lambda_\text{NP})
    \,
    \times
    \left[1+
    3.75\times10^{-2}
    \,\log\left(M_Z/\Lambda_\text{NP}\right)
    \right]
 \end{aligned}
\end{equation}

\end{itemize}

 \item Contributions to $[C_{qd}^{(1)}]_{ijij}(M_Z)$:
\begin{equation}
 \begin{aligned}
    {[}{C}_{qd}^{(1)}]_{ijij}(M_Z)
    \approx&\
    [C_{ud}^{(1)}]_{ijij}(\Lambda_\text{NP})
    \,
    \times
    \left[
    - [Y_u]_{ii}\,[Y_u^*]_{jj}
    \,
    \frac{\log\left(M_Z/\Lambda_\text{NP}\right)}{16 \pi^2}
    \right]
    \\
    +&\
    [C_{ud}^{(1)}]_{ijji}(\Lambda_\text{NP})
    \,
    \times
    \left[
    - [Y_u]_{ij}\,[Y_u^*]_{ji}
    \,
    \frac{\log\left(M_Z/\Lambda_\text{NP}\right)}{16 \pi^2}
    \right]
    \\
    +&\
    [C_{dd}]_{ijij}(\Lambda_\text{NP})
    \,
    \times
    \left[
    -\tfrac{8}{3}\, [Y_d]_{ii}\,[Y_d^*]_{jj}
    \,
    \frac{\log\left(M_Z/\Lambda_\text{NP}\right)}{16 \pi^2}
    \right]
    \\
    +&\
    [C_{qq}^{(1)}]_{ijij}(\Lambda_\text{NP})
    \,
    \times
    \left[
    -\tfrac{14}{3}\, [Y_d]_{jj}\,[Y_d^*]_{ii}
    \,
    \frac{\log\left(M_Z/\Lambda_\text{NP}\right)}{16 \pi^2}
    \right]
    \\
    +&\
    [C_{qd}^{(1)}]_{ijij}(\Lambda_\text{NP})
    \,
    \times
    \bigg[1+
    [\beta_{qd^{(1)}}^{qd^{(1)}}]_{ij}
        \,\frac{\log\left(M_Z/\Lambda_\text{NP}\right)}{16 \pi^2}
    \bigg]
 \end{aligned}
\end{equation}

\begin{itemize}
   \item $ij=12$, up-aligned flavour basis:
\begin{equation}
 \begin{aligned}
    {[}{\hat C}_{qd}^{(1)}]_{1212}(M_Z)
    \approx&\
    [\hat C_{ud}^{(1)}]_{1212}(\Lambda_\text{NP})
    \,
    \times
    \left[
    -1.10\times10^{-10}
    \,
    \log\left(M_Z/\Lambda_\text{NP}\right)
    \right]
    \\
    +&\
    [\hat C_{ud}^{(1)}]_{1221}(\Lambda_\text{NP})
    \,
    \times
    0
    \\
    +&\
    [\hat C_{dd}]_{1212}(\Lambda_\text{NP})
    \,
    \times
    \left[
    -5.19\times10^{-11}
    \,
    \log\left(M_Z/\Lambda_\text{NP}\right)
    \right]
    \\
    +&\
    [\hat C_{qq}^{(1)}]_{1212}(\Lambda_\text{NP})
    \,
    \times
    \left[
    -9.08\times10^{-11}
    \,
    \log\left(M_Z/\Lambda_\text{NP}\right)
    \right]
    \\
    +&\
    [\hat C_{qd}^{(1)}]_{1212}(\Lambda_\text{NP})
    \,
    \times
    \left[1+
    5.65\times10^{-4}
    \,\log\left(M_Z/\Lambda_\text{NP}\right)
    \right]
 \end{aligned}
\end{equation}

   \item $ij=12$, down-aligned flavour basis:
\begin{equation}
 \begin{aligned}
    {[}{C}_{qd}^{(1)}]_{1212}(M_Z)
    \approx&\
    [C_{ud}^{(1)}]_{1212}(\Lambda_\text{NP})
    \,
    \times
    \left[
    -1.05\times10^{-10}
    \,
    \log\left(M_Z/\Lambda_\text{NP}\right)
    \right]
    \\
    +&\
    [C_{ud}^{(1)}]_{1221}(\Lambda_\text{NP})
    \,
    \times
    \left[
    5.55\times10^{-12}
    \,
    \log\left(M_Z/\Lambda_\text{NP}\right)
    \right]
    \\
    +&\
    [C_{dd}]_{1212}(\Lambda_\text{NP})
    \,
    \times
    \left[
    -5.47\times10^{-11}
    \,
    \log\left(M_Z/\Lambda_\text{NP}\right)
    \right]
    \\
    +&\
    [C_{qq}^{(1)}]_{1212}(\Lambda_\text{NP})
    \,
    \times
    \left[
    -9.57\times10^{-11}
    \,
    \log\left(M_Z/\Lambda_\text{NP}\right)
    \right]
    \\
    +&\
    [C_{qd}^{(1)}]_{1212}(\Lambda_\text{NP})
    \,
    \times
    \left[1+
    5.69\times10^{-4}
    \,\log\left(M_Z/\Lambda_\text{NP}\right)
    \right]
 \end{aligned}
\end{equation}

   \item $ij=13$, down-aligned flavour basis:
\begin{equation}
 \begin{aligned}
    {[}{C}_{qd}^{(1)}]_{1313}(M_Z)
    \approx&\
    [C_{ud}^{(1)}]_{1313}(\Lambda_\text{NP})
    \,
    \times
    \left[
    -2.95\times10^{-8}
    \,
    \log\left(M_Z/\Lambda_\text{NP}\right)
    \right]
    \\
    +&\
    [C_{ud}^{(1)}]_{1331}(\Lambda_\text{NP})
    \,
    \times
    \left[
    (-6.64+8.20\,i)\times10^{-13}
    \,
    \log\left(M_Z/\Lambda_\text{NP}\right)
    \right]
    \\
    +&\
    [C_{dd}]_{1313}(\Lambda_\text{NP})
    \,
    \times
    \left[
    -2.75\times10^{-9}
    \,
    \log\left(M_Z/\Lambda_\text{NP}\right)
    \right]
    \\
    +&\
    [C_{qq}^{(1)}]_{1313}(\Lambda_\text{NP})
    \,
    \times
    \left[
    -4.81\times10^{-9}
    \,
    \log\left(M_Z/\Lambda_\text{NP}\right)
    \right]
    \\
    +&\
    [C_{qd}^{(1)}]_{1313}(\Lambda_\text{NP})
    \,
    \times
    \left[1+
    2.64\times10^{-3}
    \,\log\left(M_Z/\Lambda_\text{NP}\right)
    \right]
 \end{aligned}
\end{equation}

   \item $ij=23$, down-aligned flavour basis:
\begin{equation}
 \begin{aligned}
    {[}{C}_{qd}^{(1)}]_{2323}(M_Z)
    \approx&\
    [C_{ud}^{(1)}]_{2323}(\Lambda_\text{NP})
    \,
    \times
    \left[
    -1.47\times10^{-5}
    \,
    \log\left(M_Z/\Lambda_\text{NP}\right)
    \right]
    \\
    +&\
    [C_{ud}^{(1)}]_{1221}(\Lambda_\text{NP})
    \,
    \times
    \left[
    (2.81-0.05\,i)\times10^{-8}
    \,
    \log\left(M_Z/\Lambda_\text{NP}\right)
    \right]
    \\
    +&\
    [C_{dd}]_{2323}(\Lambda_\text{NP})
    \,
    \times
    \left[
    -5.6\times10^{-8}
    \,
    \log\left(M_Z/\Lambda_\text{NP}\right)
    \right]
    \\
    +&\
    [C_{qq}^{(1)}]_{2323}(\Lambda_\text{NP})
    \,
    \times
    \left[
    -9.81\times10^{-8}
    \,
    \log\left(M_Z/\Lambda_\text{NP}\right)
    \right]
    \\
    +&\
    [C_{qd}^{(1)}]_{2323}(\Lambda_\text{NP})
    \,
    \times
    \left[1+
    2.64\times10^{-3}
    \,\log\left(M_Z/\Lambda_\text{NP}\right)
    \right]
 \end{aligned}
\end{equation}

\end{itemize}

 \item Contributions to $[C_{qu}^{(1)}]_{ijij}(M_Z)$:
\begin{equation}
 \begin{aligned}
    {[}{C}_{qu}^{(1)}]_{ijij}(M_Z)
    \approx&\
    [C_{ud}^{(1)}]_{ijij}(\Lambda_\text{NP})
    \,
    \times
    \left[
    - [Y_d]_{ii}\,[Y_d^*]_{jj}
    \,
    \frac{\log\left(M_Z/\Lambda_\text{NP}\right)}{16 \pi^2}
    \right]
    \\
    +&\
    [C_{ud}^{(1)}]_{ijji}(\Lambda_\text{NP})
    \,
    \times
    \left[
    - [Y_d]_{ij}\,[Y_d^*]_{ji}
    \,
    \frac{\log\left(M_Z/\Lambda_\text{NP}\right)}{16 \pi^2}
    \right]
    \\
    +&\
    [C_{uu}]_{ijij}(\Lambda_\text{NP})
    \,
    \times
    \left[
    -\tfrac{8}{3}\, [Y_u]_{ii}\,[Y_u^*]_{jj}
    \,
    \frac{\log\left(M_Z/\Lambda_\text{NP}\right)}{16 \pi^2}
    \right]
    \\
    +&\
    [C_{qq}^{(1)}]_{ijij}(\Lambda_\text{NP})
    \,
    \times
    \left[
    -\tfrac{14}{3}\, [Y_u]_{jj}\,[Y_u^*]_{ii}
    \,
    \frac{\log\left(M_Z/\Lambda_\text{NP}\right)}{16 \pi^2}
    \right]
    \\
    +&\
    [C_{qu}^{(1)}]_{ijij}(\Lambda_\text{NP})
    \,
    \times
    \bigg[1+
    [\beta_{qu^{(1)}}^{qu^{(1)}}]_{ij}
        \,\frac{\log\left(M_Z/\Lambda_\text{NP}\right)}{16 \pi^2}
    \bigg]
 \end{aligned}
\end{equation}

\begin{itemize}
    \item $ij=12$, up-aligned flavour basis:
\begin{equation}
 \begin{aligned}
    {[}{\hat C}_{qu}^{(1)}]_{1212}(M_Z)
    \approx&\
    [\hat C_{ud}^{(1)}]_{1212}(\Lambda_\text{NP})
    \,
    \times
    \left[
    -1.95\times10^{-11}
    \,
    \log\left(M_Z/\Lambda_\text{NP}\right)
    \right]
    \\
    +&\
    [\hat C_{ud}^{(1)}]_{1221}(\Lambda_\text{NP})
    \,
    \times
    \left[
    1.03\times10^{-12}
    \,
    \log\left(M_Z/\Lambda_\text{NP}\right)
    \right]
    \\
    +&\
    [\hat C_{uu}]_{1212}(\Lambda_\text{NP})
    \,
    \times
    \left[
    -2.95\times10^{-10}
    \,
    \log\left(M_Z/\Lambda_\text{NP}\right)
    \right]
    \\
    +&\
    [\hat C_{qq}^{(1)}]_{1212}(\Lambda_\text{NP})
    \,
    \times
    \left[
    -5.15\times10^{-10}
    \,
    \log\left(M_Z/\Lambda_\text{NP}\right)
    \right]
    \\
    +&\
    [\hat C_{qu}^{(1)}]_{1212}(\Lambda_\text{NP})
    \,
    \times
    \left[1
    -1.13\times10^{-3}
    \,\log\left(M_Z/\Lambda_\text{NP}\right)
    \right]
 \end{aligned}
\end{equation}

    \item $ij=12$, down-aligned flavour basis:
\begin{equation}
 \begin{aligned}
    {[}{C}_{qu}^{(1)}]_{1212}(M_Z)
    \approx&\
    [C_{ud}^{(1)}]_{1212}(\Lambda_\text{NP})
    \,
    \times
    \left[
    -2.05\times10^{-11}
    \,
    \log\left(M_Z/\Lambda_\text{NP}\right)
    \right]
    \\
    +&\
    [C_{ud}^{(1)}]_{1221}(\Lambda_\text{NP})
    \,
    \times
    0
    \\
    +&\
    [C_{uu}]_{1212}(\Lambda_\text{NP})
    \,
    \times
    \left[
    -2.79\times10^{-10}
    \,
    \log\left(M_Z/\Lambda_\text{NP}\right)
    \right]
    \\
    +&\
    [C_{qq}^{(1)}]_{1212}(\Lambda_\text{NP})
    \,
    \times
    \left[
    -4.89\times10^{-10}
    \,
    \log\left(M_Z/\Lambda_\text{NP}\right)
    \right]
    \\
    +&\
    [C_{qu}^{(1)}]_{1212}(\Lambda_\text{NP})
    \,
    \times
    \left[1
    -1.13\times10^{-3}
    \,\log\left(M_Z/\Lambda_\text{NP}\right)
    \right]
 \end{aligned}
\end{equation}

    \item $ij=13$, down-aligned flavour basis:
\begin{equation}
 \begin{aligned}
    {[}{C}_{qu}^{(1)}]_{1313}(M_Z)
    \approx&\
    [C_{ud}^{(1)}]_{1313}(\Lambda_\text{NP})
    \,
    \times
    \left[
    -1.03\times10^{-9}
    \,
    \log\left(M_Z/\Lambda_\text{NP}\right)
    \right]
    \\
    +&\
    [C_{ud}^{(1)}]_{1331}(\Lambda_\text{NP})
    \,
    \times
    0
    \\
    +&\
    [C_{uu}]_{1313}(\Lambda_\text{NP})
    \,
    \times
    \left[
    -7.86\times10^{-8}
    \,
    \log\left(M_Z/\Lambda_\text{NP}\right)
    \right]
    \\
    +&\
    [C_{qq}^{(1)}]_{1313}(\Lambda_\text{NP})
    \,
    \times
    \left[
    -1.38\times10^{-7}
    \,
    \log\left(M_Z/\Lambda_\text{NP}\right)
    \right]
    \\
    +&\
    [C_{qu}^{(1)}]_{1313}(\Lambda_\text{NP})
    \,
    \times
    \left[1
    +6.47\times10^{-3}
    \,\log\left(M_Z/\Lambda_\text{NP}\right)
    \right]
 \end{aligned}
\end{equation}

    \item $ij=23$, down-aligned flavour basis:
\begin{equation}
 \begin{aligned}
    {[}{C}_{qu}^{(1)}]_{2323}(M_Z)
    \approx&\
    [C_{ud}^{(1)}]_{2323}(\Lambda_\text{NP})
    \,
    \times
    \left[
    -2.10\times10^{-8}
    \,
    \log\left(M_Z/\Lambda_\text{NP}\right)
    \right]
    \\
    +&\
    [C_{ud}^{(1)}]_{2332}(\Lambda_\text{NP})
    \,
    \times
    0
    \\
    +&\
    [C_{uu}]_{2323}(\Lambda_\text{NP})
    \,
    \times
    \left[
    -3.93\times10^{-5}
    \,
    \log\left(M_Z/\Lambda_\text{NP}\right)
    \right]
    \\
    +&\
    [C_{qq}^{(1)}]_{2323}(\Lambda_\text{NP})
    \,
    \times
    \left[
    -6.87\times10^{-5}
    \,
    \log\left(M_Z/\Lambda_\text{NP}\right)
    \right]
    \\
    +&\
    [C_{qu}^{(1)}]_{2323}(\Lambda_\text{NP})
    \,
    \times
    \left[1
    +6.48\times10^{-3}
    \,\log\left(M_Z/\Lambda_\text{NP}\right)
    \right]
 \end{aligned}
\end{equation}

\end{itemize}

 \item Contributions to $[C_{qd}^{(8)}]_{ijij}(M_Z)$:
\begin{equation}
 \begin{aligned}
    {[}{C}_{qd}^{(8)}]_{ijij}(M_Z)
    \approx&\
    [C_{dd}]_{ijij}(\Lambda_\text{NP})
    \,
    \times
    \left[
    -4\, [Y_d]_{ii}\,[Y_d^*]_{jj}
    \,
    \frac{\log\left(M_Z/\Lambda_\text{NP}\right)}{16 \pi^2}
    \right]
    \\
    +&\
    [C_{qq}^{(1)}]_{ijij}(\Lambda_\text{NP})
    \,
    \times
    \left[
    -4\, [Y_d]_{jj}\,[Y_d^*]_{ii}
    \,
    \frac{\log\left(M_Z/\Lambda_\text{NP}\right)}{16 \pi^2}
    \right]
    \\
    +&\
    [C_{qd}^{(1)}]_{ijij}(\Lambda_\text{NP})
    \,
    \times
    \bigg[
     [\beta_{qd^{(8)}}^{qd^{(1)}}]_{ij}
        \,\frac{\log\left(M_Z/\Lambda_\text{NP}\right)}{16 \pi^2}
    \bigg]
 \end{aligned}
\end{equation}

\begin{itemize}
    \item $ij=12$, up-aligned flavour basis:
\begin{equation}
 \begin{aligned}
    {[}{\hat C}_{qd}^{(8)}]_{1212}(M_Z)
    \approx&\
    [\hat C_{dd}]_{1212}(\Lambda_\text{NP})
    \,
    \times
    \left[
    -7.78\times10^{-11}
    \,
    \log\left(M_Z/\Lambda_\text{NP}\right)
    \right]
    \\
    +&\
    [\hat C_{qq}^{(1)}]_{1212}(\Lambda_\text{NP})
    \,
    \times
    \left[
    -7.78\times10^{-11}
    \,
    \log\left(M_Z/\Lambda_\text{NP}\right)
    \right]
    \\
    +&\
    [\hat C_{qd}^{(1)}]_{1212}(\Lambda_\text{NP})
    \,
    \times
    \left[
    -7.4\times10^{-2}
    \,\log\left(M_Z/\Lambda_\text{NP}\right)
    \right]
 \end{aligned}
\end{equation}

    \item $ij=12$, down-aligned flavour basis:
\begin{equation}
 \begin{aligned}
    {[}{C}_{qd}^{(8)}]_{1212}(M_Z)
    \approx&\
    [C_{dd}]_{1212}(\Lambda_\text{NP})
    \,
    \times
    \left[
    -8.20\times10^{-11}
    \,
    \log\left(M_Z/\Lambda_\text{NP}\right)
    \right]
    \\
    +&\
    [C_{qq}^{(1)}]_{1212}(\Lambda_\text{NP})
    \,
    \times
    \left[
    -8.20\times10^{-11}
    \,
    \log\left(M_Z/\Lambda_\text{NP}\right)
    \right]
    \\
    +&\
    [C_{qd}^{(1)}]_{1212}(\Lambda_\text{NP})
    \,
    \times
    \left[
    -7.4\times10^{-2}
    \,\log\left(M_Z/\Lambda_\text{NP}\right)
    \right]
 \end{aligned}
\end{equation}

    \item $ij=13$, down-aligned flavour basis:
\begin{equation}
 \begin{aligned}
    {[}{C}_{qd}^{(8)}]_{1313}(M_Z)
    \approx&\
    [C_{dd}]_{1313}(\Lambda_\text{NP})
    \,
    \times
    \left[
    -4.12\times10^{-9}
    \,
    \log\left(M_Z/\Lambda_\text{NP}\right)
    \right]
    \\
    +&\
    [C_{qq}^{(1)}]_{1313}(\Lambda_\text{NP})
    \,
    \times
    \left[
    -4.12\times10^{-9}
    \,
    \log\left(M_Z/\Lambda_\text{NP}\right)
    \right]
    \\
    +&\
    [C_{qd}^{(1)}]_{1313}(\Lambda_\text{NP})
    \,
    \times
    \left[
    -7.4\times10^{-2}
    \,\log\left(M_Z/\Lambda_\text{NP}\right)
    \right]
 \end{aligned}
\end{equation}

    \item $ij=23$, down-aligned flavour basis:
\begin{equation}
 \begin{aligned}
    {[}{C}_{qd}^{(8)}]_{2323}(M_Z)
    \approx&\
    [C_{dd}]_{2323}(\Lambda_\text{NP})
    \,
    \times
    \left[
    -8.41\times10^{-8}
    \,
    \log\left(M_Z/\Lambda_\text{NP}\right)
    \right]
    \\
    +&\
    [C_{qq}^{(1)}]_{2323}(\Lambda_\text{NP})
    \,
    \times
    \left[
    -8.41\times10^{-8}
    \,
    \log\left(M_Z/\Lambda_\text{NP}\right)
    \right]
    \\
    +&\
    [C_{qd}^{(1)}]_{2323}(\Lambda_\text{NP})
    \,
    \times
    \left[
    -7.4\times10^{-2}
    \,\log\left(M_Z/\Lambda_\text{NP}\right)
    \right]
 \end{aligned}
\end{equation}

\end{itemize}

 \item Contributions to $[C_{qu}^{(8)}]_{ijij}(M_Z)$:
\begin{equation}
 \begin{aligned}
    {[}{C}_{qu}^{(8)}]_{ijij}(M_Z)
    \approx&\
    [C_{uu}]_{ijij}(\Lambda_\text{NP})
    \,
    \times
    \left[
    -4\, [Y_u]_{ii}\,[Y_u^*]_{jj}
    \,
    \frac{\log\left(M_Z/\Lambda_\text{NP}\right)}{16 \pi^2}
    \right]
    \\
    +&\
    [C_{qq}^{(1)}]_{ijij}(\Lambda_\text{NP})
    \,
    \times
    \left[
    -4\, [Y_u]_{jj}\,[Y_u^*]_{ii}
    \,
    \frac{\log\left(M_Z/\Lambda_\text{NP}\right)}{16 \pi^2}
    \right]
    \\
    +&\
    [C_{qu}^{(1)}]_{ijij}(\Lambda_\text{NP})
    \,
    \times
    \bigg[
    [\beta_{qu^{(8)}}^{qu^{(1)}}]_{ij}
   \,\frac{\log\left(M_Z/\Lambda_\text{NP}\right)}{16 \pi^2}
    \bigg]
 \end{aligned}
\end{equation}

\begin{itemize}
     \item $ij=12$, up-aligned flavour basis:
\begin{equation}
 \begin{aligned}
    {[}{\hat C}_{qu}^{(8)}]_{1212}(M_Z)
    \approx&\
    [\hat C_{uu}]_{1212}(\Lambda_\text{NP})
    \,
    \times
    \left[
    -4.42\times10^{-10}
    \,
    \log\left(M_Z/\Lambda_\text{NP}\right)
    \right]
    \\
    +&\
    [\hat C_{qq}^{(1)}]_{1212}(\Lambda_\text{NP})
    \,
    \times
    \left[
    -4.42\times10^{-10}
    \,
    \log\left(M_Z/\Lambda_\text{NP}\right)
    \right]
    \\
    +&\
    [\hat C_{qu}^{(1)}]_{1212}(\Lambda_\text{NP})
    \,
    \times
    \left[
    -7.4\times10^{-2}
    \,\log\left(M_Z/\Lambda_\text{NP}\right)
    \right]
 \end{aligned}
\end{equation}

     \item $ij=12$, down-aligned flavour basis:
\begin{equation}
 \begin{aligned}
    {[}{C}_{qu}^{(8)}]_{1212}(M_Z)
    \approx&\
    [C_{uu}]_{1212}(\Lambda_\text{NP})
    \,
    \times
    \left[
    -4.19\times10^{-10}
    \,
    \log\left(M_Z/\Lambda_\text{NP}\right)
    \right]
    \\
    +&\
    [C_{qq}^{(1)}]_{1212}(\Lambda_\text{NP})
    \,
    \times
    \left[
    -4.19\times10^{-10}
    \,
    \log\left(M_Z/\Lambda_\text{NP}\right)
    \right]
    \\
    +&\
    [C_{qu}^{(1)}]_{1212}(\Lambda_\text{NP})
    \,
    \times
    \left[
    -7.4\times10^{-2}
    \,\log\left(M_Z/\Lambda_\text{NP}\right)
    \right]
 \end{aligned}
\end{equation}

     \item $ij=13$, down-aligned flavour basis:
\begin{equation}
 \begin{aligned}
    {[}{C}_{qu}^{(8)}]_{1313}(M_Z)
    \approx&\
    [C_{uu}]_{1313}(\Lambda_\text{NP})
    \,
    \times
    \left[
    -1.18\times10^{-7}
    \,
    \log\left(M_Z/\Lambda_\text{NP}\right)
    \right]
    \\
    +&\
    [C_{qq}^{(1)}]_{1313}(\Lambda_\text{NP})
    \,
    \times
    \left[
    -1.18\times10^{-7}
    \,
    \log\left(M_Z/\Lambda_\text{NP}\right)
    \right]
    \\
    +&\
    [C_{qu}^{(1)}]_{1313}(\Lambda_\text{NP})
    \,
    \times
    \left[
    -6.57\times10^{-2}
    \,\log\left(M_Z/\Lambda_\text{NP}\right)
    \right]
 \end{aligned}
\end{equation}

     \item $ij=23$, down-aligned flavour basis:
\begin{equation}
 \begin{aligned}
    {[}{C}_{qu}^{(8)}]_{2323}(M_Z)
    \approx&\
    [C_{uu}]_{2323}(\Lambda_\text{NP})
    \,
    \times
    \left[
    -5.89\times10^{-5}
    \,
    \log\left(M_Z/\Lambda_\text{NP}\right)
    \right]
    \\
    +&\
    [C_{qq}^{(1)}]_{2323}(\Lambda_\text{NP})
    \,
    \times
    \left[
    -5.89\times10^{-5}
    \,
    \log\left(M_Z/\Lambda_\text{NP}\right)
    \right]
    \\
    +&\
    [C_{qu}^{(1)}]_{2323}(\Lambda_\text{NP})
    \,
    \times
    \left[
    -6.57\times10^{-2}
    \,\log\left(M_Z/\Lambda_\text{NP}\right)
    \right]
 \end{aligned}
\end{equation}

\end{itemize}

\end{itemize}

\FloatBarrier

\bibliographystyle{JHEP}
\bibliography{Bookallrefs}
\end{document}